\documentclass[useAMS,usenatbib]{mn2e}

\usepackage{color}
\usepackage{colortbl}
\usepackage{multirow}
\usepackage{dcolumn}
\usepackage{amsmath}
\usepackage{amssymb}
\usepackage{graphicx}
\usepackage{epsfig}
\usepackage{changebar}
\usepackage{natbib}
\usepackage{multirow}
\usepackage{subfigure}
\usepackage{txfonts}
\usepackage{relsize}

\usepackage{aalongtable}
\usepackage{longtable,lscape}
\usepackage[figuresright]{rotating}

\newcolumntype{d}[1]{D{.}{\cdot}{#1}}

\newcolumntype{.}{D{.}{.}{-1}}

\newcommand{\lsun}{L$_\odot$}

\newcommand{\mum}{$\mu$m}

\newcommand{\kms}{km~s$^{-1}$}

\newcommand{\tastar}{$T_{\mathrm{A}}^{*}$}
\newcommand{\nh}{NH$_3$}

\newcommand{\Tk}{$T_{\mathrm{k}}$}
\newcommand{\Tex}{$T_{\mathrm{ex}}$}
\newcommand{\Tr}{$T_{\mathrm{r}}$}
\newcommand{\Vlsr}{$V_{\mathrm{LSR}}$}

\title[Ammonia observations of bright-rimmed clouds]{Ammonia observations of bright-rimmed clouds: establishing a sample of triggered protostars}
\author[L. K. Morgan et al.]{L. K. Morgan$^{1}$\thanks{E-mail:
lkm@astro.livjm.ac.uk}, C. C .Figura$^{2}$, J. S. Urquhart$^{3}$ and M. A. Thompson$^{4}$\\
$^{1}$Astrophysics Research Institute, Liverpool John Moores University, Twelve Quays House, Egerton Wharf, Birkenhead CH41 1LD \\ 
$^{2}$Wartburg College, 100 Wartburg Blvd. Waverley, IA 50677, USA\\
$^{3}$Australia Telescope National Facility, CSIRO, Sydney, NSW 2052, Australia\\
$^{4}$Centre for Astrophysics Research, Science and Technology Research Institute,\\
University of Hertfordshire, College Lane, Hatfield AL10 9AB}
\begin{document}

\date{Accepted ??. Received ??; in original form ??}

\pagerange{\pageref{firstpage}--\pageref{lastpage}} \pubyear{2010}

\maketitle

\label{firstpage}

\begin{abstract}
We observed 42 molecular condensations within previously identified bright-rimmed clouds in the ammonia rotational inversion lines NH$_{3}$(1,1), (2,2), (3,3) and (4,4) using the Green Bank Telescope in Green Bank, West Virginia. Using the relative peaks of the ammonia lines and their hyperfine satellites we have determined important parameters of these clouds, including rotational temperatures and column densities. \\
These observations confirm the presence of dense gas towards IRAS point sources detected at submillimetre wavelengths. Derived physical properties allow us to refine the sample of bright-rimmed clouds
into those likely to be sites of star formation, triggered via the process of radiatively-driven implosion.
An investigation of the physical properties of our sources show that triggered sources are host to greater turbulent velocity dispersions, likely indicative of shock motions within the cloud material. These may be attributed to the passage of triggered shocks or simply the association of outflow activity with the sources.

In all, we have refined the \citet{Sugitani1991} catalogue to 15 clouds which are clearly star-forming and influenced by external photoionisation-induced shocks. These sources may be said, with high confidence, to represent the best examples of triggering within bright-rimmed clouds.
\end{abstract}


\section{Introduction}
\label{sec:introduction}
Bright-rimmed clouds (BRCs) are small molecular clouds associated with old ($>$1 Myr) H\textsc{II} regions and are potential examples of the radiatively driven implosion (RDI) mode of triggered star formation. In the RDI model the UV flux of associated OB star(s) powering the H\textsc{II} regions ionizes the external layers of small clouds. The ionisation of the cloud's surface leads to the formation of an ionized boundary layer (IBL), which subsequently expands due to the increased pressure of the ionised gas. The IBL expands into the intercloud medium and an ionisation front preceded by a shock in the neutral gas propagates into the cloud \citep{Lefloch1994,White1997}. The associated pressure often sweeps the molecular material of the cloud into a cometary morphology with a dense core located at the `head' of the cometary globule.

If the IBL has a pressure greater than, or equal to, the turbulent and thermal pressure of the cloud then photo-ionisation shocks and a D-critical ionisation front propagate into the cloud's interior. These shocks are driven into the clouds; compressing and heating the molecular gas, leading to the formation of dense cores and potentially causing their collapse \citep{Bertoldi1989,Lefloch1994}. This mechanism could be responsible for triggering the formation of several hundred stars for each HII region, possibly accounting for as much as 10-15\% of the low to intermediate mass stars (\citealt{Sugitani1991}). The morphologies and orientations with respect to nearby OB stars and optically bright rims identify BRCs as potential sites of triggered star formation. Moreover, their relative isolation and simple geometry make BRCs an ideal laboratory in which to study this mode of triggered star formation.

Sugitani et al. (1991, 1994) created a catalogue of 89 BRCs found to be associated with IRAS point sources that have IR colours consistent with being protostellar in nature. There have been a number of recent studies of small sub-samples (\citealt{devries2002,urquhart2004,urquhart2006a}) and individual sources (\citealt{Thompson2004a,urquhart2007d}) taken from this catalogue. All of these have found evidence consistent with the hypothesis that the star formation within the observed sources has been triggered; however, this tells us little about the more global aspects of this mode of triggered star formation.

We have therefore designed a multi-wavelength programme of observations to investigate the current level of star formation taking place within the whole sample of clouds, and to ascertain whether the interaction with the HII region has been a contributing factor in that formation. The programme includes the use of radio continuum observations (\citealt{Thompson2004b}) and NVSS archival data (\citealt{Morgan2004}) to trace the ionised gas associated with the IBL, CO molecular line observations  to probe the kinematics of the protostellar cores (\citealt{urquhart2009,Morgan2009}), and submillimetre observations to trace the dust associated with the protostellar envelopes and obtain core temperatures and masses (\citealt{Morgan2008}). In addition to our programme of observations we have obtained archival near- to far-infrared imaging and photometry (2MASS, MSX, IRAS, Spitzer IRAC and MIPS) in an effort to build up a detailed picture of the structure within BRCs and a census of the current state of star formation within them.

\begin{figure}
\begin{center} 
\includegraphics*[width=0.45\textwidth]{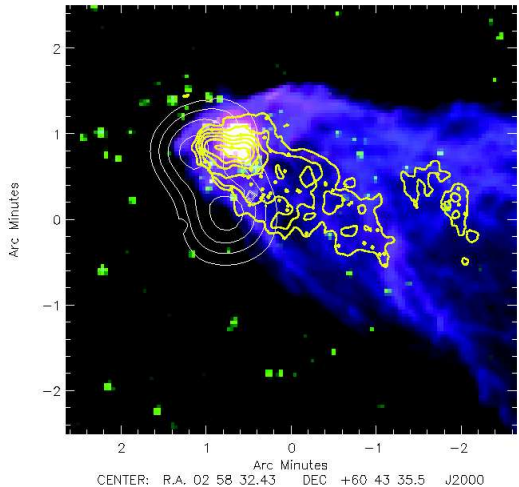}
\caption{Multi-wavelength composite image of SFO 13, Spitzer IRAC bands at 4.5 and 8.0~\mum\ are coloured green and blue respectively, a Spitzer MIPS 24~\mum\ image is incorporated in red. 21~cm NVSS contours are overlaid in white, beginning at a level of two times the r.m.s. noise ($\sigma$) of the radio image and increasing in units of 1~$\sigma$. SCUBA 850~\mum\ contours are overlaid in yellow, beginning at a level of 6~$\sigma$ and increasing in units of 3~$\sigma$.}
\label{fig:SFO_13_3col}
\end{center}
\end{figure}

A key element of our campaign is to answer the question of whether the observed star formation within BRCs has been triggered by the action of the OB star or is pre-existing and is simply being unveiled by the dispersal of the cloud. In order to address this question, we must first confirm the existence of protostellar cores within BRCs and characterise their properties. The sample of 44 BRCs catalogued by \citet{Sugitani1991} forms the basis for our campaign in the northern hemisphere. A high proportion of these clouds have been found to be undergoing a strong interaction with their surrounding HII region \citep{Morgan2004} and are associated with embedded protostellar cores \citep{Morgan2008}. Furthermore, a significant proportion of these sources have been found to be associated with water masers \citep{Cesaroni1988,Henning1992,Wouterloot1993,Xiang1995,Claussen1996,Valdettaro2005,Valdettaro2008} and kinematic line profiles indicative of molecular outflows (see \citealt{Morgan2009}; and references therein). 

In Fig.~\ref{fig:SFO_13_3col} we present an example of a BRC that appears to be a good candidate for triggered star formation. In this figure we have combined two Spitzer IRAC bands and a MIPS image to create a three colour image of SFO~13 (IRAC 4.5 and 8.0~\mum\ and MIPS 24~\mum\ bands are coloured in green, blue and red respectively). We have over-plotted contours of the NVSS radio continuum emission (white) and the SCUBA 850~\mum\ emission (yellow) which trace ionised gas and thermal dust emission respectively.

The multiple wavelength data that comprise Fig.~\ref{fig:SFO_13_3col} show many different features of this small molecular cloud on the edge of an HII region. The 4.5~\mum\ emission in green likely indicates shocked H$_2$ gas (e.g. \citealt{Teixeira2008}), possibly associated with the jets/outflows of young stars.
  Probing the interface between the neutral molecular material of the cloud itself and the ionized gas of the HII region, the 8.0~\mum\ emission, traced in blue, indicates the presence of polycyclic aromatic hydrocarbons (PAHs), complex molecules which absorb far UV photons and re-emit in the IR regime \citep{Leger1984}. Dominant lines of these PAHs are at 7.7 and 8.6~\mum\ which are within the passband of the Spitzer 8.0~\mum\ band. This reveals the photon-dominated region (PDR), which is preceded by the ionization front at the head of the cloud. It is here that the cloud is actively supporting the presence of an established IBL (as described in e.g. \citet{Lefloch1994}), emitting free-free emission which is traced by the 20~cm NVSS contours in white.
  The 24~\mum\ emission, shown in red in Fig.~\ref{fig:SFO_13_3col}, traces small grains of heated dust (e.g. \citealt{Koenig2008}). This emission peaks in a centrally condensed region at the head of the cloud, coincident with the peak of 850~\mum\ contours in yellow, which show submillimetre-emission associated with warm dust within the cloud and surrounding the protostellar core at the head of the cloud.

In order to explore the possible origins of these protostars, it is essential to explore the physical properties of the protostellar cores themselves and their environs. In this paper we present a set of ammonia observations made towards 42 BRCs in an effort to confirm the presence of protostellar cores and to probe their physical conditions. Analysis of the line properties of ammonia rotational transitions at $
\sim$ 23 GHz allows for determinations of kinetic temperature, column density and velocity dispersion. These properties will allow the definitive identification of protostellar cores within our sample and allow us to determine the global properties of molecular material within BRCs.

The structure of the paper is as follows: in Sect.~2 we describe the observational set up, data reduction procedures and derive the physical parameters. In Sect.~3 we present the observational results, in Sect.~4 we discuss these results with respect to triggered star formation. In Sect.~5 we present a summary and highlight our main findings.

\section{GBT observations}
\subsection{Observational set up}

Observations of the \nh\ (1,1) and (2,2), (3,3) and (4,4) rotational transitions were made over three sessions from the 11$^{th}$ of February to the 20$^{th}$ of March 2005 on the Green Bank telescope operated by the National Radio Astronomy Observatory\footnote{The National Radio Astronomy Observatory is a facility of the National Science Foundation operated under cooperative agreement by Associated Universities, Inc.}.

Observations were performed in `nod' mode in order to remove sky contributions and a noise diode was observed in switching mode throughout the observations in order to achieve absolute flux calibration on the \tastar\ scale to an accuracy of $\sim$10\%. Atmospheric opacity values (at zenith) were determined from local weather models, collated in the archive maintained by R.Maddalena. Two spectrometer set ups were used during the observations due to hardware issues at the time of observation, which resulted in the first two sessions being observed with poorer spectral resolution than the third. Spectrometer problems meant that only two spectral windows could be simultaneously observed in the first two sessions. The observed bandwidth was thus increased so that the (1,1) and (2,2) transitions could be observed in the same 50~MHz ($\sim$650 \kms) bandpass with the (3,3) transition in another spectral window (the (4,4) transition was not observed in these sessions). The first two sessions were observed with a spectral channel width of 6.1 kHz ($\sim$0.08 \kms). By the time of the third observing session, hardware problems had been resolved and all four transitions were observed at the optimum spectral resolution of $\sim$3 kHz ($\sim$0.04 \kms) in 12.5 MHz bandwidth ($\sim$160 \kms) spectral windows. All transitions were observed simultaneously using the configurations presented in Table~\ref{tbl:Obssetup}.

\begin{table}
\caption{Observational setup}

\begin{tabular}{cccc}
\hline
\hline

{Date}	& Centre Frequencies (GHz)   & RMS$^{a}$ (K)	    & T$_{\mathrm{sys}}^{\mathrm{a}}$ (K) \\
\hline
11$^{th}$ Feb 2005 &  23.710, 23.877                 & 0.05 & 41.0\\
12$^{th}$ Feb 2005 &  23.710, 23.877                 & 0.04 & 31.2\\
20$^{th}$ Mar 2005 &  23.695, 23.724, 23.871, 24.140 & 0.07 & 38.4\\
\hline
\end{tabular}	
$^{a}$-RMS and Tsys values represent mean values of measurements taken from single two minute integrations at intervals over the entire observing period. 
\label{tbl:Obssetup}\\
\end{table}	

Weather conditions were stable across all three observing sessions, typical pointing offsets were $\sim$5\arcsec\ in azimuth and elevation with a beamsize of $\sim$29\arcsec. Sources were observed for periods ranging between four and 66 minutes, dependent upon the signal strength of observed lines. The median observing time was 20 minutes per object.

\begin{table}

\caption{Summary of source names, observed positions and observation dates.}
\begin{tabular}{lcccc}
\hline
\hline
{}	& IRAS   & {$\alpha$} & {$\delta$} & Obs. Date \\
{Source} & Source & {(J2000)}   & {(J2000)}  & \\
\hline
SFO 01     & 23568+6706    & 23:59:32.5	   &  67:24:06   &  20/3/2005\\
SFO 04     & 00560+6037    & 00:59:01.6	   &  60:53:27   &  11/2/2005\\
SFO 05     & 02252+6120    & 02:29:01.6	   &  61:33:33   &  11/2/2005\\
SFO 06     & 02309+6034    & 02:34:45.1	   &  60:47:49   &  20/3/2005\\
SFO 07     & 02310+6133    & 02:34:48.0	   &  61:46:36   &  11/2/2005\\
	   & $\cdots$      & $\cdots$      &  $\cdots$   &  20/3/2005\\
SFO 08     & 02318+6106    & 02:35:37.5	   &  61:19:37   &  20/3/2005\\
SFO 09     & 02326+6110    & 02:36:27.6	   &  61:24:01   &  11/2/2005\\
SFO 10     & 02443+6012    & 02:48:12.2	   &  60:24:36   &  11/2/2005\\
SFO 11     & 02476+5950    & 02:51:33.7	   &  60:03:54   &  20/3/2005\\
SFO 11E    & $\cdots$      & 02:51:57.1	   &  60:06:55   &  20/3/2005\\
SFO 11NE   & $\cdots$      & 02:52:14.9	   &  60:02:54   &  20/3/2005\\
SFO 12     & 02511+6023    & 02:55:01.1	   &  60:35:45   &  12/2/2005\\
SFO 13     & 02570+6028    & 03:00:55.4	   &  60:40:18   &  11/2/2005\\
SFO 14     & 02575+6017    & 03:01:31.5	   &  60:29:20   &  11/2/2005\\
	   & $\cdots$      & $\cdots$      &  $\cdots$   &  20/3/2005\\
SFO 15     & 05202+3309    & 05:23:28.3	   &  33:11:49   &  11/2/2005\\
SFO 16     & 05173$-$0555  & 05:19:48.1	   & $-$05:52:04 &  20/3/2005\\
SFO 17     & 05286+1203    & 05:31:28.1	   &  12:05:21   &  20/3/2005\\
SFO 18     & 05417+0907    & 05:44:29.8	   &  09:08:55   &  12/2/2005\\
	   & $\cdots$      & $\cdots$      &  $\cdots$   &  20/3/2005\\
SFO 19     & 05320$-$0300  & 05:34:30.6    & $-$02:58:15 &  20/3/2005\\
SFO 20     & 05355$-$0146  & 05:38:04.8    & $-$01:45:10 &  20/3/2005\\
SFO 21     & 05371$-$0338  & 05:39:41.2    & $-$03:37:12 &  20/3/2005\\
SFO 22     & 05359$-$0515  & 05:38:26.2	   & $-$05:14:06 &  20/3/2005\\
SFO 23	   & 06199+2311    & 06:22:58.2    &  23:10:11   &  12/2/2005\\
	   & $\cdots$      & $\cdots$      &  $\cdots$   &  20/3/2005\\
SFO 24	   & 06322+0427    & 06:34:52.7    &  04:25:21   &  12/2/2005\\
SFO 25	   & 06382+1017    & 06:41:03.3    &  10:15:07   &  12/2/2005\\
SFO 26	   & 07014$-$1141  & 07:03:47.2    & $-$11:45:46 &  20/3/2005\\
SFO 28	   & 07023$-$1017  & 07:04:43.6    & $-$10:21:58 &  20/3/2005\\
SFO 29	   & 07025$-$1204  & 07:04:52.5    & $-$12:09:25 &  20/3/2005\\
SFO 30	   & 18159$-$1346  & 18:18:46.6    & $-$13:44:26 &  12/2/2005\\
SFO 31	   & 20489+4410    & 20:50:42.9    &  44:21:56   &  12/2/2005\\
SFO 32	   & 21308+5710    & 21:32:29.5    &  57:24:33   &  20/3/2005\\
SFO 33	   & 21316+5716    & 21:33:13.7    &  57:30:03   &  12/2/2005\\
 	   & $\cdots$      & $\cdots$      &  $\cdots$   &  20/3/2005\\
SFO 34	   & 21320+5750    & 21:33:32.2    &  58:03:34   &  12/2/2005\\
	   & $\cdots$      & $\cdots$      &  $\cdots$   &  20/3/2005\\
SFO 35	   & 21345+5818    & 21:36:05.5    &  58:31:38   &  20/3/2005\\
SFO 36	   & 21346+5714    & 21:36:07.0    &  57:26:40   &  12/2/2005\\
SFO 37	   & 21388+5622    & 21:40:28.9    &  56:35:54   &  12/2/2005\\
SFO 38	   & 21391+5802    & 21:40:41.8    &  58:16:12   &  12/2/2005\\
 	   & $\cdots$      & $\cdots$      &  $\cdots$   &  20/3/2005\\
SFO 39	   & 21445+5712    & 21:46:06.7    &  57:26:38   &  12/2/2005\\
SFO 41	   & 21448+5704    & 21:46:29.3    &  57:18:41   &  20/3/2005\\
SFO 42	   & 21450+5658    & 21:46:36.8    &  57:12:25   &  20/3/2005\\
SFO 43	   & 22458+5746    & 22:47:50.1    &  58:02:47   &  12/2/2005\\
SFO 44	   & 22272+6358A   & 22:28:51.5    &  64:13:37   &  12/2/2005\\

\hline
\end{tabular}
\label{tbl:Sources}
\end{table}

In total, observations were made towards 42 BRCs, including SFO 11E and 11NE, identified by \citet{Thompson2004a} but not included in the original SFO catalogue. Of the 44 sources listed in the northern SFO catalogue, four (SFO 2, 3, 27 and 40) were not observed due to time and hour angle limitations. The data of two sources (SFO 11 and 32) were later found to be corrupt, resulting in a total source sample size of 40 out of a potential 46. In Table~\ref{tbl:Sources} we present a summary of the BRCs observed and the pointing centres used for the observations. Source positions were largely determined from examination of SCUBA observed dust emission maps, presented in \citet{Morgan2008} and \citet{Thompson2004a}. For sources not observed in those works, the coordinates of the IRAS source associated with the relevant object were used.

\begin{table}
\begin{center}
\caption{\label{tbl:non-detections}Sources not detected in our survey with RMS values of the observations.}
\begin{tabular}{cc}
\hline
\hline
{Source} & RMS (mK) \\
\hline
SFO 04	& 14.8  \\
SFO 06	&  9.6  \\
SFO 08	& 22.3  \\
SFO 10	& 12.8  \\
SFO 15	& 15.3  \\
SFO 19	& 33.8  \\
SFO 21	& 20.6  \\
SFO 22	& 19.4  \\
SFO 28	& 19.6  \\
\hline
\end{tabular}	
\end{center}
\end{table}	

\subsection{Data reduction}

Data were reduced using the GBTIDL\footnote{http://gbtidl.nrao.edu/} data analysis package. Bad scans were removed and channels outside the region of interest were discarded. In reduction, use was made of the GBTIDL procedural ability to smooth the `off'. This involves the smoothing of the reference spectra before the usual spectral subtraction and calibration are performed. This results in higher signal to noise ratios for the final calibrated spectrum with no loss of spectral resolution. After careful testing of this procedure, the default smoothing ratio of 16 channels was used for all sources except SFO 31. This source showed emission in the reference spectrum which resulted in negative emission seen in the final calibrated spectrum. Through extensive smoothing of the reference spectrum (by 250 channels) this `off' emission was greatly minimised. Temperature scale corrections for atmospheric opacity were made using the zenith values provided from local weather models.

As data from different observing sessions were taken in different configurations, in certain cases, individual objects may have been observed in multiple configurations.  Spectra for these objects were smoothed to a common resolution before combination. All (1,1) and (2,2) spectra were ultimately smoothed to a resolution of 7.4 kHz ($\sim$0.09 \kms) while (3,3) and (4,4) spectra were smoothed to a resolution of 50.2 kHz (due to the expected lower signal to noise ratio). A polynomial baseline was subtracted from all spectra for normalisation. Several sources were detected in the (3,3) and (4,4) transitions. However, analyses of these data contribute little to the scope of this paper and so the spectra and related parameters are presented in appendix \ref{ap}.

The corrected antenna temperature \tastar, hyperfine linewidth $\Delta v$, and central velocity \Vlsr\ values for each source in the (1,1) and (2,2) inversion transitions were determined by fitting with the NH3(1,1) METHOD process in the CLASS data analysis package\footnote{http://www.iram.fr/IRAMFR/GILDAS}. In some cases (SFO 09, 11E, 17, 26, 35 and 42) the main quadrupole of ammonia was detected but the hyperfines were not seen at a level required for CLASS to fit them. For these sources, and the (3,3) and (4,4) transitions presented in Appendix~\ref{ap}, the main quadrupoles were fitted by a single Gaussian using the `fitgauss' procedure in GBTIDL.

\begin{table*}
\caption{\label{tbl:Detections} Observationally measured and derived physical source parameters.}
\footnotesize
\begin{tabular}{lcccccccrrrcrc} 
\hline\hline
{}	 & \multicolumn{2}{c}{ NH$_3$ (1,1)}		&	       & \multicolumn{2}{c}{NH$_3$ (2,2)} 			   & {} 	& {} \\
\cline{2-3} \cline{5-6}
{}	 & $T_{\mathrm{A}}^{*}$   & {$\Delta v$ }	 && $T_{\mathrm{A}}^{*}$ & {$\Delta v$}  &  {\Vlsr} & {$\tau$} & {$T_{\mathrm{ex}}$} & {$\eta_\mathrm{f}$} & {$T_{\mathrm{k}}$} & {$N_{\mathrm{NH}_3}$} & $T_{\rm{dust}}$ & {$N_{\mathrm{H}_2}^\mathrm{b}$}\\
{Source} & (K) & {(km~s$^{-1}$)}  & & (K)& {(km~s$^{-1}$)} & {(km~s$^{-1}$)} &{} & (K) &  & (K) & {($10^{13} $cm$^{-2}$)} & (K) & {($10^{21} $cm$^{-2}$)}\\
\hline
 \multicolumn{13}{c}{\textbf{Triggered Sources}} \\
\hline
SFO 01			& 0.52$\pm$0.01 & 0.58 & & 0.21$\pm$0.01     & 0.78     &  -13.5 & 1.44$\pm$0.184  & 2.9$\pm$0.1   & 0.07    	   & 19.2$\pm$0.5  & 26.6$\pm$1.8  & 26       & 11.0	 \\
SFO 05			& 0.37$\pm$0.01 & 1.27 & & 0.18$\pm$0.01     & 1.48     &  -51.5 & 0.20$\pm$0.027  & 4.2$\pm$0.2   & 0.21    	   & 23.8$\pm$1.3  & 9.6$\pm$0.8   & 24       & 25.3	 \\
SFO 07			& 0.58$\pm$0.01 & 1.54 & & 0.22$\pm$0.01     & 1.48     &  -39.9 & 0.64$\pm$0.006  & 3.5$\pm$0.0   & 0.14    	   & 19.8$\pm$0.1  & 32.0$\pm$0.2  & 21       & 26.5	    \\
SFO 11E$^{\mathrm{a}}$	& 0.10$\pm$0.02 & 1.21 & & $\cdots$	     & $\cdots$ &  -40.3 & $\cdots$	   & $\cdots$	   & $\cdots$	   & $\cdots$	   & $\cdots$	   & $\cdots$ & $\cdots$  \\
SFO 12  		& 0.33$\pm$0.01 & 1.14 & & 0.12$\pm$0.01     & 1.34     &  -38.3 & 0.32$\pm$0.258  & 3.5$\pm$2.3   & 0.15    	   & 19.9$\pm$13.0 & 11.9$\pm$10.0 & 21       & 16.5	 \\
SFO 13  		& 0.30$\pm$0.01 & 1.08 & & 0.15$\pm$0.01     & 1.12     &  -38.9 & 0.42$\pm$0.070  & 3.2$\pm$0.2   & 0.09    	   & 23.8$\pm$1.3  & 17.1$\pm$1.7  & 23       & 17.5	  \\
SFO 14  		& 1.09$\pm$0.01 & 1.62 & & 0.65$\pm$0.01     & 1.83     &  -38.2 & 1.08$\pm$0.042  & 3.7$\pm$0.0   & 0.13    	   & 25.6$\pm$0.3  & 71.1$\pm$1.6  & 27       & 47.9	    \\
SFO 25  		& 0.45$\pm$0.01 & 1.93 & & 0.14$\pm$0.01     & 2.13     &   6.77 & 1.42$\pm$0.095  & 2.8$\pm$0.0   & 0.07    	   & 16.8$\pm$0.2  & 81.5$\pm$2.6  & 21       & 29.3	    \\
SFO 30  		& 0.88$\pm$0.01 & 1.43 & & 0.52$\pm$0.01     & 1.69     &   19.7 & 3.00$\pm$0.046  & 3.0$\pm$0.0   & 0.07    	   & 21.8$\pm$0.1  & 149.9$\pm$1.3 & 34       & 18.9	    \\
SFO 31  		& 0.30$\pm$0.01 & 1.13 & & 0.09$\pm$0.01     & 1.62     &  -2.47 & 0.68$\pm$0.168  & 2.9$\pm$0.2   & 0.08    	   & 17.5$\pm$1.1  & 23.3$\pm$3.1  & 27       & 7.6	    \\
SFO 35$^{\mathrm{a}}$  	& 0.10$\pm$0.02 & 0.88 & & 0.07$\pm$0.02     & 0.56     &  -4.17 & $\cdots$	   & $\cdots$	   & $\cdots$	   & $\cdots$	   & $\cdots$	   & $\cdots$ & $\cdots$  \\
SFO 36  		& 0.38$\pm$0.01 & 1.15 & & 0.16$\pm$0.01     & 1.25     &  -8.36 & 0.75$\pm$0.119  & 3.0$\pm$0.1   & 0.08    	   & 20.8$\pm$0.9  & 29.0$\pm$2.6  & $\cdots$ & $\cdots$  \\
SFO 37  		& 0.23$\pm$0.01 & 0.83 & & 0.06$\pm$0.01     & 0.93     &  0.611 & 1.95$\pm$0.252  & 2.5$\pm$0.0   & 0.03    	   & 14.9$\pm$0.2  & 46.6$\pm$2.5  & 20       & 22.5	    \\
SFO 38  		& 1.10$\pm$0.01 & 1.86 & & 0.48$\pm$0.01     & 1.94     &  0.181 & 0.61$\pm$0.001  & 4.3$\pm$0.0   & 0.25    	   & 21.5$\pm$0.1  & 39.2$\pm$0.1  & $\cdots$ & $\cdots$  \\
SFO 41  		& 0.12$\pm$0.01 & 1.26 & & 0.06$\pm$0.01     & 1.22     &  -3.33 & 0.36$\pm$0.611  & 2.7$\pm$0.3   & 0.04    	   & 23.9$\pm$2.5  & 17.2$\pm$15.1 & 24       & 4.7	   \\
SFO 42$^{\mathrm{a}}$  	& 0.10$\pm$0.02 & 0.87 & & $\cdots$	     & $\cdots$ &  -2.61 & $\cdots$	   & $\cdots$	   & $\cdots$	   & $\cdots$	   & $\cdots$	   & 22       & $\cdots$  \\
SFO 43  		& 0.22$\pm$0.01 & 1.12 & & 0.12$\pm$0.01     & 1.16     &  -42.1 & 0.51$\pm$0.214  & 2.9$\pm$0.4   & 0.05    	   & 25.1$\pm$3.2  & 22.8$\pm$6.0  & 26       & 13.2	  \\
SFO 44  		& 0.79$\pm$0.01 & 1.10 & & 0.34$\pm$0.01     & 1.31     &  -10.1 & 1.22$\pm$0.085  & 3.3$\pm$0.1   & 0.11    	   & 20.3$\pm$0.4  & 44.4$\pm$1.7  & $\cdots$ & $\cdots$  \\
\hline
 \multicolumn{13}{c}{\textbf{Non-Triggered Sources}} \\ 							      
\hline														      
SFO 09$^{\mathrm{a}}$   & 0.07$\pm$0.01 & 1.52 & & $\cdots$	     & $\cdots$ &  -48.3 & $\cdots$	   & $\cdots$	   & $\cdots$	   & $\cdots$	   & $\cdots$	   & $\cdots$ & $\cdots$ \\
SFO 11NE		& 0.39$\pm$0.01 & 0.98 & & 0.19$\pm$0.01     & 0.91     &  -35.4 & 0.32$\pm$0.158  & 3.7$\pm$0.9   & 0.14    	   & 23.6$\pm$5.6  & 11.8$\pm$4.4  & $\cdots$ & $\cdots$ \\
SFO 16  		& 2.35$\pm$0.01 & 0.52 & & 0.42$\pm$0.01     & 0.57     &   7.85 & 3.37$\pm$0.034  & 3.8$\pm$0.0   & 0.35    	   & 11.9$\pm$0.0  & 51.3$\pm$0.2  & 17       & 31.8	  \\
SFO 17$^{\mathrm{a}}$  	& 0.24$\pm$0.02 & 0.56 & & $\cdots$	     & $\cdots$ &  10.6  & $\cdots$	   & $\cdots$	   & $\cdots$	   & $\cdots$	   & $\cdots$	   & $\cdots$ & $\cdots$ \\
SFO 18  		& 1.48$\pm$0.01 & 0.64 & & 0.28$\pm$0.01     & 0.87     &   11.3 & 3.55$\pm$0.079  & 3.3$\pm$0.0   & 0.22    	   & 12.0$\pm$0.0  & 66.4$\pm$0.4  & 18       & 23.6	  \\
SFO 20  		& 0.79$\pm$0.01 & 0.48 & & 0.14$\pm$0.01     & 0.58     &   12.8 & 1.86$\pm$0.174  & 3.1$\pm$0.1   & 0.14    	   & 12.9$\pm$0.2  & 25.6$\pm$0.9  & $\cdots$ & $\cdots$ \\
SFO 23  		& 0.52$\pm$0.01 & 0.78 & & 0.09$\pm$0.01     & 0.82     &  -5.43 & 1.68$\pm$0.091  & 2.9$\pm$0.0   & 0.10    	   & 13.0$\pm$0.1  & 37.6$\pm$0.8  & 19       & 7.0	 \\
SFO 24  		& 0.20$\pm$0.01 & 0.77 & & 0.07$\pm$0.01     & 1.14     &   8.75 & 0.88$\pm$0.547  & 2.6$\pm$0.5   & 0.04    	   & 18.6$\pm$3.3  & 21.2$\pm$8.1  & $\cdots$ & $\cdots$ \\
SFO 26$^{\mathrm{a}}$  	& 0.06$\pm$0.02 & 1.06 & & $\cdots$	     & $\cdots$ &   11.5 & $\cdots$	   & $\cdots$	   & $\cdots$	   & $\cdots$	   & $\cdots$	   & 24       & $\cdots$ \\
SFO 29  		& 0.26$\pm$0.01 & 1.27 & & 0.08$\pm$0.01     & 0.97     &   14.8 & 0.48$\pm$0.216  & 3.0$\pm$0.5   & 0.09    	   & 18.0$\pm$2.9  & 18.7$\pm$5.3  & 22       & 8.4     \\
SFO 33  		& 0.22$\pm$0.01 & 0.81 & & 0.13$\pm$0.01     & 0.85     &  -5.97 & 0.20$\pm$0.294  & 3.6$\pm$0.7   & 0.11    	   & 27.3$\pm$5.3  & 7.1$\pm$6.0   & 24       & 1.2     \\
SFO 34  		& 0.59$\pm$0.01 & 0.57 & & 0.10$\pm$0.01     & 0.88     &  -5.34 & 1.65$\pm$0.125  & 2.9$\pm$0.0   & 0.12    	   & 12.9$\pm$0.2  & 27.0$\pm$0.8  & 21       & 9.0     \\
SFO 39  		& 0.14$\pm$0.01 & 1.11 & & 0.06$\pm$0.01     & 1.31     &  -2.11 & 0.20$\pm$0.071  & 3.2$\pm$0.5   & 0.09    	   & 22.0$\pm$3.1  & 7.8$\pm$1.9   & 20.5     & 16.1	\\
\hline\\
\end{tabular}
\footnotesize
a-Optical depth, temperature and column density are not determinable for these sources as satellite (1,1) quadrupole hyperfines are too weak for detection and/or the (2,2) main line was not detected.
b-Determined from the observations of \citet{Morgan2008}.\\
\end{table*}	

\subsection{Deriving physical parameters}
\label{sec:parameters}
The optical depth ($\tau$) associated with ammonia emission in our sources may be determined through the ratio of main to satellite antenna temperatures of the (1,1) inversion transition \citep{Ho1983}:

\begin{equation}\label{eq:opticalDepth}
\frac{\Delta T_{\mathrm{A}}^{*} (J,K)_\mathrm{m}}{\Delta T_{\mathrm{A}}^{*} (J,K)_\mathrm{s}} = \frac {1-\mathrm{e}^{-\tau(J,K)_\mathrm{m}}}{1-\mathrm{e}^{-\mathrm{a}\tau(J,K)_\mathrm{m}}}
\end{equation}

\noindent where m and s subscripts indicate quantities associated with the main and satellite quadrupole lines, respectively, and a is the ratio of intensities of satellite to main lines (0.278 and 0.221 for inner and outer satellites, respectively). Optical depth values were determined through use of the NH3(1,1) METHOD process in the CLASS data analysis package.

Given the optical depth associated with the (1,1) transition, corrected antenna temperatures can be used to calculate the rotational temperature (\Tr) associated with the (2,2) and (1,1) transitions \citep{Ho1983}:
\begin{equation}
 T_\mathrm{r} = \frac{-T_0}{\ln \left\{\frac{-0.282}{\tau _{\mathrm{m} (1,1)}} \ln \left[1-\frac{\Delta T_{\mathrm{Am}(2,2)}^{*}}{\Delta T_{\mathrm{Am}(1,1)}^{*}}\left(1-\mathrm{e}^{-\tau _{\mathrm{m}(1,1)}} \right) \right] \right\}}
\end{equation}

\noindent where $T_0 = \frac{E_{(2,2)} - E_{(1,1)}}{\mathrm{k_B}} \approx 41.5$K.

For \Tk\ $< T_0$, a relationship between rotational and kinetic temperature may be calculated by consideration of (1,1), (2,2), and (2,1) states only \citep{Walmsley1983,Swift2005}:
\begin{equation}
T_\mathrm{r} = \frac{T_\mathrm{k}} {1+\frac{T_\mathrm{k}}{T_0}\ln\left[ 1+0.6\exp\left(-\frac{15.7}{T_\mathrm{k}}\right)\right]},
\end{equation}

\noindent although kinetic temperatures calculated by this method associated with \Tr\ $ \approx T_0$ may be overestimated. Calculated \Tk\ values range from 11.9 to 27.3K, suggesting that this approximation is reasonable for these source objects. 

In order to calculate the column density of our sources, it is necessary to calculate the excitation temperature of each source. This may be found via

\begin{equation}
T_{\mathrm{A}}^{*} = \eta_{\mathrm{mb}} \eta_{\mathrm{f}} [J(T_{\mathrm{ex}})-J(T_{\mathrm{bg}})][1-e^{-\tau_m}],
\end{equation}

where the quantity \mbox{$J_\nu(T)$} is defined as \mbox{$\frac{\mathrm{h} \nu/\mathrm{k}}{(e^{\mathrm{h} \nu/\mathrm{k} T}-1)}$}, $\eta_{\mathrm{mb}}$ is the main beam efficiency of the GBT (0.89\footnote{from `The Proposer's Guide for the Green Bank Telescope'}) and $\eta_{\mathrm{f}}$ is the filling factor of the source in the relevant observation. The filling factor of each source is somewhat hard to determine; values of \Tex\ derived assuming a value of $\eta_{\mathrm{f}}$ equal to unity are presented in Table~\ref{tbl:Detections} along with values of $\eta_{\mathrm{f}}$ determined by assuming LTE in our sources and letting \Tex = \Tk. This assumption is supported by the rough equivalence of \Tr, \Tk\ and the values of $T_{\mathrm{dust}}$ determined by \citet{Morgan2008}. A typical value of a filling factor in a similar study was determined by \citet{Rosolowsky2008} in their study of dense cores in Perseus to be $\sim$0.3. Given the relatively large distances of our sources compared to Perseus, our values of $\eta_{\mathrm{f}}$ appear reasonable, ranging from 0.03 to 0.35 with a mean of 0.12.

If excitation conditions are homogeneous along the beam and all hyperfine lines have the same excitation temperature, then the column densities at a given (J,K=J) transition can be written as \citep{Rosolowsky2008}
\begin{equation}\label{eq:coldensity}
 N(1,1) = \frac{4 \pi ^{3/2} \nu _0 ^3}{\sqrt{\ln 2} \mathrm{c}^3 \mathrm{A}}  \left[1-\mathrm{e}^{-\frac{\mathrm{h} \nu _0}{\mathrm{k} T_{\mathrm{ex}}}} \right] ^{-1} \Delta v \tau,
\end{equation}
where A is the Einstein spontaneous emission coefficient. Our determination of the filling factors associated with each source indicate that the \Tex\ of our sources may be significantly underestimated. This is likely due to beam dilution or clumping within the beam of our observations, in order to account for this effect we have assumed LTE in our sources and set \Tex = \Tk\ in our determinations of column density.

Overall column density is obtained from N(1,1) and/or N(2,2) and the partition function: 
\begin{equation}
 \frac{N}{Z} = \frac{N(i,i)}{Z(i)} \longrightarrow N = N(i,i) \frac{Z}{Z(i)} ; Z = \sum_i Z_i
\end{equation}

where the partition function,
\begin{equation}
Z_\mathrm{J} = (2\mathrm{J}+1) \mathrm{~S(J) ~exp}\frac{-\mathrm{h~[B~J~(J+1)+(C-B)~J^2}]}{\mathrm{k} T_\mathrm{k}}.
\end{equation}

The values of the rotational constants B and C are 298117 and 186726 MHz respectively and the function S(J) is 2 for J=3,6,9,... and 1 for all other J.

Values for optical depth, \Tex\ and column density are listed for each source in Table~\ref{tbl:Detections}.

\section{Results}

We detected ammonia (1,1) emission towards 31 of the 40 successfully observed sources.  We present spectral plots of both the NH$_3$~(1,1) and (2,2) lines towards all of these in Fig.~\ref{img:Spectra}. In these plots we show the baseline corrected data (black histogram) and the model fits (red line) to the data. The measured properties of the main NH$_3$~(1,1) and NH$_3$~(2,2) lines are given in Table~\ref{tbl:Detections}. In this table we have separated the sample into those clouds considered to be good triggered star formation candidates and those in which the star formation is unlikely to have been induced. The determination of which sources are triggered candidates and which not is described in \citet{Morgan2009} and is based upon the detection of a strong IBL in radio and PDR in IR emission.

The hyperfine structure of the NH$_3$~(1,1) transition is clearly resolved towards 25 BRCs, as is the main quadrupole of the NH$_3$~(2,2) transition, allowing estimates of the optical depth, kinematic and excitation temperatures and column densities to be obtained (Several sources also show hyperfine structure in the NH$_3$~(2,2) transition 07, 14, 16, 18, 30, 38, 44). Towards the remaining six sources only the main NH$_3$~(1,1) quadrupole line is visible above the noise and of these the NH$_3$~(2,2) transition is only visible in three. Our results can thus be separated into three groups, strong detections, weak detections and non-detections, each of which contain 25, 6 and 9 clouds, respectively. Non-detections with respective RMS noise levels are listed in Table~\ref{tbl:non-detections}.

The overall detection rate is 75\%, with both transitions being strongly detected towards roughly $\sim$60\% of the sources. Ammonia is a high density tracer needing a critical density of \mbox{$n_{crit}$ $=$10$^4$ cm$^{-3}$} before becoming thermally excited \citep{Swade1989}. Due to the requirement for this density before ammonia becomes thermally excited, ammonia is better able to probe the properties of the dense protostellar core surrounding the accreting protostar than CO or dust continuum emission, which tend to probe the whole column of the gas along the line of sight.

As previously mentioned, the ammonia observations were centred on the position of either the dust cores revealed by the SCUBA observations (\citealt{Morgan2008}), or towards the IRAS point source thought to be protostellar in nature. It is somewhat surprising then to find 25\% of sources observed resulted in non-detections, which would seem to rule out the presence of a dense core and subsequently call into question the association of these clouds with protostars. Moreover, the weaker emission detected towards another six clouds may indicate that conditions are not conducive for star formation. We will investigate these weak and non-detections in more detail in Sect.~\ref{sect:ammonia_detections}.  

\begin{figure*}
\begin{center}
\includegraphics*[width=0.49\textwidth, trim= 50 0 20 0]{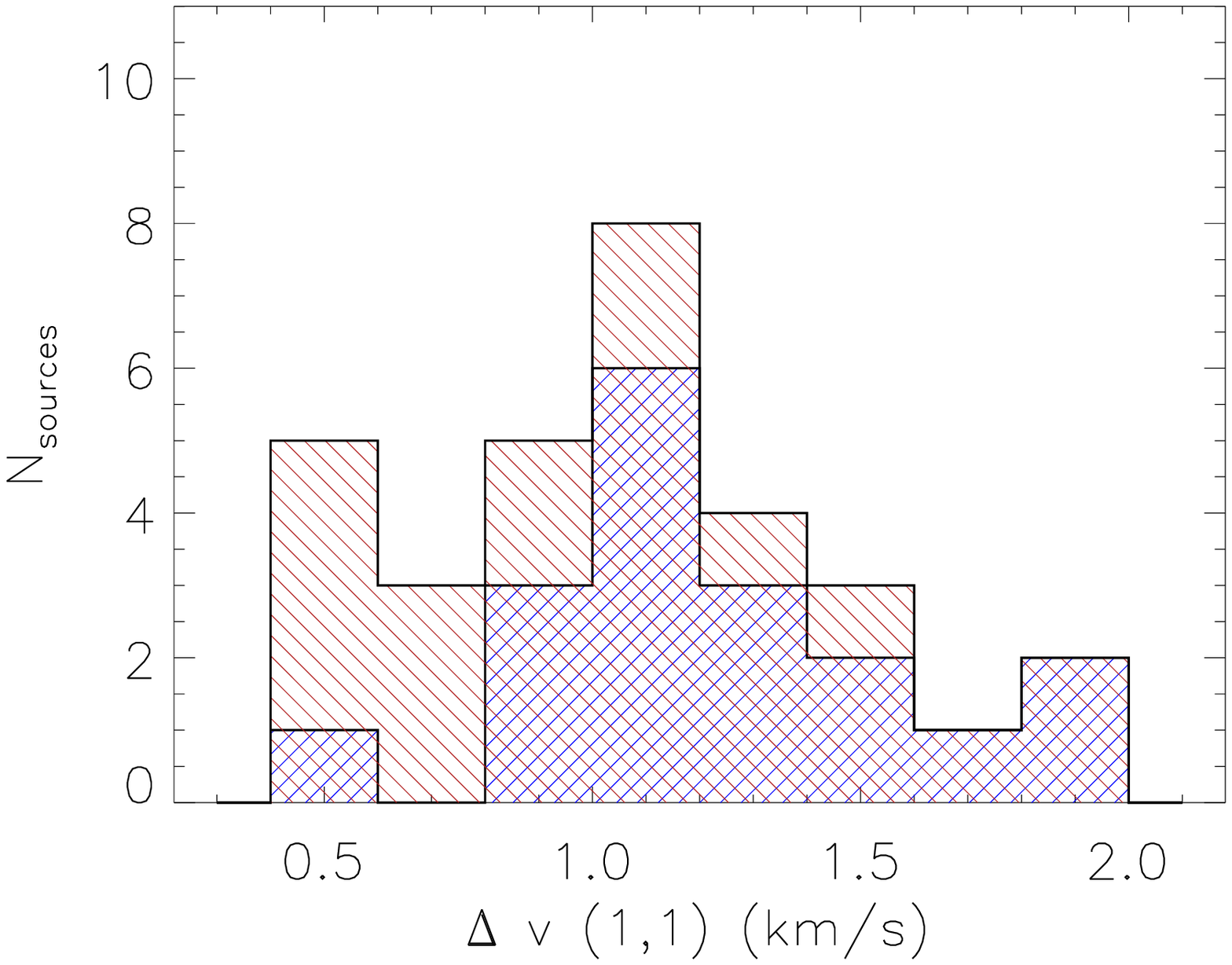}
\includegraphics*[width=0.49\textwidth, trim= 50 0 20 0]{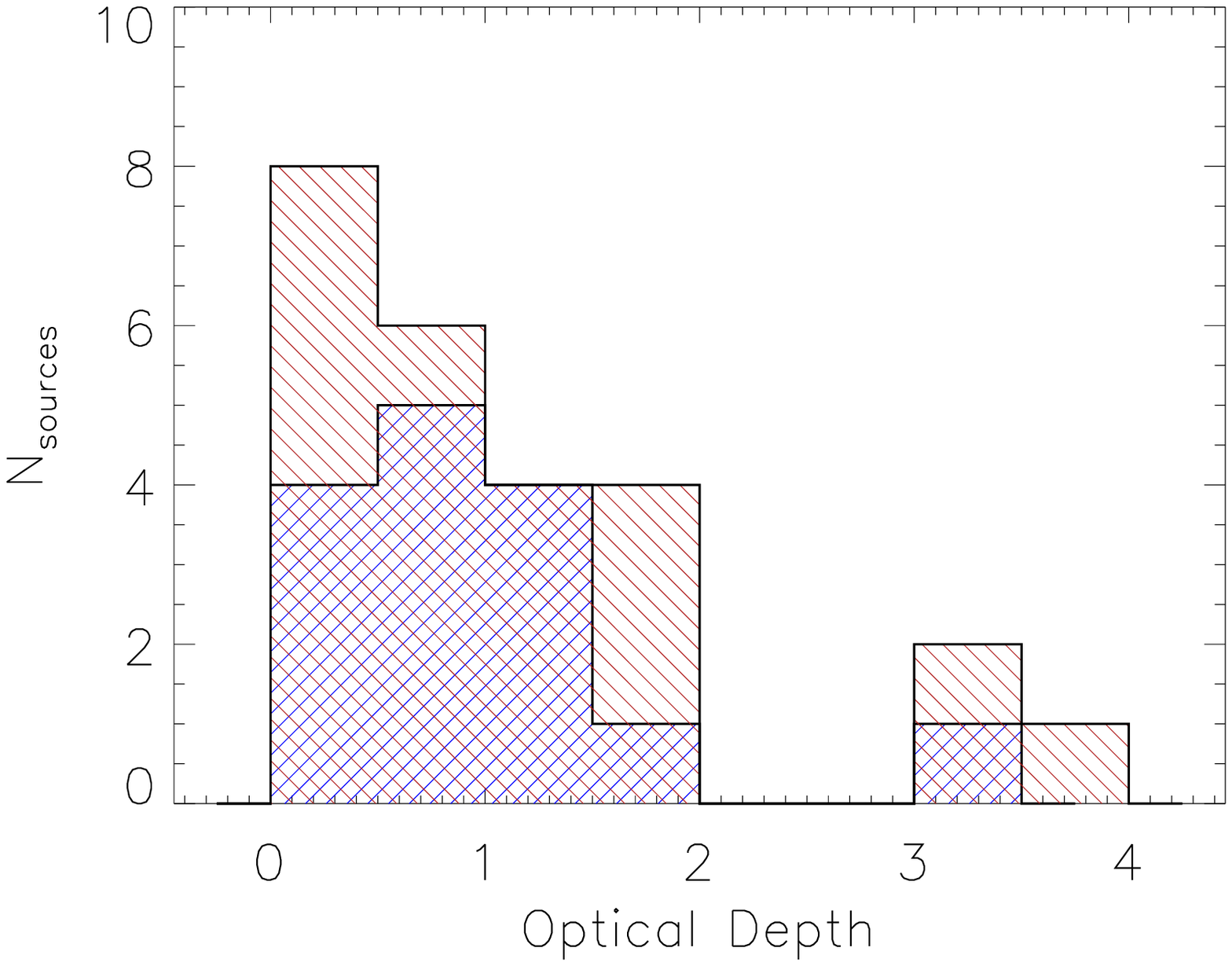}\\
\includegraphics*[width=0.49\textwidth, trim= 50 0 20 0]{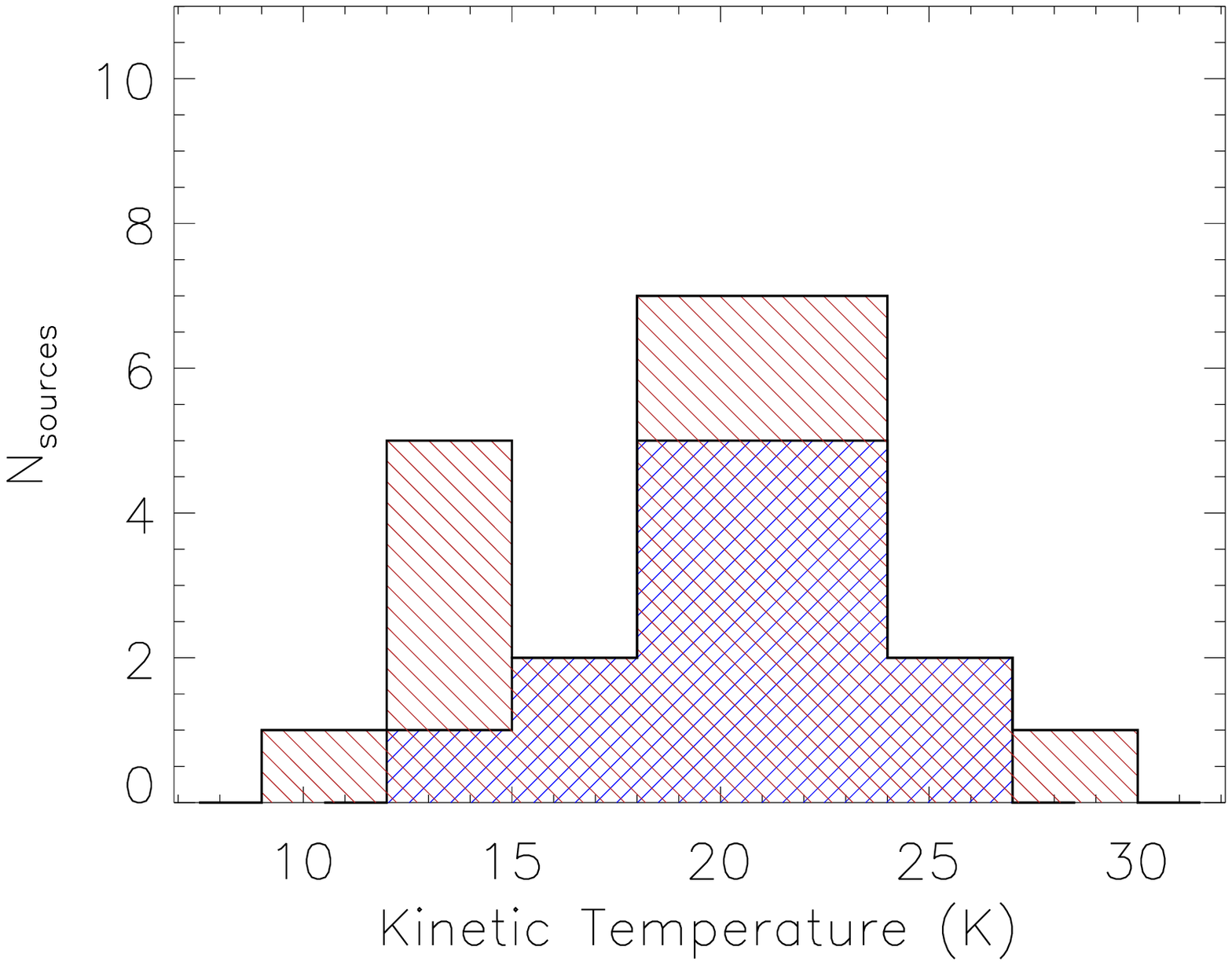}
\includegraphics*[width=0.49\textwidth, trim= 50 0 20 0]{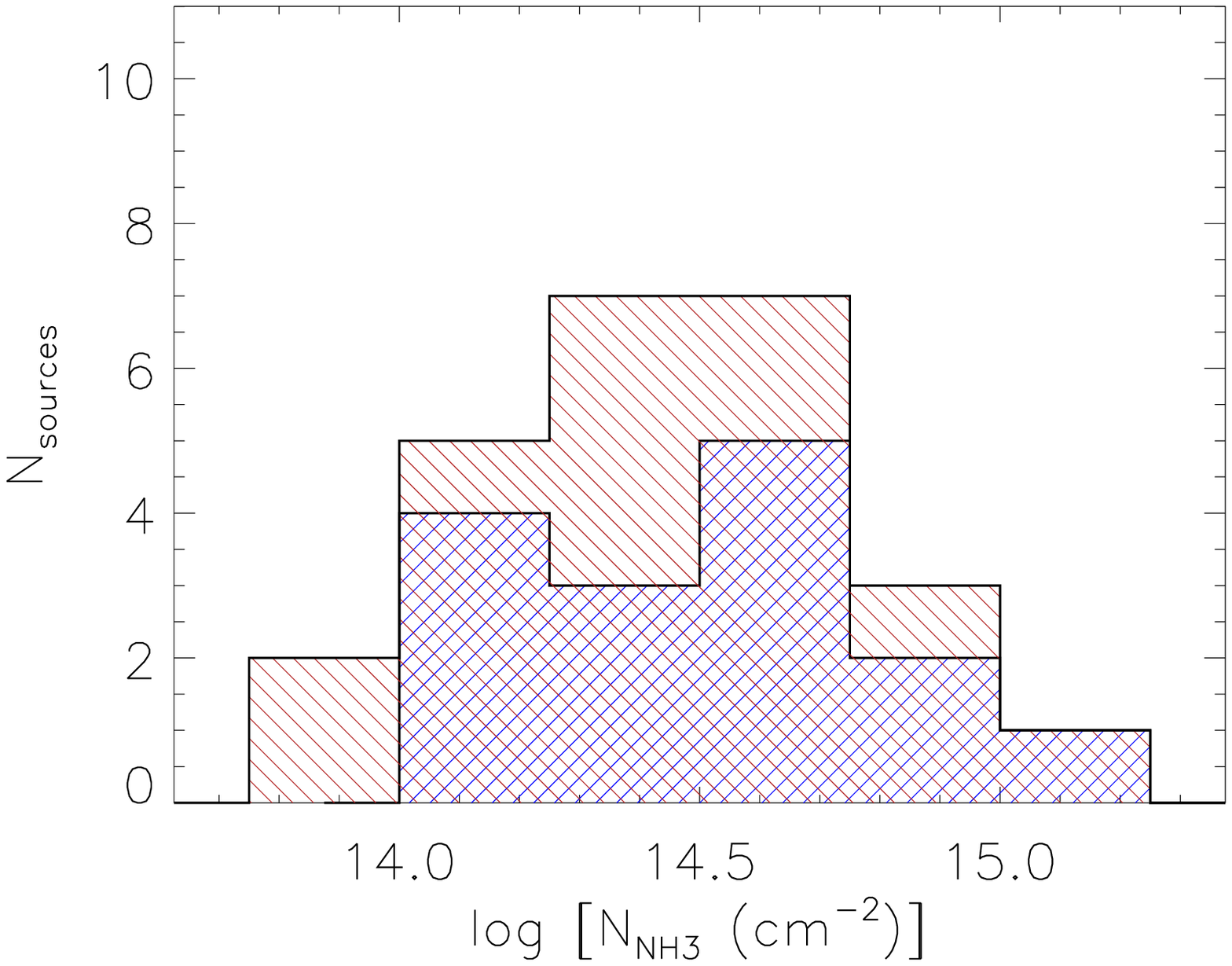}

\end{center}
\caption{\label{img:histograms} Histograms showing the distributions of the physical parameters derived from the ammonia detections. Filled red histograms represent our entire sample while blue represents our triggered sample only.}
\end{figure*}

\begin{table}
\begin{center}
\caption{Summary of measured and derived parameters.}
\label{tbl:parameter_summary}
\begin{minipage}{\linewidth}
\begin{tabular}{lccccc}
\hline
\hline
Parameter			& Min	& Max	& Mean	& Median & Std. dev. \\
\hline
Excitation Temp. (K)		& 2.5  & 4.3  & 3.2  & 3.1  & 0.5  \\
Kinetic Temp. (K)		& 11.9 & 27.3 & 19.5 & 19.9 & 4.6  \\
$\tau$				& 0.20 & 3.55 & 1.15 & 0.75 & 0.98 \\
Log[N(NH$_3$) (cm$^{-2}$)] 	& 13.9 & 15.2 & 14.4 & 14.4 & 0.3  \\
$\Delta v$ (km s$^{-1}$)		& 0.48 & 1.93 & 1.07 & 1.10 & 0.38 \\
\hline
\end{tabular}\\
\end{minipage}
\end{center}
\end{table}

In Fig.~\ref{img:histograms} we present a set of four histograms showing the distribution of ammonia (1,1) linewidth, optical depth, kinetic temperature and column density. In these plots we show the distribution for all of the detections (filled red) and the parameters associated with BRCs identified as triggered candidates by \citet{Morgan2009} (filled blue). The ranges and averages of the properties of all detected sources are presented in Table~\ref{tbl:parameter_summary}. 

The histograms of Fig.~\ref{img:histograms} show that the temperatures of our sources do not show any large variation dependent upon their triggered status. It would appear that the fundamental observational properties of ammonia emission are somewhat insensitive to environmental factors. Temperatures and linewidths of protostellar sources, as observed in ammonia, appear fairly constant across many sample bases. Some examples drawn from the literature include \citet{Molinari1996} \mbox{($\overline{\Delta v}$=1.76 \kms, $\overline{T_{\mathrm{k}}}$=22.0 K,} 260 IRAS sources), \citet{Jijina1999} \mbox{($\overline{\Delta v}$=0.74 \kms, $\overline{T_{\mathrm{r}}}$=14.7 K,} 264 dense cores drawn from the literature, 1971--1999) and \citet{Wu2006} \mbox{($\overline{\Delta v}$=1.54 \kms, $\overline{T_{\mathrm{k}}}$=19.0 K,} 27 IRAS sources associated with water masers).

\section{Discussion}

We used the results of our infrared, submillimetre, molecular line and radio analyses (\citealt{Morgan2004,Morgan2008},  2009) to identify 26 BRCs that show strong evidence they are 1) undergoing recently initiated star formation and 2) are being subjected to intense levels of ionizing radiation. These clouds are therefore considered to be excellent candidates in which the observed star formation may have been triggered via RDI. Although there is evidence that star formation is present in many of the remaining 18 clouds of the SFO catalogue there is little or no evidence that the ionisation from the nearby OB stars is having a significant dynamical impact on these clouds. In this section therefore we will only concern ourselves with the BRCs that remain in the refined triggered BRC catalogue presented in Morgan et al. (2009).

\subsection{Ammonia detections towards triggered sources}
\label{sect:ammonia_detections}

In Morgan et al. (2009)  we identified 26 clouds from the SFO catalogue in which there is a strong likelihood that observed star formation is the result of the RDI triggering process. Twenty-one of these have been observed in ammonia here. Four of the triggered candidates are non-detections (SFO 4, 6, 10, 15)  and a further two fall into the weak category (SFO 35 and 42). 
These six clouds were either not detected or only marginally detected by \citet{Morgan2008} at submillimetre wavelengths. SFO 6 was not detected in that study and the remaining five sources were marginally detected at 850~\micron\ but not detected at 450~\micron. These marginal/non-detections suggest that these clouds are not host to protostars, though they may still be at an early point in their evolution with respect to the observed ionization fronts and may, in the future, develop protostellar cores. These sources shall be discarded from further consideration of triggered protostellar souces.

Ten of the sources in our observational sample have been previously associated with water maser activity (see Table~\ref{tbl:OF_Masers}). These account for 40\% of our `strongly detected' sources, a figure consistent with the 40\% detection rate for Class 0 protostars found by \citet{Furuya2001}. All ten of the sources associated with water masers are also associated with outflow activity (see Table~\ref{tbl:OF_Masers}), supporting the general association of jets and outflows with water masers (\citealt{Codella2004} and references therein). Of these ten sources, only two are found within our non-triggered sample. While this finding is perhaps not statistically significant, given the small sample size, it does indicate that high radiation environments are unlikely to suppress maser emission, as suggested by \citet{Valdettaro2008}.
The presence of molecular outflows and water masers towards so many triggered candidates is a strong indication that star formation is currently taking place within them. 
  Observed linewidths within sources host to outflows and masers are typically greater than in other sources (median linewidths are 1.4 \kms\ for maser sources, compared to 1.1 \kms\ for other sources), indicating that the high density gas surrounding the protostellar cores traced by ammonia emission is dynamically linked to the collisional processes associated with the maser emission.

\begin{table}
\caption{Outflow and water maser activity in our sources}
\begin{tabular}{lcccc} 
\hline
{Source} & $\Delta v$ (\kms) & Outflow & Maser \\
\hline
SFO 05 		& 1.27 & 1 & 3  \\
SFO 07 		& 1.54 & 1 & 4  \\
SFO 14 		& 1.62 & 2 & 5  \\
SFO 18$^\star$	& 0.64 & 1 & 6  \\
SFO 30		& 1.43 & 1 & 7  \\
SFO 31		& 1.13 & 1 & 8  \\
SFO 36		& 1.15 & 1 & 9  \\
SFO 37		& 0.83 & 2 & 9  \\
SFO 38		& 1.86 & 2 & 7  \\
SFO 39$^\star$	& 1.11 & 2 & 9  \\
\hline\\
\end{tabular}\\
$^\star$ BRCs not considered to be triggered candidates.\\
References (NB, only most recent reference cited): 1-\citet{Morgan2009}, 2-\citet{Wu2004}, 3-\citet{Xiang1995}, 4-\citet{Wouterloot1993}, 5-\citet{Henning1992}, 6-\citet{Claussen1996}, 7-\citet{Valdettaro2005}, 8-\citet{Cesaroni1988}, 9-\citet{Valdettaro2008}\\
\label{tbl:OF_Masers}
\end{table}	

\subsection{Comparing protostellar properties derived from different tracers}

In this section we will compare the derived ammonia and dust properties to identify correlations and anti-correlations in the data and to check for inconsistencies between the different tracers. In Section~\ref{sec:parameters} we derived the kinetic temperatures and NH$_3$ column densities which are the most readily available quantities we have at hand to compare with quantities derived from the dust emission. In the next section we will describe how the dust temperatures were determined and estimate the H$_2$ column densities.

\subsubsection{Dust properties}

In an earlier paper (\citealt{Morgan2008}) we presented observations of submillimetre emission which tracing the distribution of warm dust. Spectral energy distributions (SEDs) were determined by fitting greybody functions to the measured submillimetre fluxes and mid- and far-infrared fluxes. Fits were presented for all sources for which good quality data was available.

The H$_2$ column density associated with each SFO object detected by \citet{Morgan2008} may be calculated using

\begin{equation}
N(\mathrm{H}_{2})=S_{\nu}/[\Omega \mu \mathrm{m_H} \kappa_{\nu} B_{\nu}(T_d)]
\label{eqn:NH2}
\end{equation}

\noindent where $S_{\nu}$ is the 850~\micron\ flux density, $\Omega$ is the solid angle associated with the 30\arcsec ~aperture used to observe each core in \citet{Morgan2008}, $\mu$ = 2.3 is the mean molecular weight, $\mathrm{m_H}$ is the mass of a hydrogen atom, $\kappa_{\nu}$ is the dust opacity per unit mass at 850~\micron\ (0.02 cm$^2$ g$^{-1}$, following \citet{Morgan2008}) and $\mathrm{B}_{\nu}(T_\mathrm{d})$ is the Planck function, evaluated at dust temperature $T_\mathrm{d}$. Resulting values of H$_{2}$ column density are presented in the final column of Table~\ref{tbl:Detections}.

\subsubsection{Dust and gas temperatures and column densities}
\begin{figure}
\begin{center}
\includegraphics*[width=0.45\textwidth]{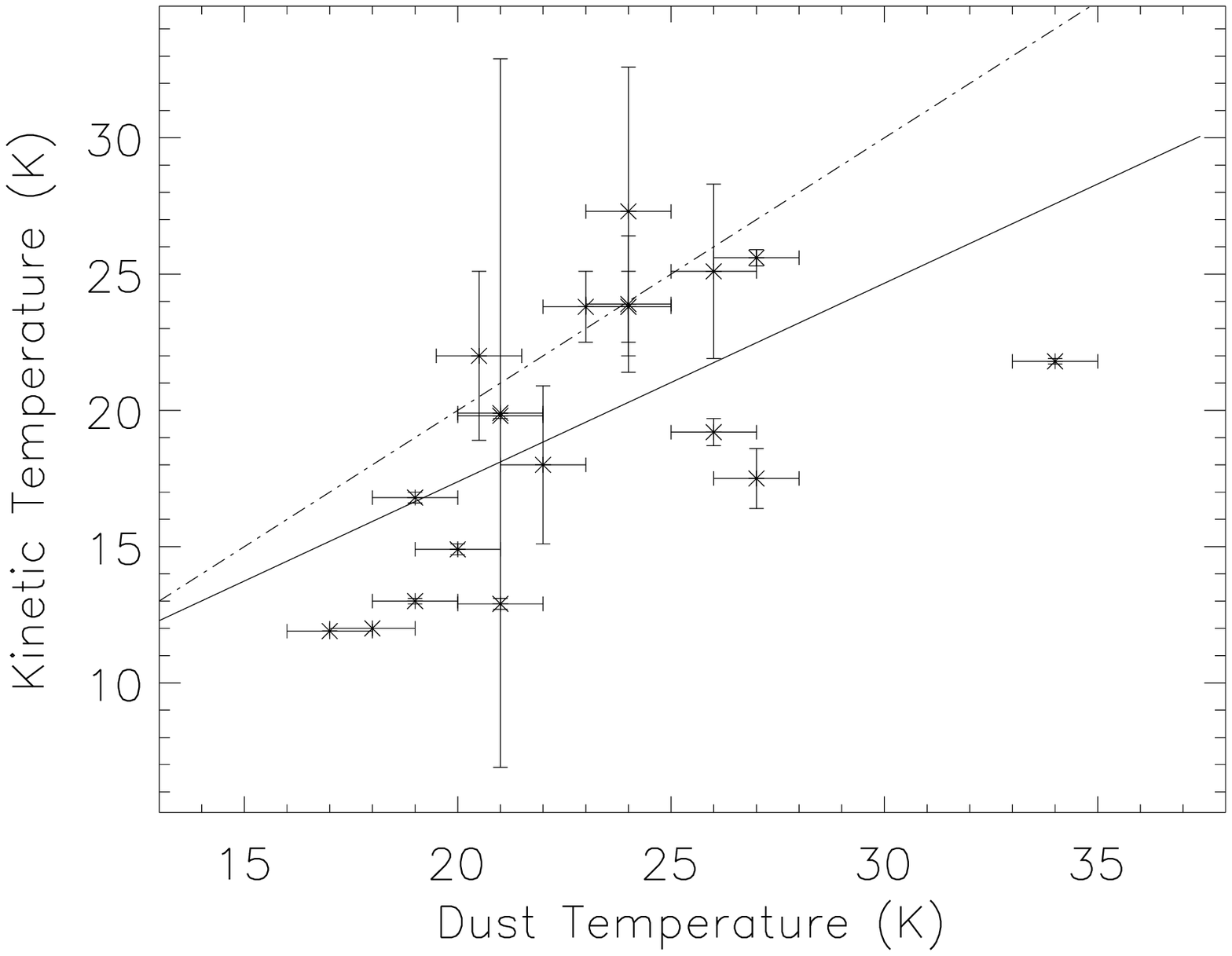}
\includegraphics*[width=0.45\textwidth]{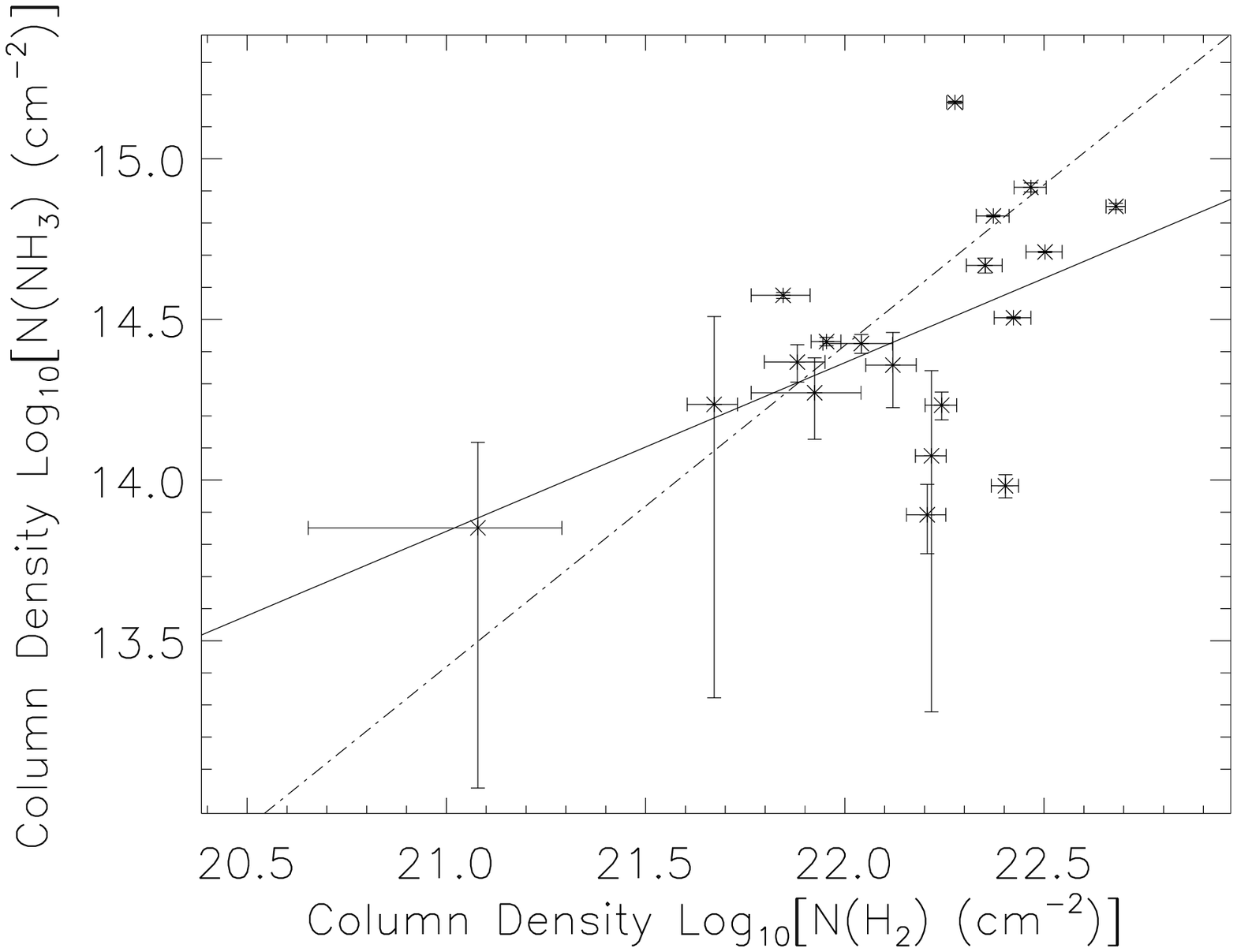}
\caption{Upper panel: Dust temperature vs. kinetic temperature, a line of best fit is shown as a solid line while the dotted line indicates the line of equality. Lower panel: \nh\ vs. H$_{2}$ column density, a line of best fit is shown as a solid line while the dotted line indicates the line of constant fractional abundance.}
\label{fig:Temp_Graph}
\end{center}
\end{figure}

In Fig.~\ref{fig:Temp_Graph} we present two scatter plots comparing the temperatures (upper panel) and column densities (lower panel) derived from the two tracers of ammonia and submillimetre emission. We show the linear-square fit to the data as a solid line. Additionally, in the temperature plot, we include a dashed line indicating the position of the data if both temperatures were equal. There is quite a lot of scatter in the distributions seen in both plots. However, there is a general correlation between the kinetic and dust temperatures and the H$_2$ and NH$_3$ column densities. The dust temperatures are generally slightly higher than the observed kinetic temperatures. The ratio of dust temperature to \Tk\ ranges from 0.9 to 1.6 with a mean of 1.2, this slightly higher dust temperature may be attributed to the fact that submillimetre emission is associated with a wide range of densities, covering the star-forming core itself as well as the warm envelope surrounding the core (and the interface between the two). As ammonia emission requires a critical density of \mbox{$\sim$10$^4$ cm$^{-3}$}, it is likely to trace the inner, more dense regions of the protostellar core. It should be noted that the dust temperatures were derived using fluxes from the IRAS with a significantly larger beam than the present observations. These observations are therefore likely to incorporate more of the hotter dust at the edges of the BRCs. We therefore conclude that the two tracers are probing material at similar temperatures (difference between median averages is $<$3 K), though the submillimetre observations may incorporate some additional material associated with the warm protostellar envelopes and cloud rims.

Although there is some correlation seen in the comparison plot of the H$_2$ and NH$_3$ column densities the scatter is significant. The fractional abundance of \nh\ may be found through simple comparison of the derived H$_2$ and \nh\ column densities\footnote{Note that the beam size of the SCUBA observations was $\sim$ 15\arcsec, significantly smaller than the $\sim$ 29\arcsec\ beamsize of the ammonia observations presented here. However, the fluxes taken from \citet{Morgan2008} were summed over a 30\arcsec\ aperture, thus making a direct comparison valid. The disparity between the pointing positions of each set of observations is typically small ($\sim$1-2\arcsec) and only exceeds a half-beamwidth for one source, SFO 29.}. Looking at individual sources we find the fractional abundances range from a few times 10$^{-9}$ to a few times 10$^{-8}$. The mean fractional abundance of \nh\ to H$_2$ is 2.6 $\times$\ 10$^{-8}$, this is the value used in Fig. \ref{fig:Temp_Graph} to illustrate a line of constant fractional abundance. The scatter in the plot of \nh\ vs. H$_2$ column density reflects the variation of fractional abundance from source to source. Overall, the determined values of fractional abundance are typical across a wide range of protostellar environments, from low-mass starless cores \citep{Tafalla2006,Crapsi2007} and low to intermediate mass dense cores \citep{Hotzel2001,Friesen2009}, to complex, PDR-associated regions \citep{Larsson2003} and high-mass star forming regions \citep{Kuiper1995,Pillai2006}.\\

The fractional abundances found here reflect a more general trend in the properties of ammonia in star forming regions. The physical properties of our sources, as determined from our ammonia observations, are typical in most star forming environments, with only very hot cores showing any significant variation in column density or temperature (c.f.~\citealt{Longmore2007,Pillai2007}). The implication of our analysis is that ammonia, once excited beyond its critical threshold, is insensitive to environmental circumstances, i.e. resistant to depletion in cold, dense cores and likely shielded from photoionisation in high-radiation environments.
 
\subsubsection{Turbulence in BRCs}
The contribution of turbulent motions to the observed linewidth of each source was estimated by removing linewidth contributions from other broadening agents. Thermal contributions to each line were calculated via \mbox{$\mathrm{d}V_{\mathrm{therm}}=\sqrt{8\mathrm{k_{B}}T_\mathrm{k} \mathrm{ln 2/m_{NH_3}}}$}, where $T_\mathrm{k}$ is the kinetic temperature of the source in question and $\mathrm{m_{NH_3}}$ is the mean molecular mass of an ammonia molecule (17.03 amu).

\begin{figure}
\begin{center}
\caption{Histogram of turbulent velocity dispersion for triggered and non-triggered Sources, Group I (triggered) sources are shown in blue and Group II sources are shown in red. Binsize is 0.1 \kms.}
\label{fig:tvd}
\includegraphics*[width=0.4\textwidth]{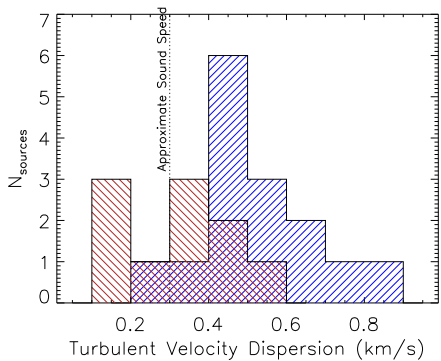}
\end{center}
\end{figure}

  After removing the thermal broadening contribution to our observed linewidths we may evaluate the contribution of turbulent motions in our clouds.
An analysis of the turbulent velocity dispersions ($\sigma=\sqrt{<\Delta v>^2/(8\mathrm{ln}2)}$) of our sources, separated based upon their triggered status (see \mbox{Sect. \ref{sec:introduction}}), reveals some interesting differences between the two samples. A histogram of the velocity dispersions of the potentially triggered and non-triggered samples shows that our non-triggered sources have typically lower velocity dispersions than our triggered candidates \mbox{(Fig.~\ref{fig:tvd})}. A Kolmogorov-Smirnov test of the $\sigma$ of each of the two samples indicates that the two distributions are drawn from separate populations with a probability of 99.8\%. An interesting point to note in the distributions of the $\sigma$ of each sample is that the non-triggered sources are largely subsonic, based upon an estimate of the sound speed in our sources of \mbox{$\sim$0.3 \kms} (e.g. \citealt{Thompson2004a}). In contrast, the triggered sample typically exhibit supersonic velocities.

Several explanations of the difference in $\sigma$ between our two samples exist, the sources in our triggered sample are likely undergoing the progression of shocks through the host clouds. As this sample was selected on the basis of the presence of an ionisation front, the presence of shocks in the observed material would thus be consistent with the models of triggering put forward by \citet{Bertoldi1989,Lefloch1994,Miao2006,Gritschneder2009,Miao2009}. An alternative explanation would be that the increased turbulent motions within our triggered sample are due to increased systematic motions within the relevant clouds (due to larger numbers of outflows, for example). Alternatively, these clouds may simply be larger in extent and/or mass. However, we found no trend toward higher dust mass for these clouds from the submillimetre observations of \citet{Morgan2008}. The source luminosities from those observations show that the most luminous sources of the entire sample are those of the triggered sub-sample, as suggested by \citet{Morgan2008}. The median luminosity of the triggered sample is 201 \lsun\ while the non-triggered sample median average is almost an order of magnitude lower at 28 \lsun. While this may go some way to explain the higher velocity dispersions within the triggered sample, no direct correlation between source luminosity and linewidth can be drawn. Such a correlation might be expected if the higher turbulent velocity dispersions of the triggered sample were due to outflow momentum (e.g. \citealt{Cabrit1992,Bontemps1996}). The ambiguity in the true nature of the increased velocity dispersions in these clouds cannot be resolved with single pointing data and must be addressed with maps of the molecular emission associated with each region.

\section{Summary and conclusions}
We present observations made with the GBT of the ammonia (1,1) and (2,2) inversion transitions towards 40 bright-rimmed clouds taken from the sample compiled by  \citet{Sugitani1991}. We detected the (1,1) transition towards 31 of the sources observed, with the (2,2) transitions seen towards 28 of these. We use these emission lines to derive the optical depths, kinetic temperatures and NH$_3$ column densities towards the embedded protostars located within these clouds.

Across the entire sample, the ammonia kinetic temperatures of our sources correlate reasonably well with the dust temperatures seen towards the same cores by \citet{Morgan2008}. Dust temperatures are typically slightly higher than the observed kinetic temperatures, likely due to the slightly different material traced by the two sets of observations. A correlation between H$_2$ and NH$_3$ column densities is also seen, although there is somewhat more scatter in the plotted points. This likely reflects variation in the fractional abundance of ammonia from source to source. A mean fractional abundance of \nh\ to H$_2$ is 2.6 $\times$\ 10$^{-8}$, typical for most observations of protostellar regions. 

Using a combination of mid-infrared, submillimetre and radio images, and CO molecular line data presented in earlier works \citep{Morgan2004,Morgan2008,Morgan2009}, we have refined the original sample of 44 BRCs. Our efforts have identified 26 bright-rimmed clouds in which the data are consistent with the hypothesis that any observed star formation is likely to have been triggered. Within this refined sample we have detected strong ammonia emission towards 15. In combination with the submillimetre continuum and CO line emission results of \citet{Morgan2008} and \citet{Morgan2009} respectively, in addition to outflow and maser detections from the literature, our ammonia detections leave little doubt of the star-forming nature of these sources. Having already established the likelihood that these sources represent photoionisation-triggering processes in progress, these 15 sources are some of the best examples yet known of the RDI process.

An investigation of our samples, separated based upon triggered status, indicates a bimodality within observations of the turbulent velocity dispersion. Those sources which have been identified as likely triggered in nature show typically supersonic turbulent velocity dispersions. While non-triggered sources are more often found to be subsonic. This disparity may stem from the presence of shocks, traversing the clouds in the triggered sample. It is also possible that the higher observed velocity dispersion in the triggered sample is simply due to the higher occurence of outflows found in that sample. It is tempting to draw conclusions of differing physical processes occuring within our triggered and non-triggered samples. However, we are not able to determine the true cause of this finding without mapping of molecular emission in these clouds.

\section*{Acknowledgments}

The authors would like to thank an anonymous referee for a careful examination of this work which has resulted in considerable improvements.
We would like to thank the helpful staff of the Green Bank Telescope and Bill Saxton for his assistance in creating Fig.1. LKM is supported by a STFC postdoctoral grant (ST/G001847/1). JSU is
supported by a CSIRO OCE postdoctoral grant. This research would not have been possible without the SIMBAD astronomical database service operated at CDS, Strasbourg, France and the NASA Astrophysics Data System Bibliographic Services. This research makes use of data products from the MSX and 2MASS and GLIMPSE Surveys, which is are joint projects of the University of Massachusetts and the Infrared Processing and Analysis Center/California Institute of Technology, funded by the National Aeronautics and Space Administration and the National Science
Foundation.

\bibliography{References}

\bibliographystyle{mn2e}

\begin{center}
\begin{figure*}
\begin{center}
\includegraphics*[width=0.24\textwidth]{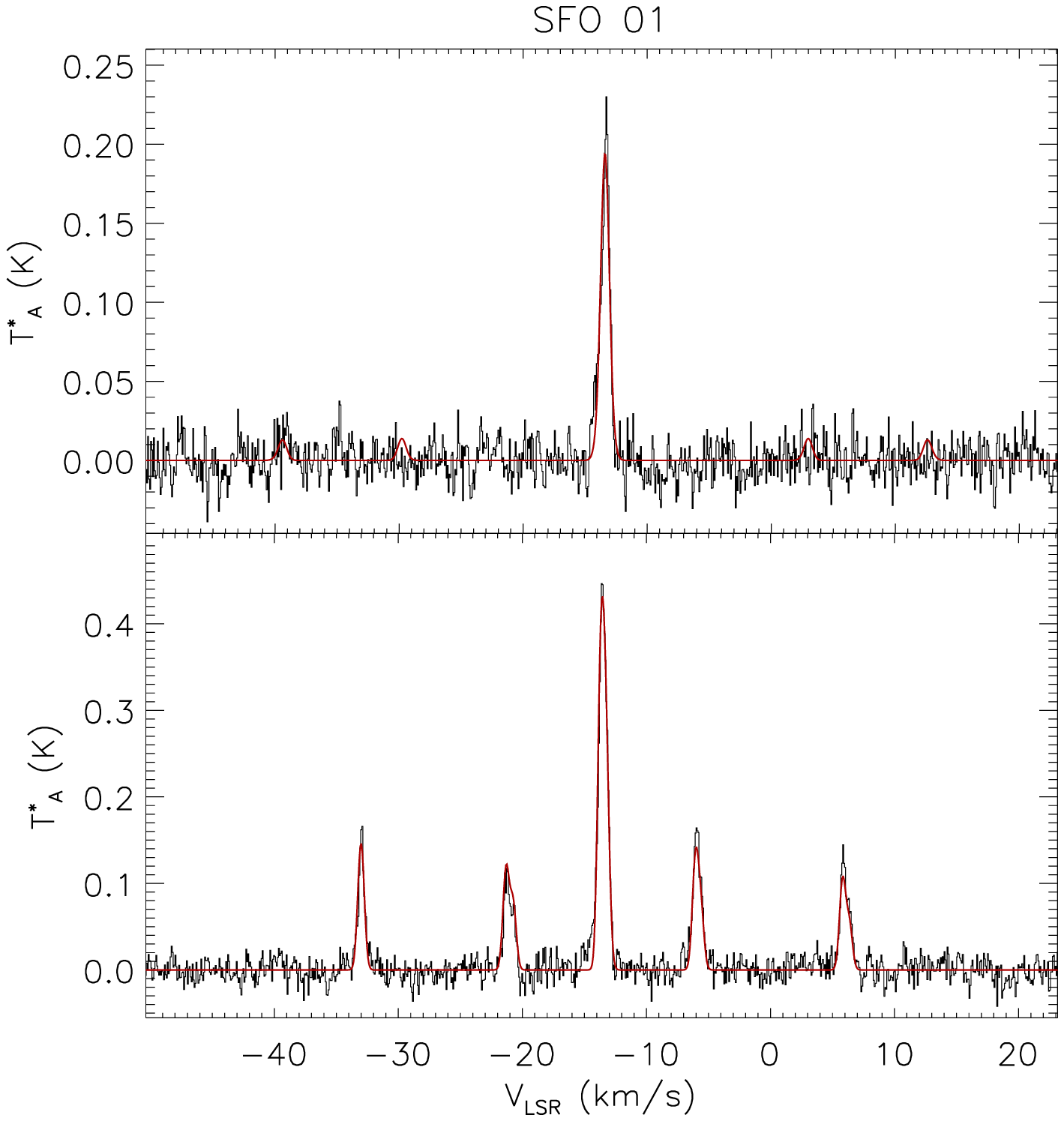}
\includegraphics*[width=0.24\textwidth]{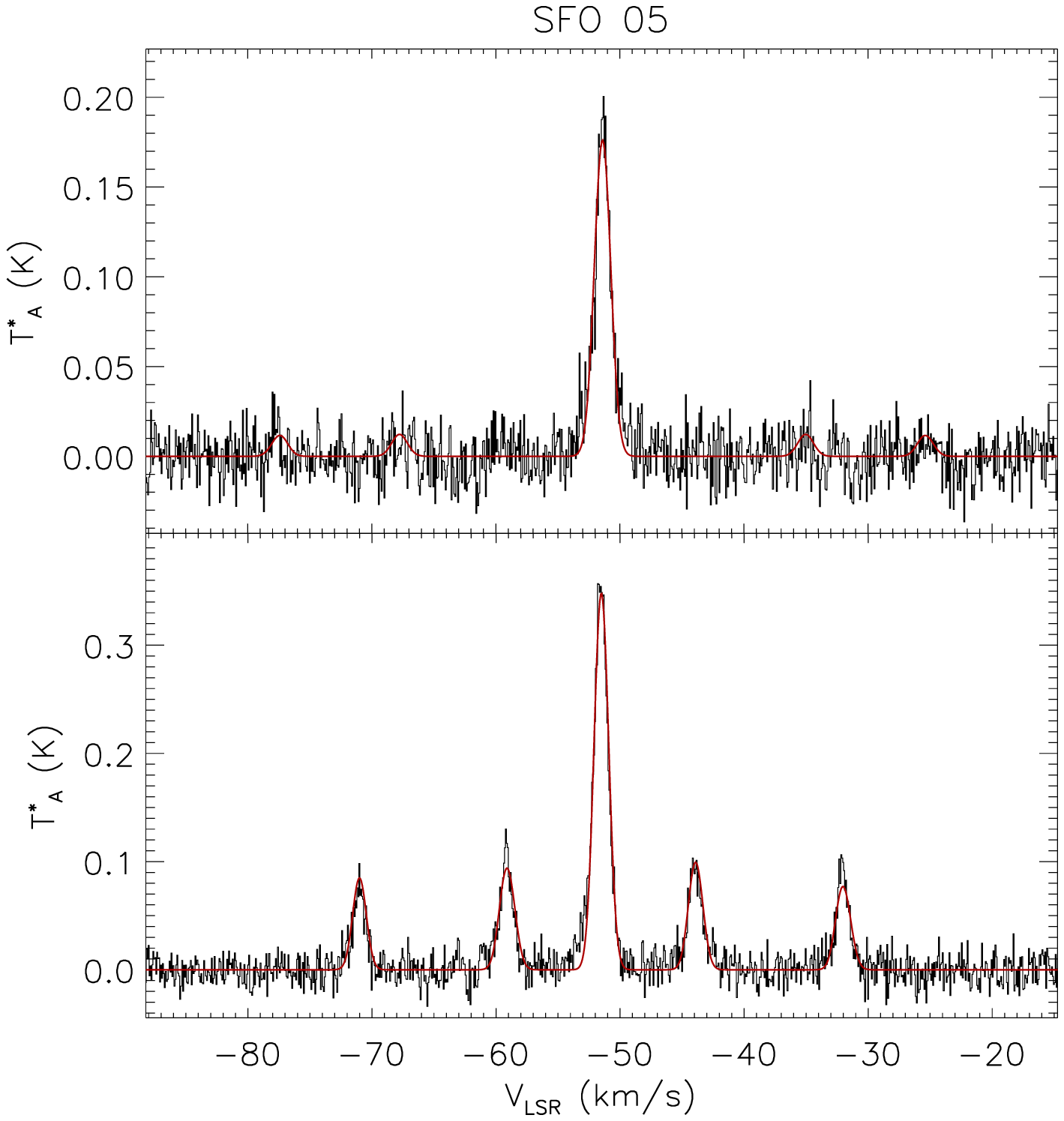}
\includegraphics*[width=0.24\textwidth]{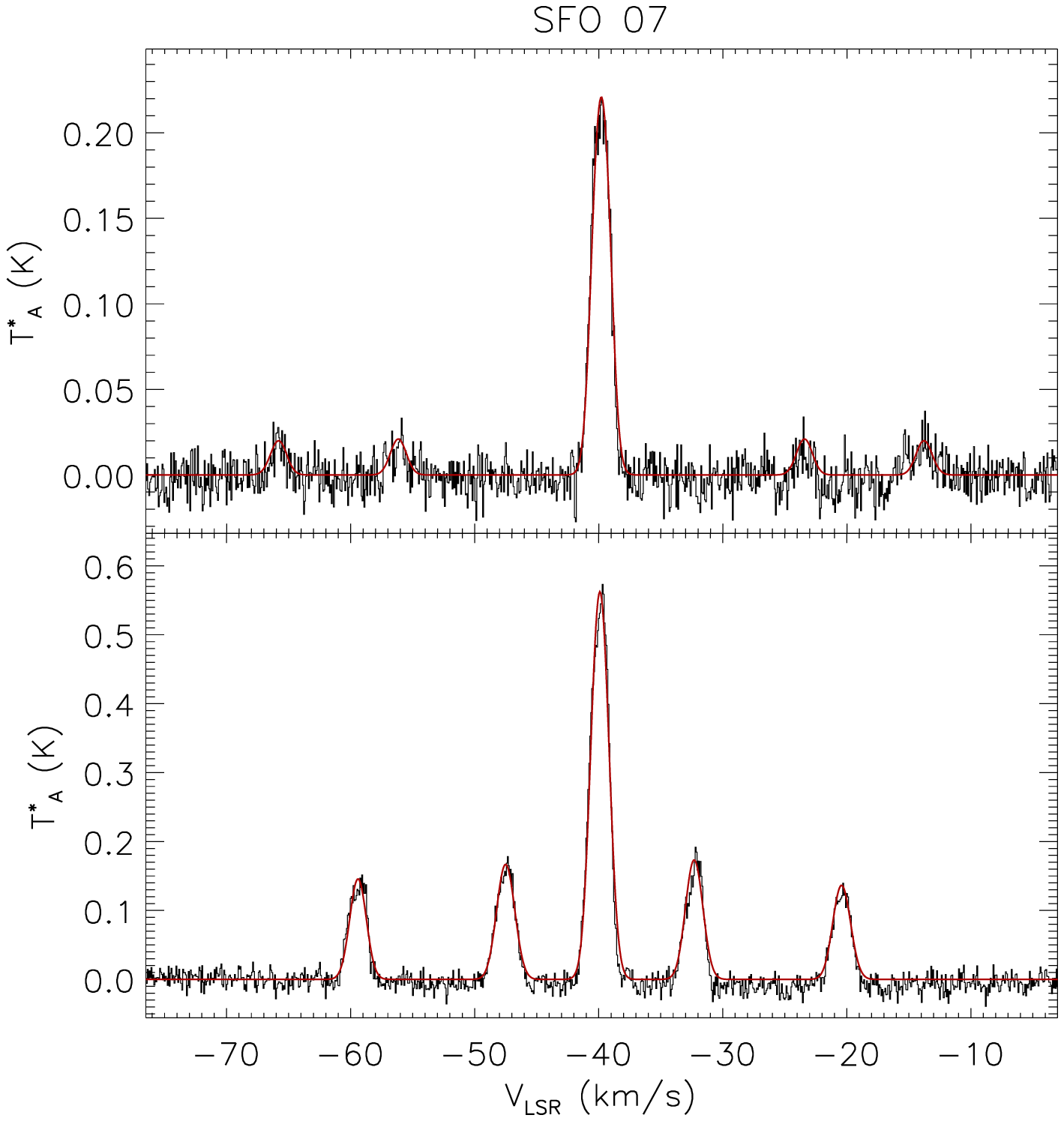}
\includegraphics*[width=0.24\textwidth]{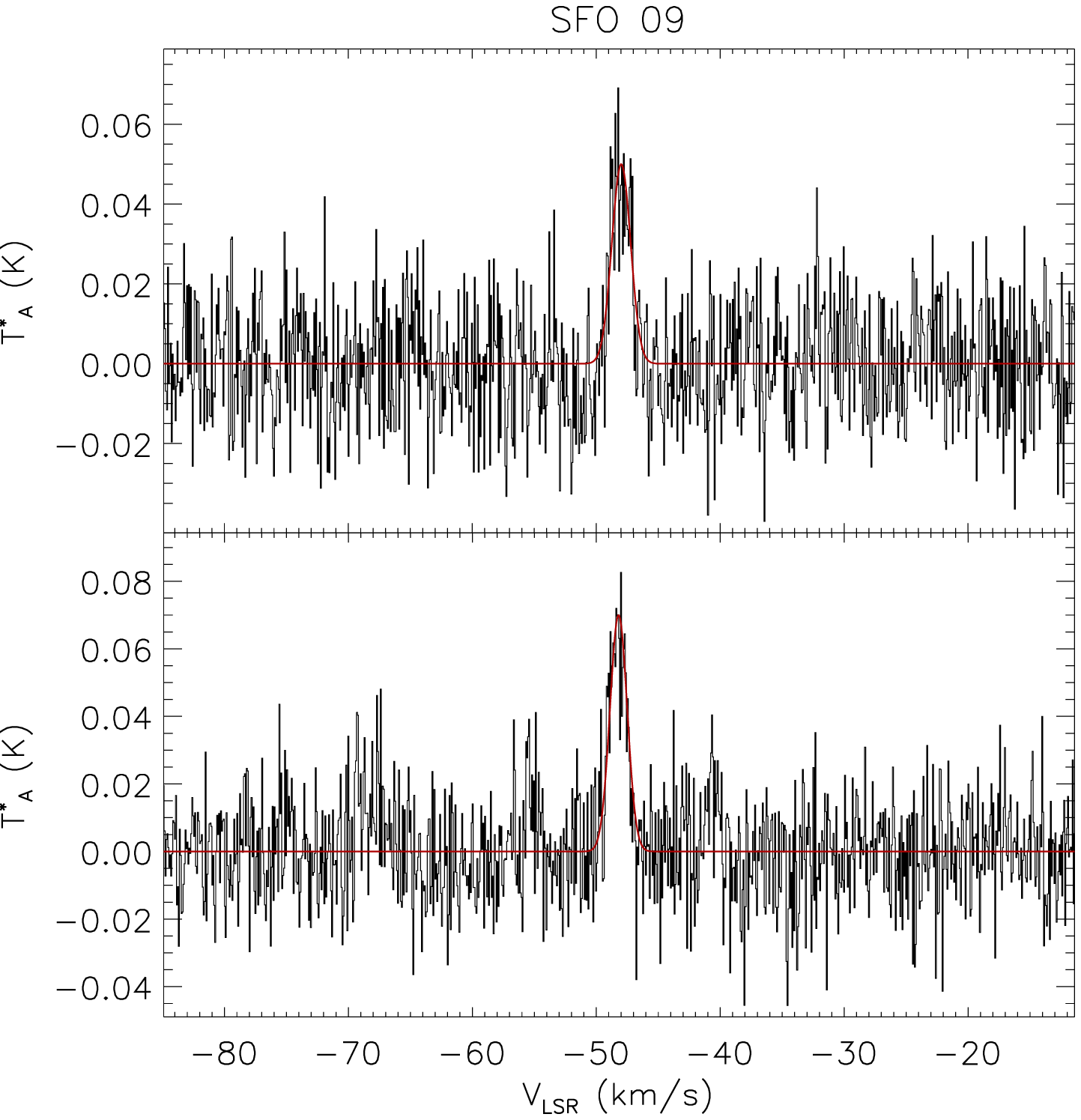}\\
\includegraphics*[width=0.24\textwidth]{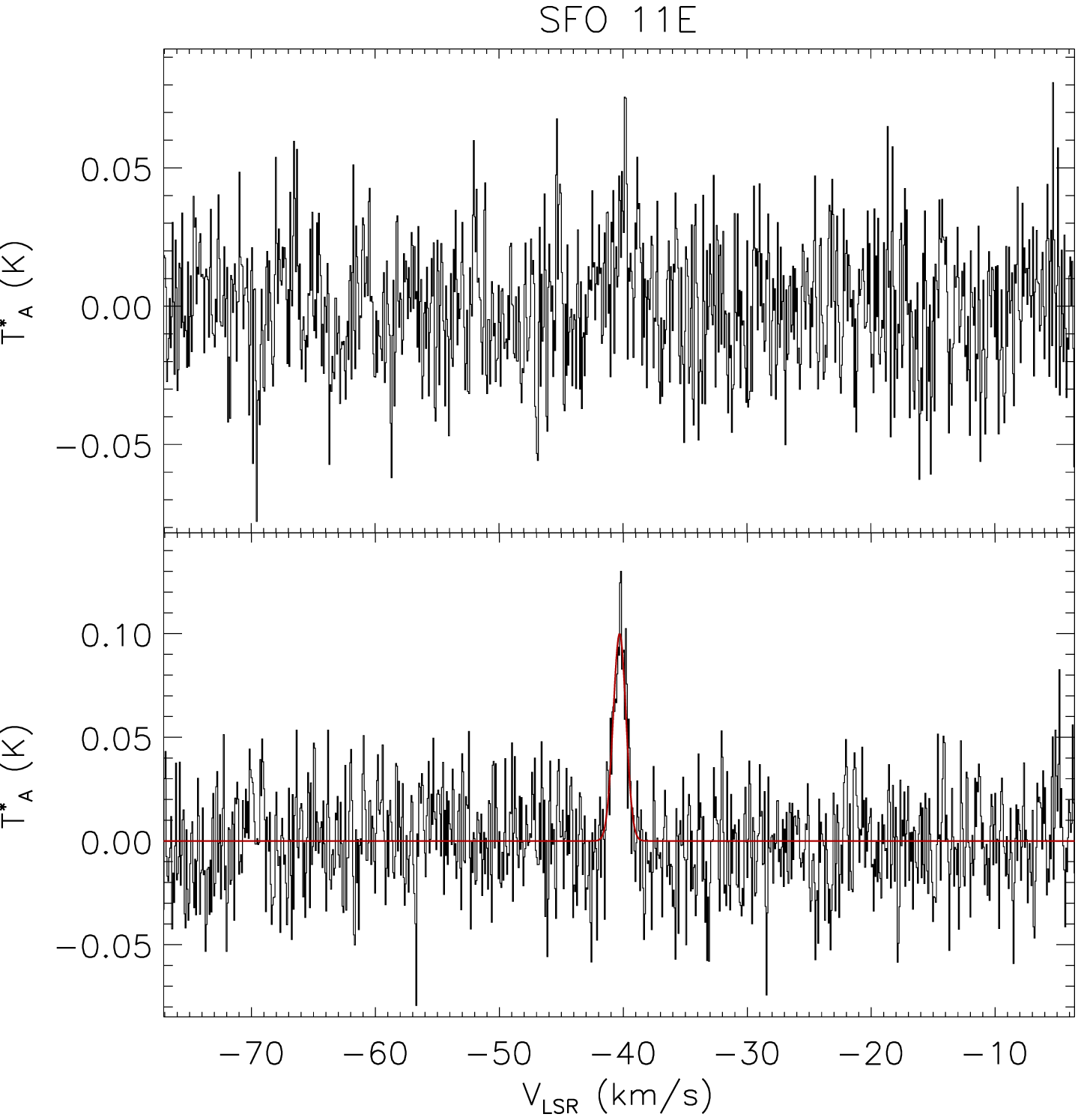}
\includegraphics*[width=0.24\textwidth]{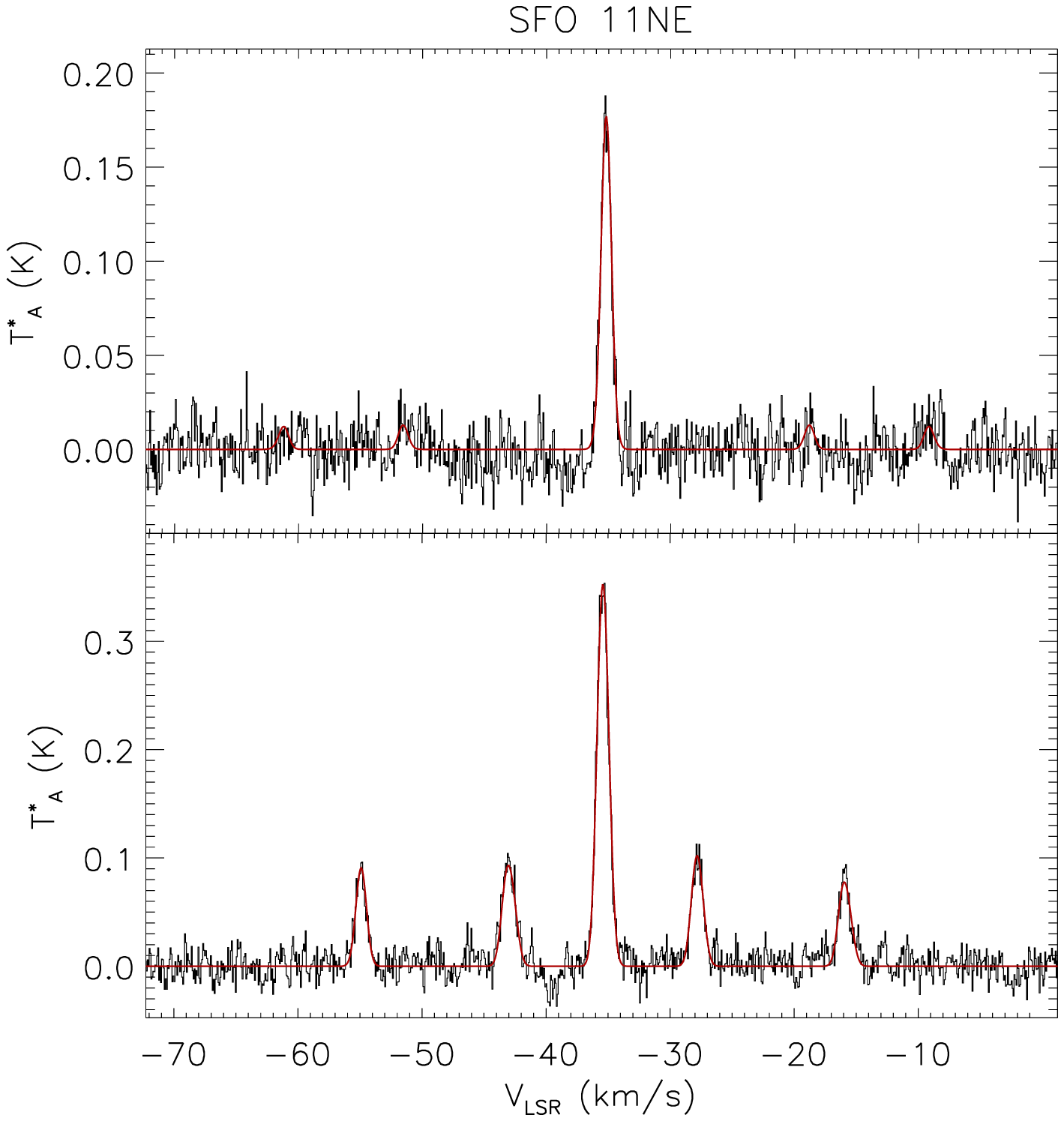}
\includegraphics*[width=0.24\textwidth]{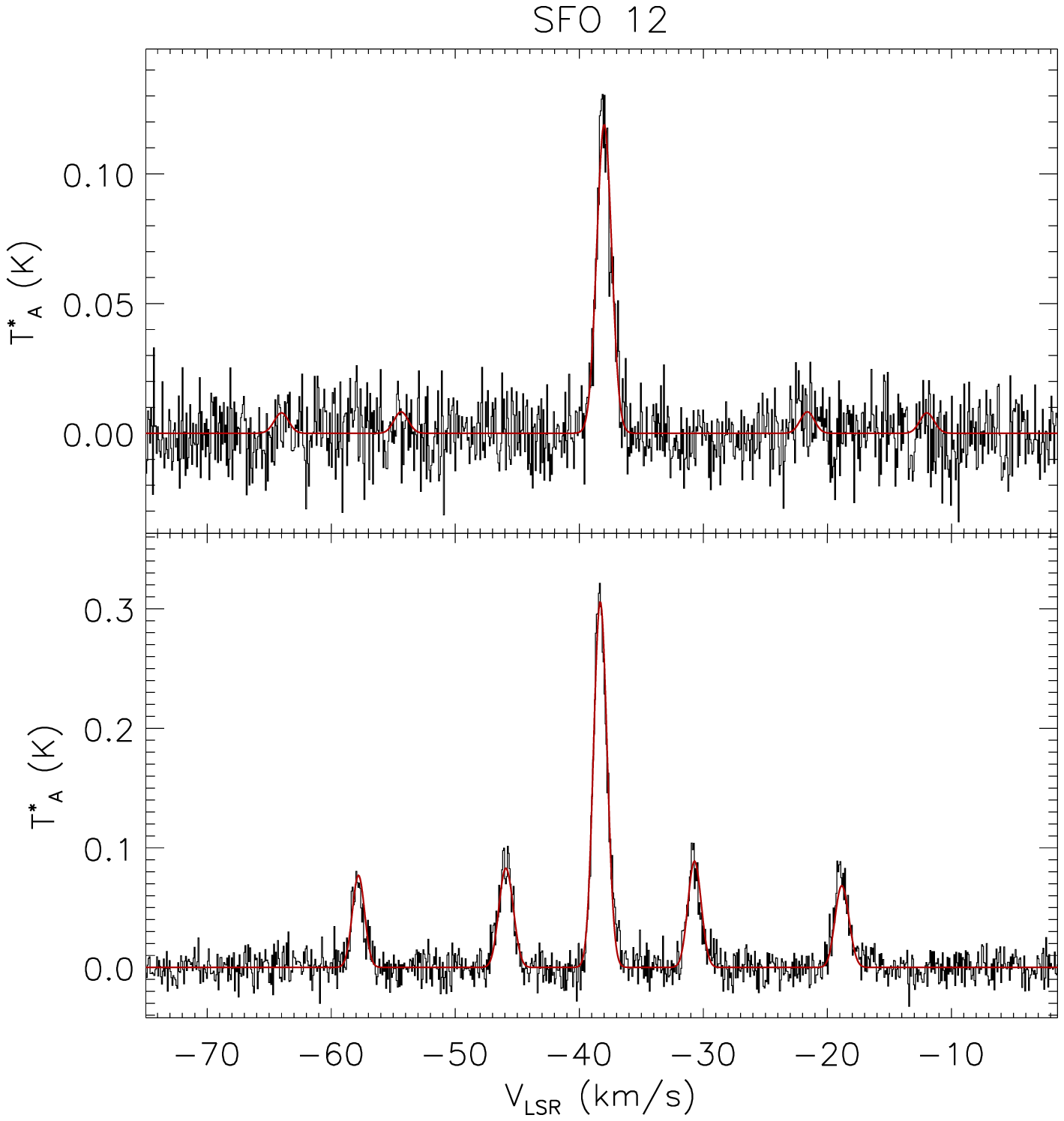}
\includegraphics*[width=0.24\textwidth]{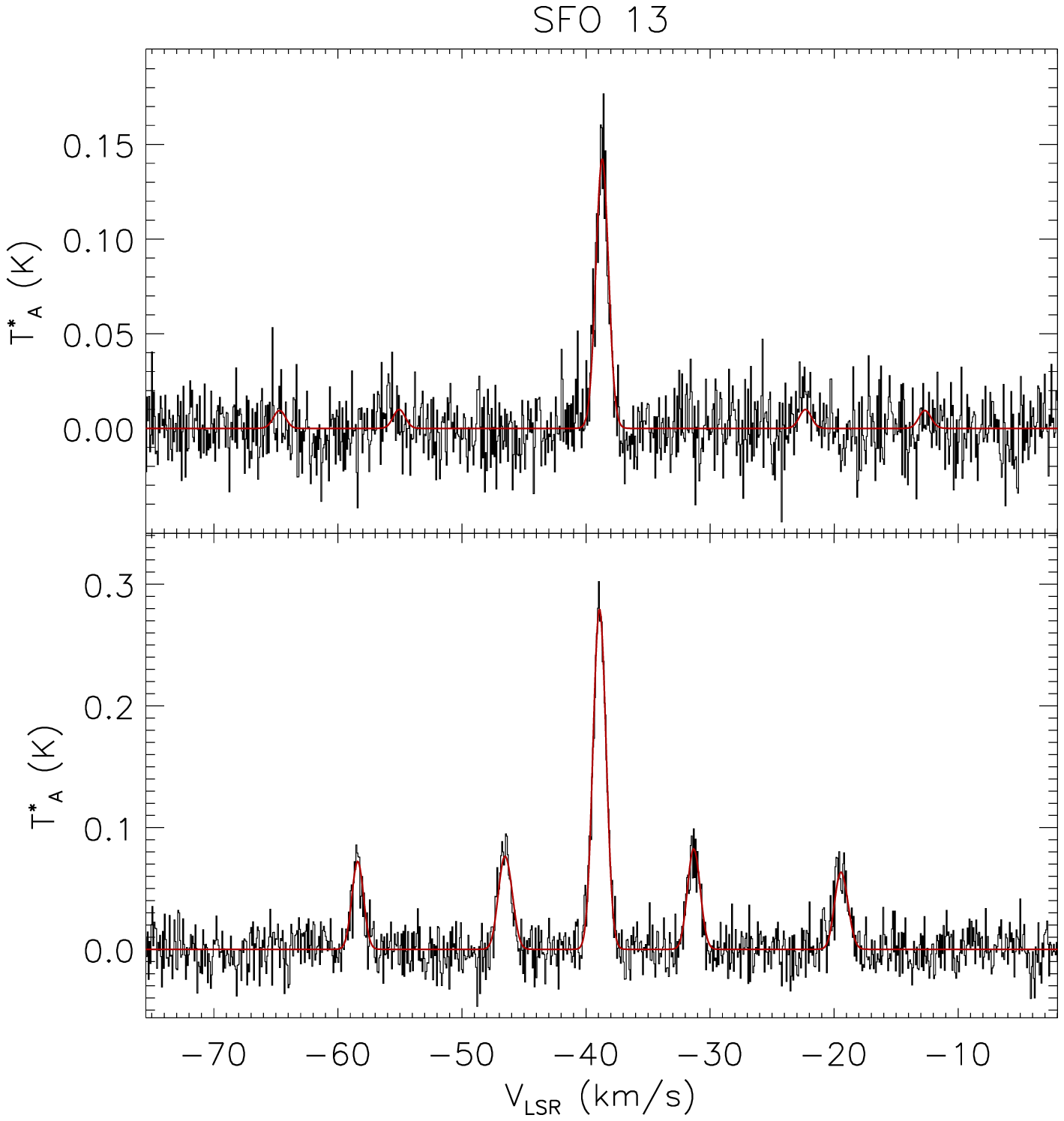}\\
\includegraphics*[width=0.24\textwidth]{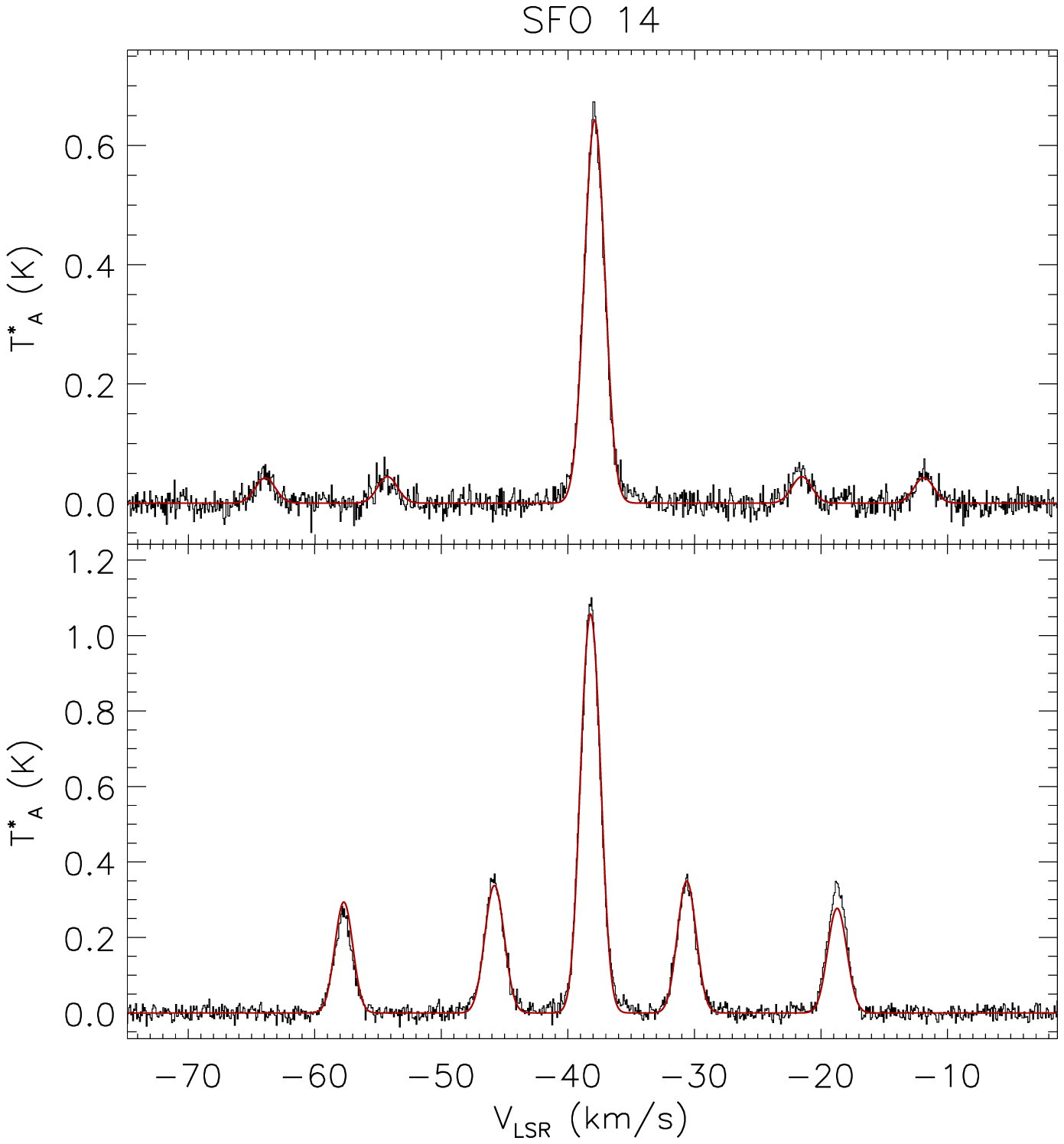}
\includegraphics*[width=0.24\textwidth]{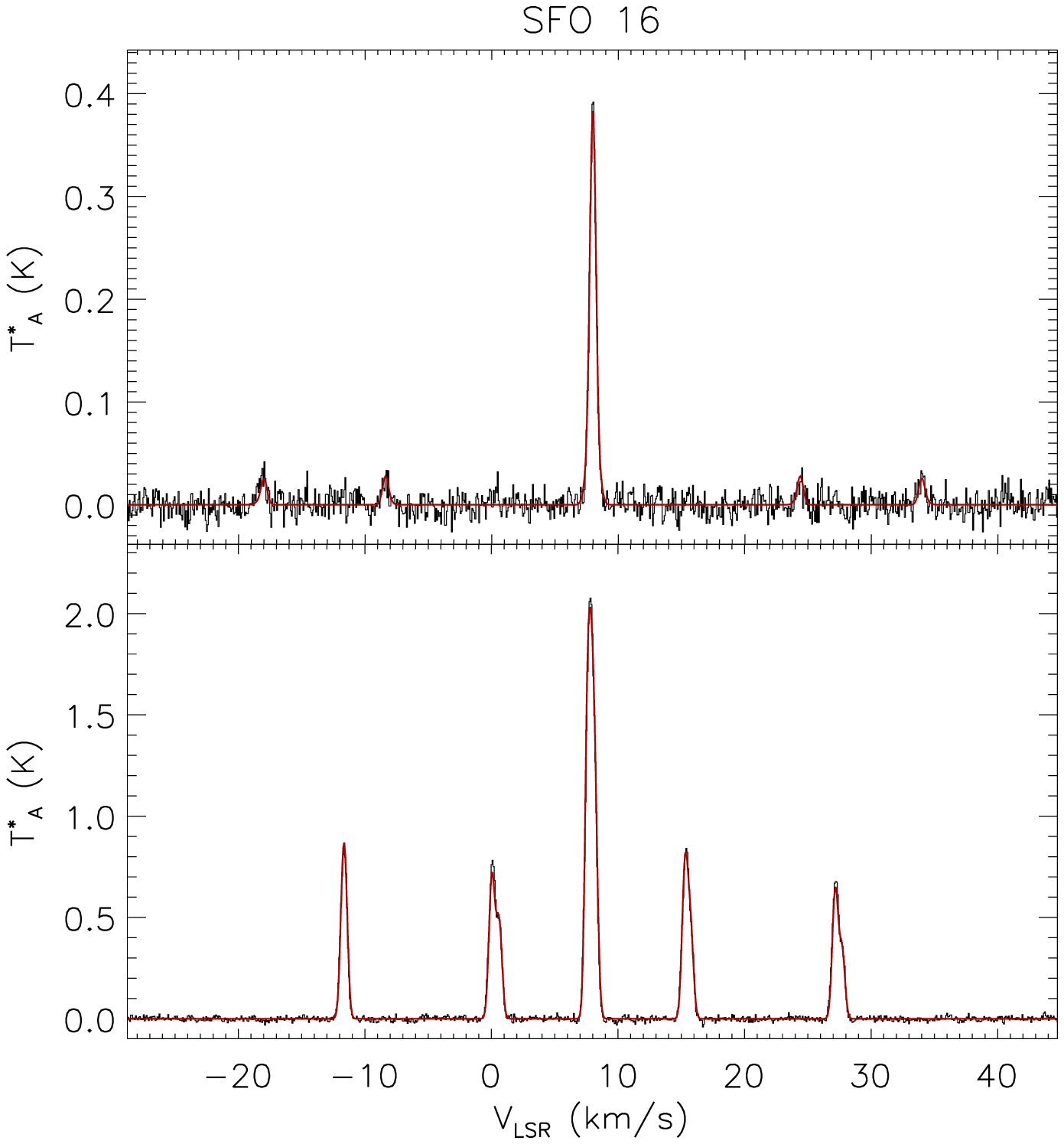}
\includegraphics*[width=0.24\textwidth]{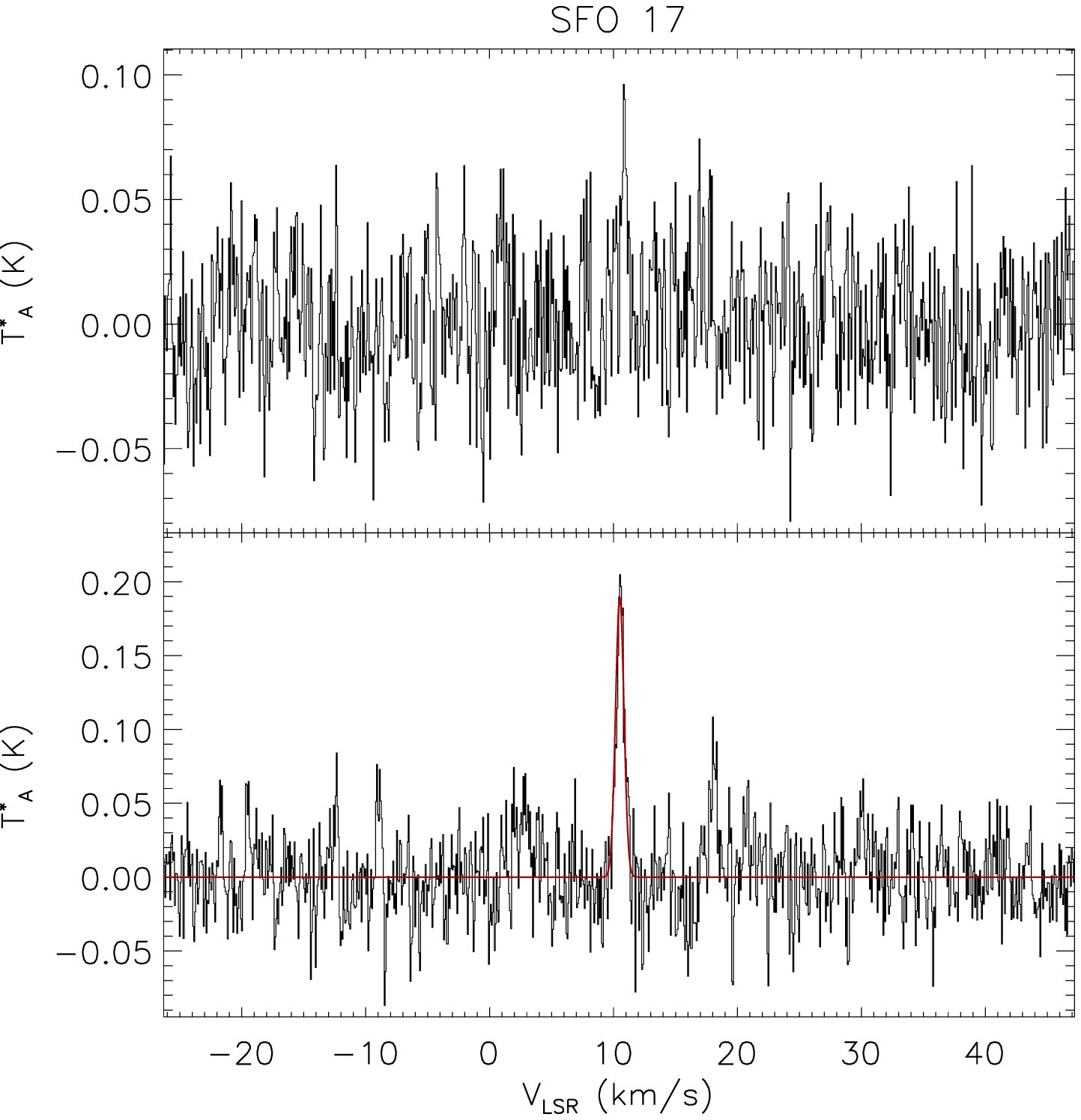}
\includegraphics*[width=0.24\textwidth]{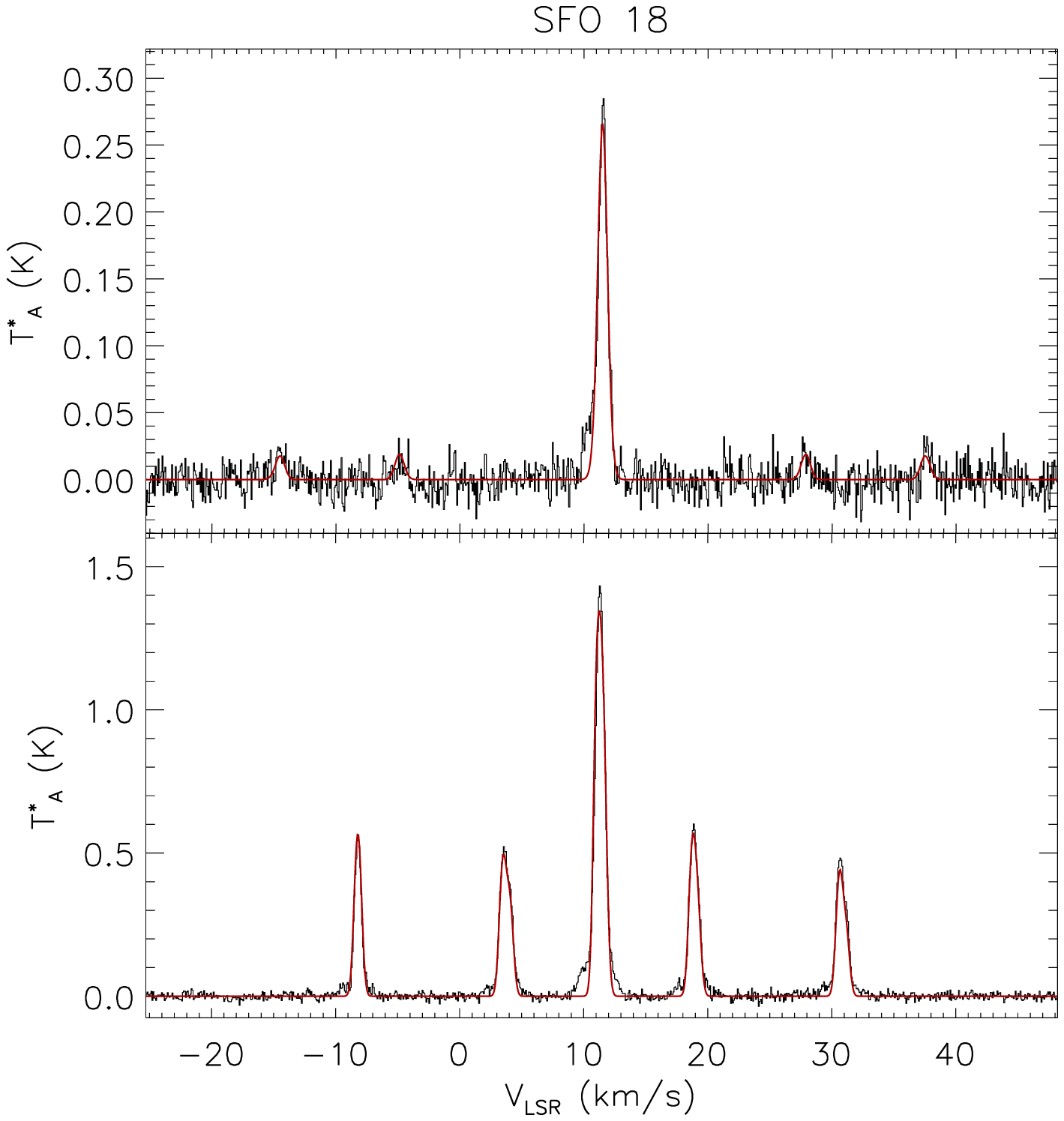}\\
\includegraphics*[width=0.24\textwidth]{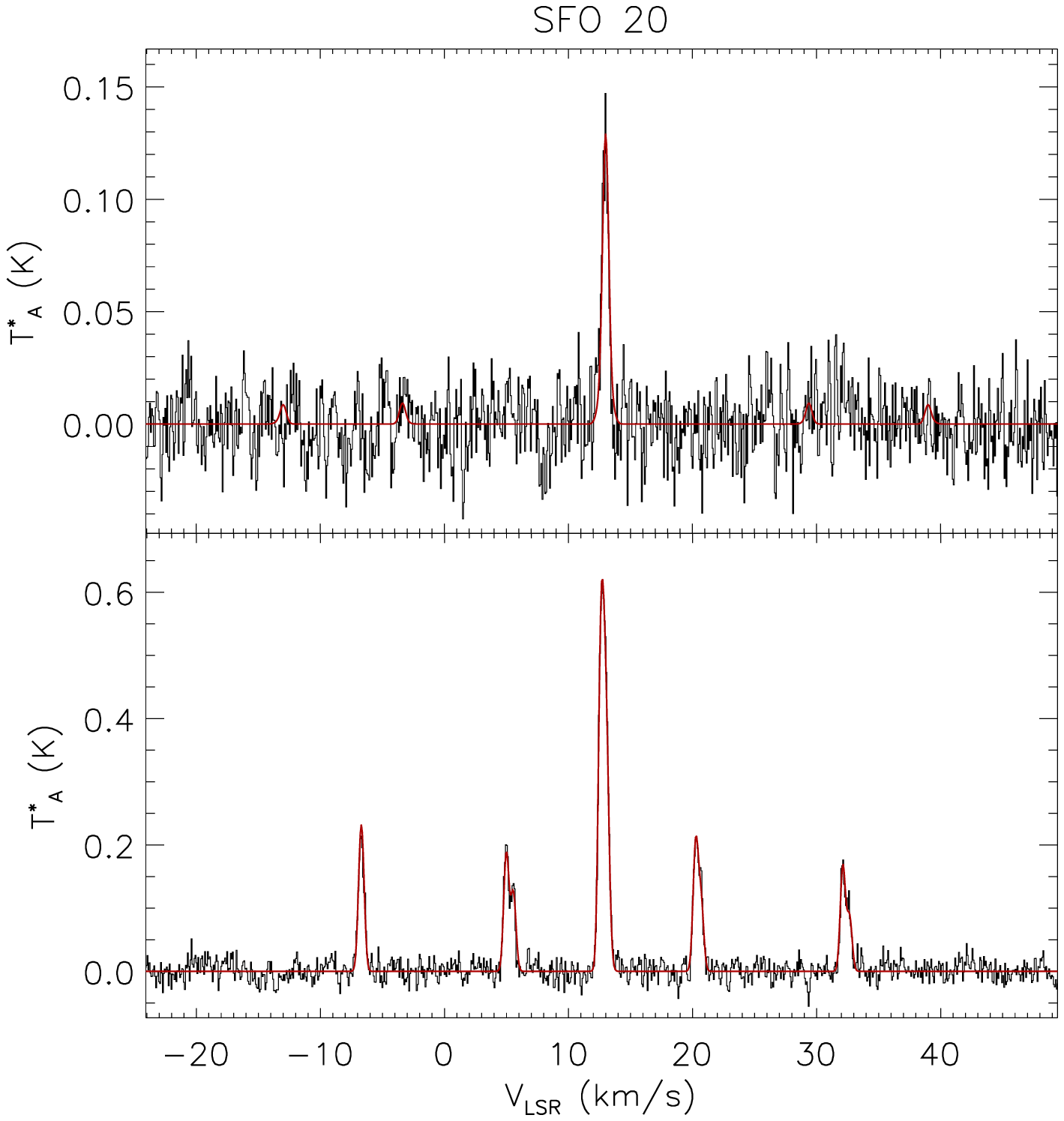}
\includegraphics*[width=0.24\textwidth]{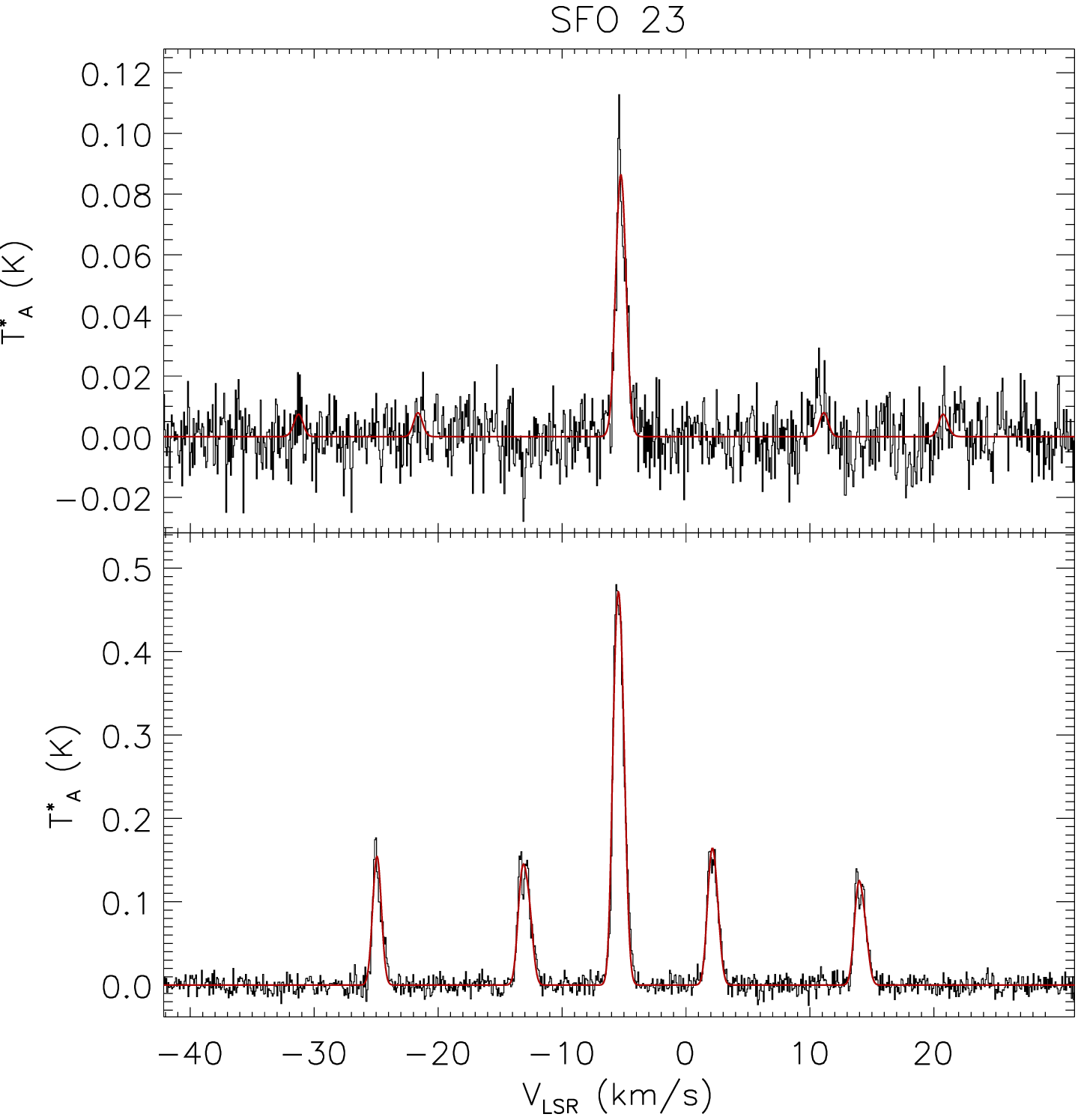}
\includegraphics*[width=0.24\textwidth]{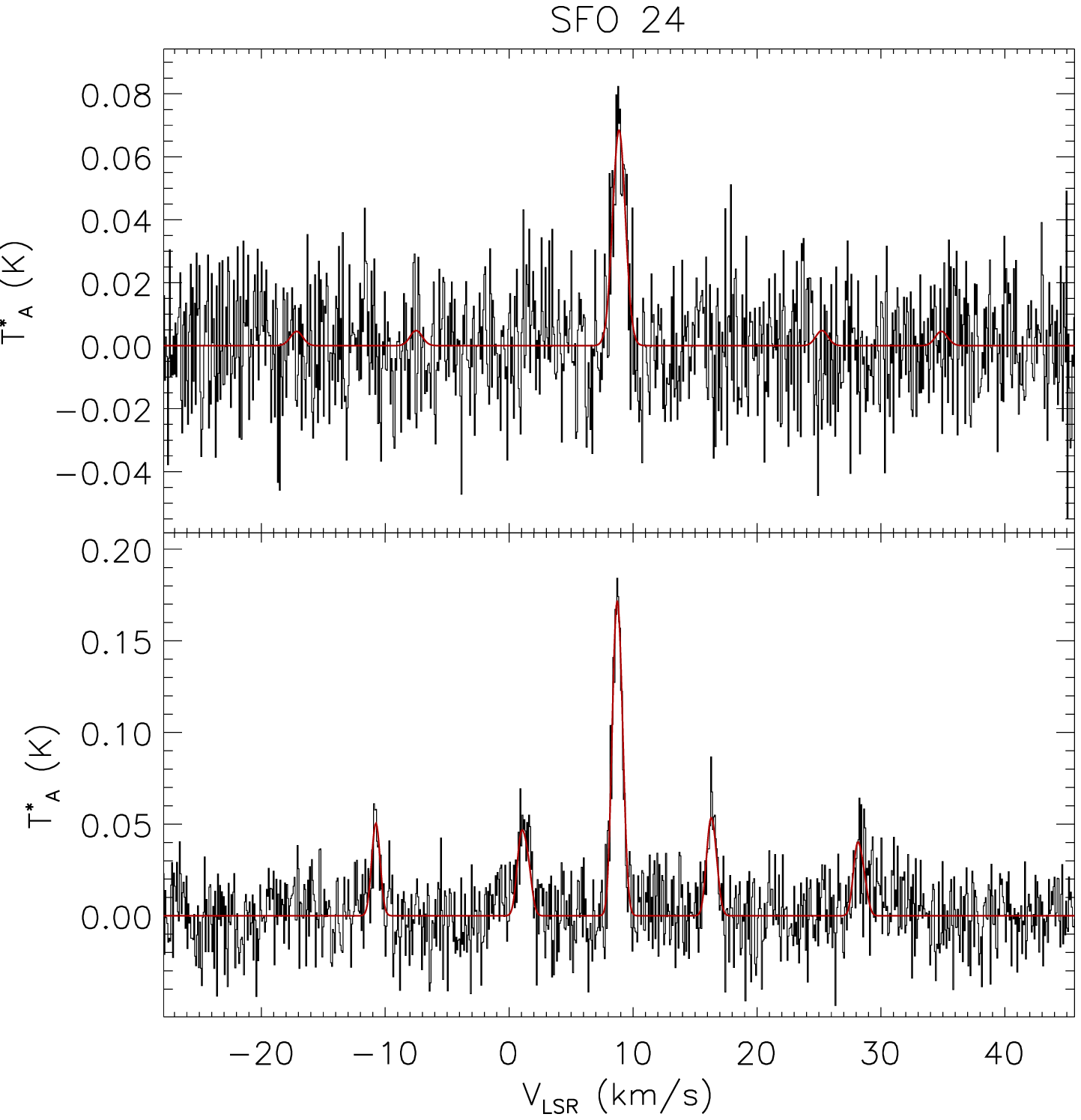}
\includegraphics*[width=0.24\textwidth]{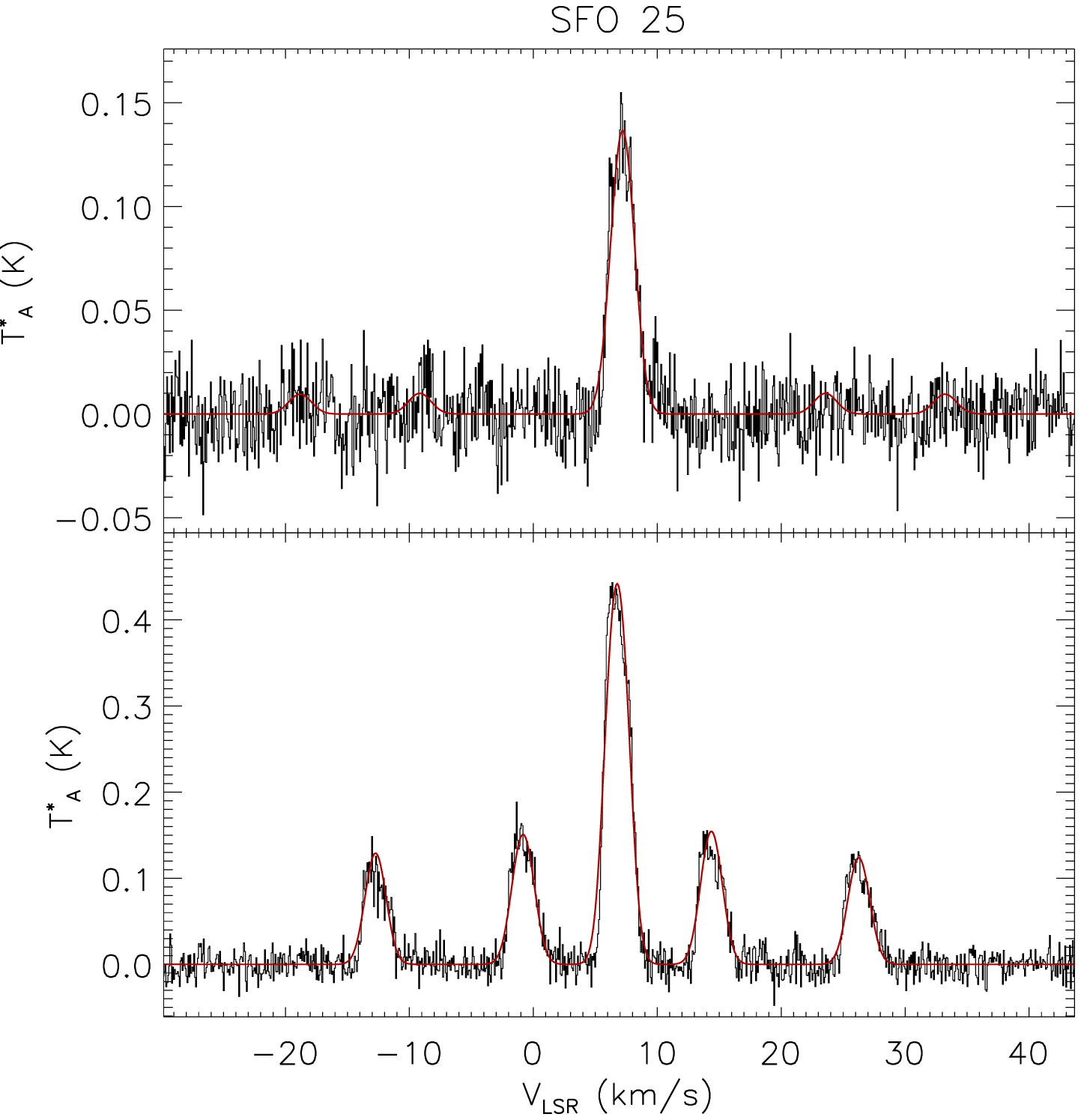}\\
\includegraphics*[width=0.24\textwidth]{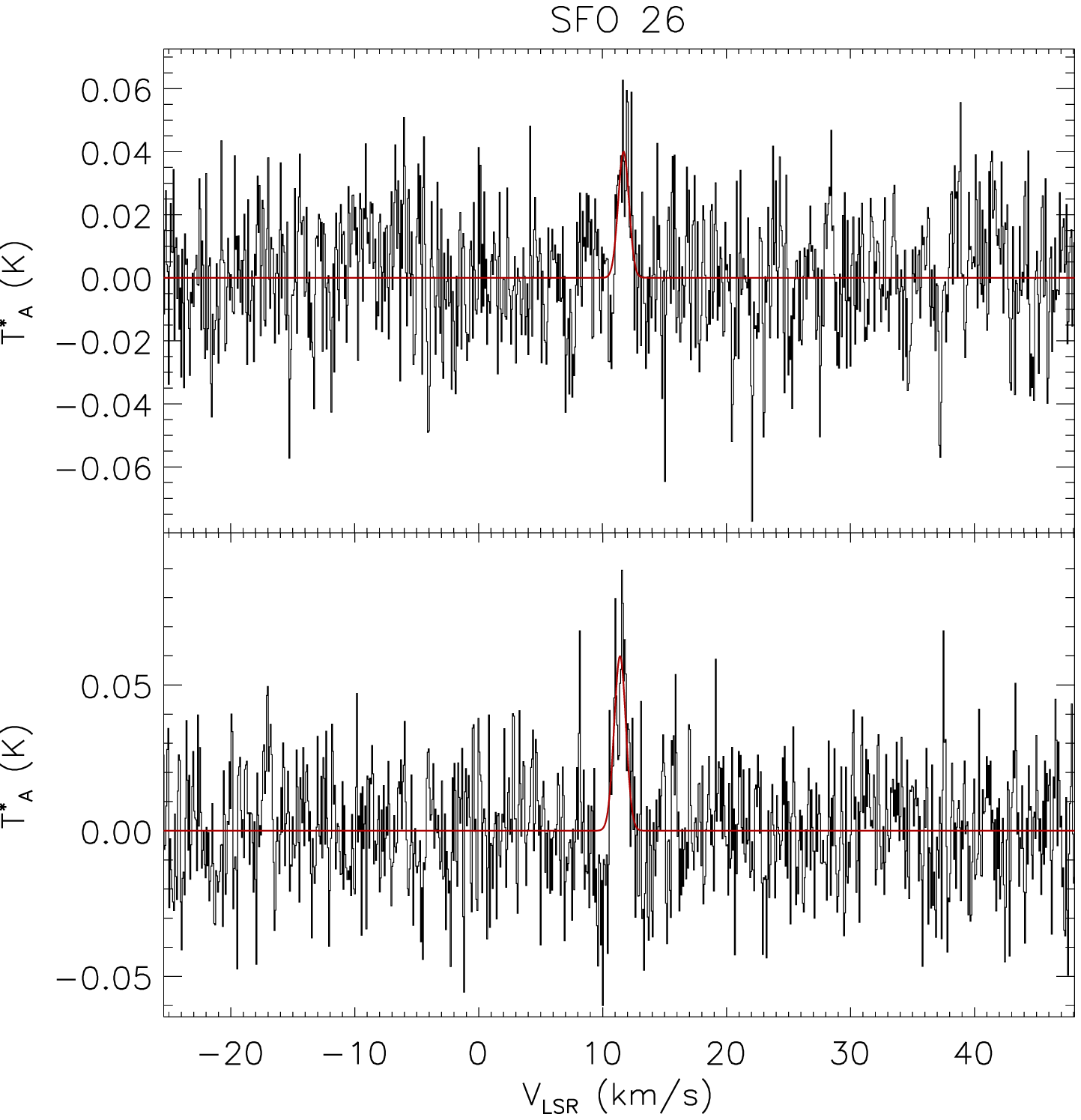}
\includegraphics*[width=0.24\textwidth]{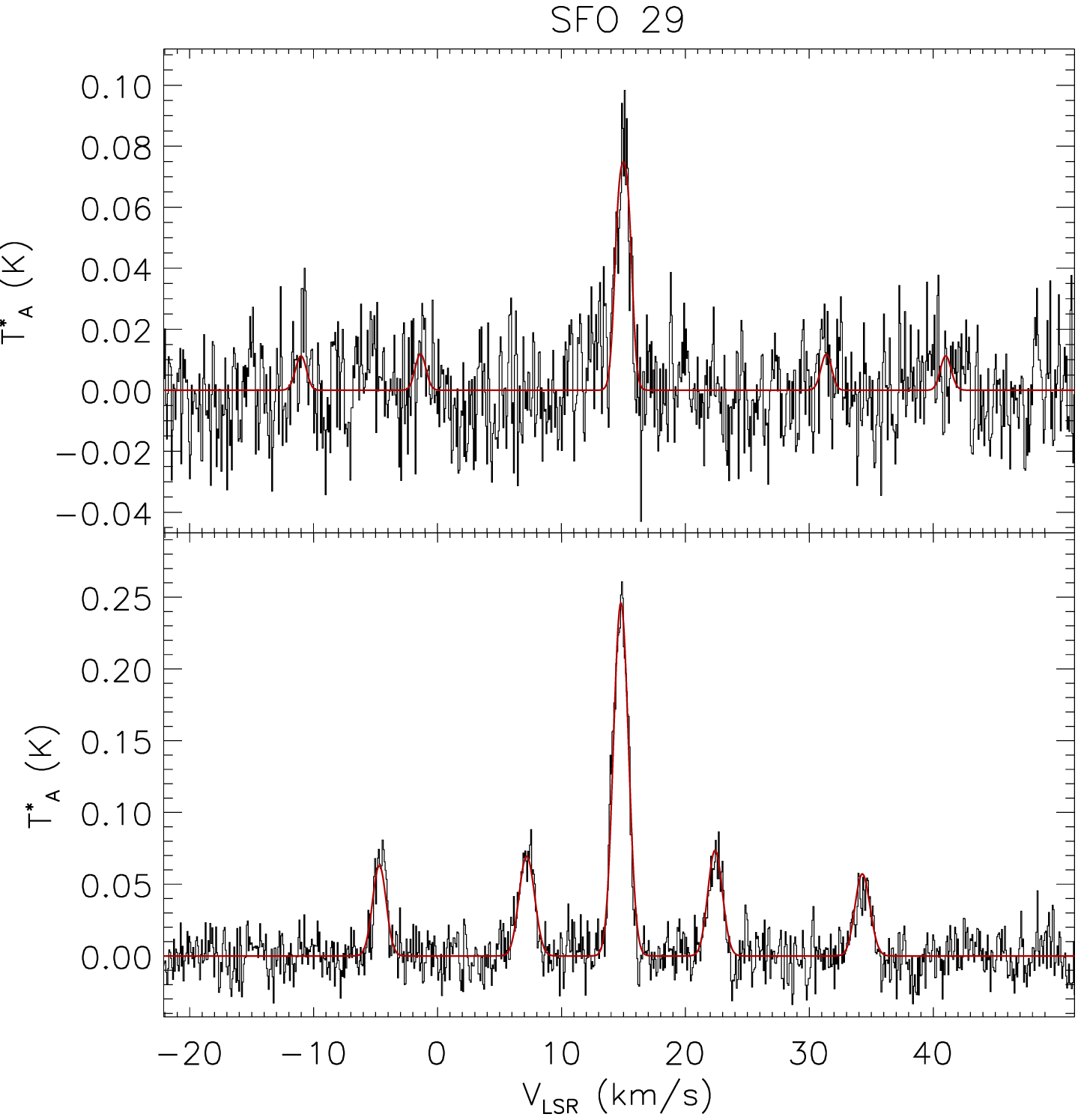}
\includegraphics*[width=0.24\textwidth]{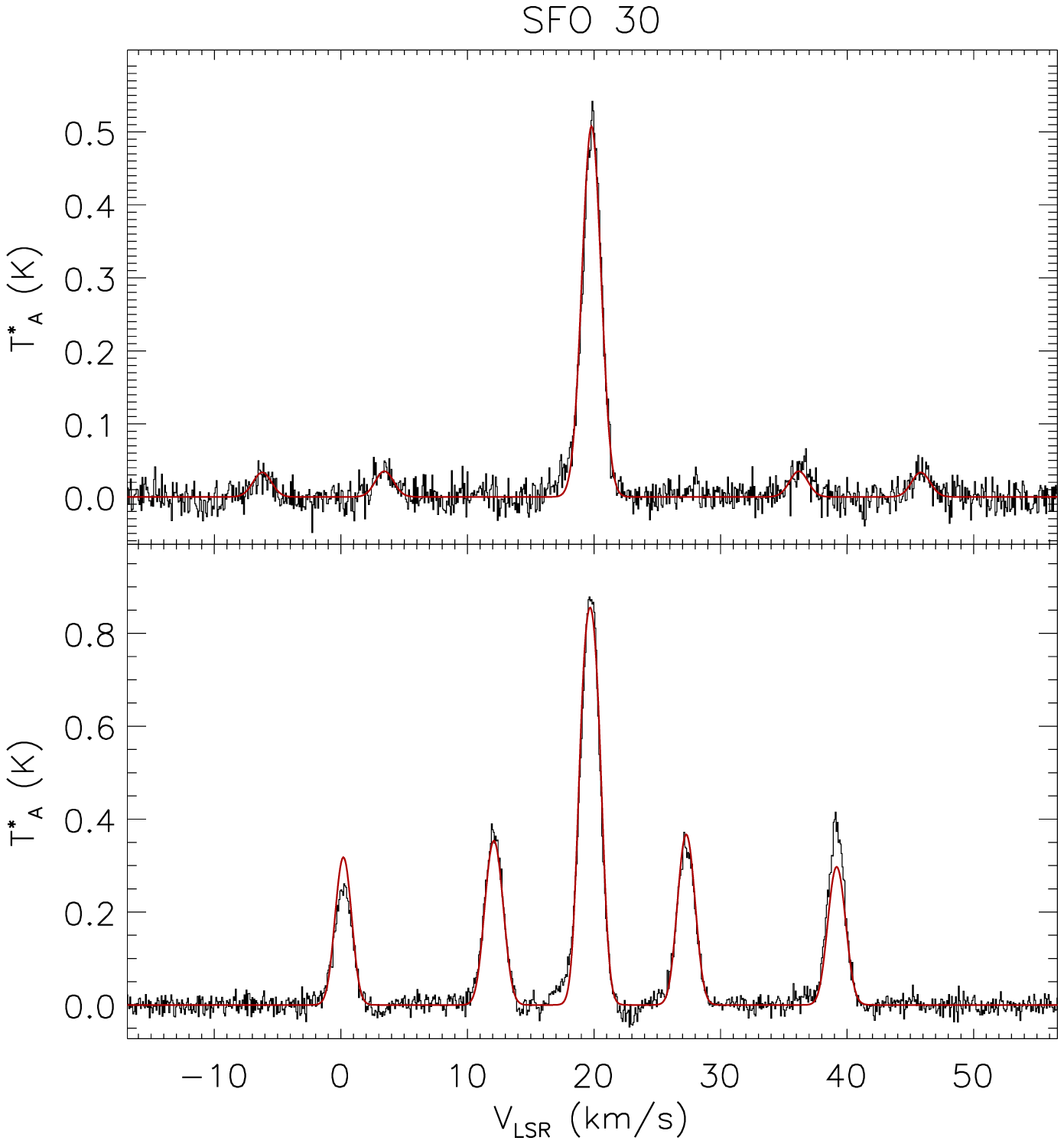}
\includegraphics*[width=0.24\textwidth]{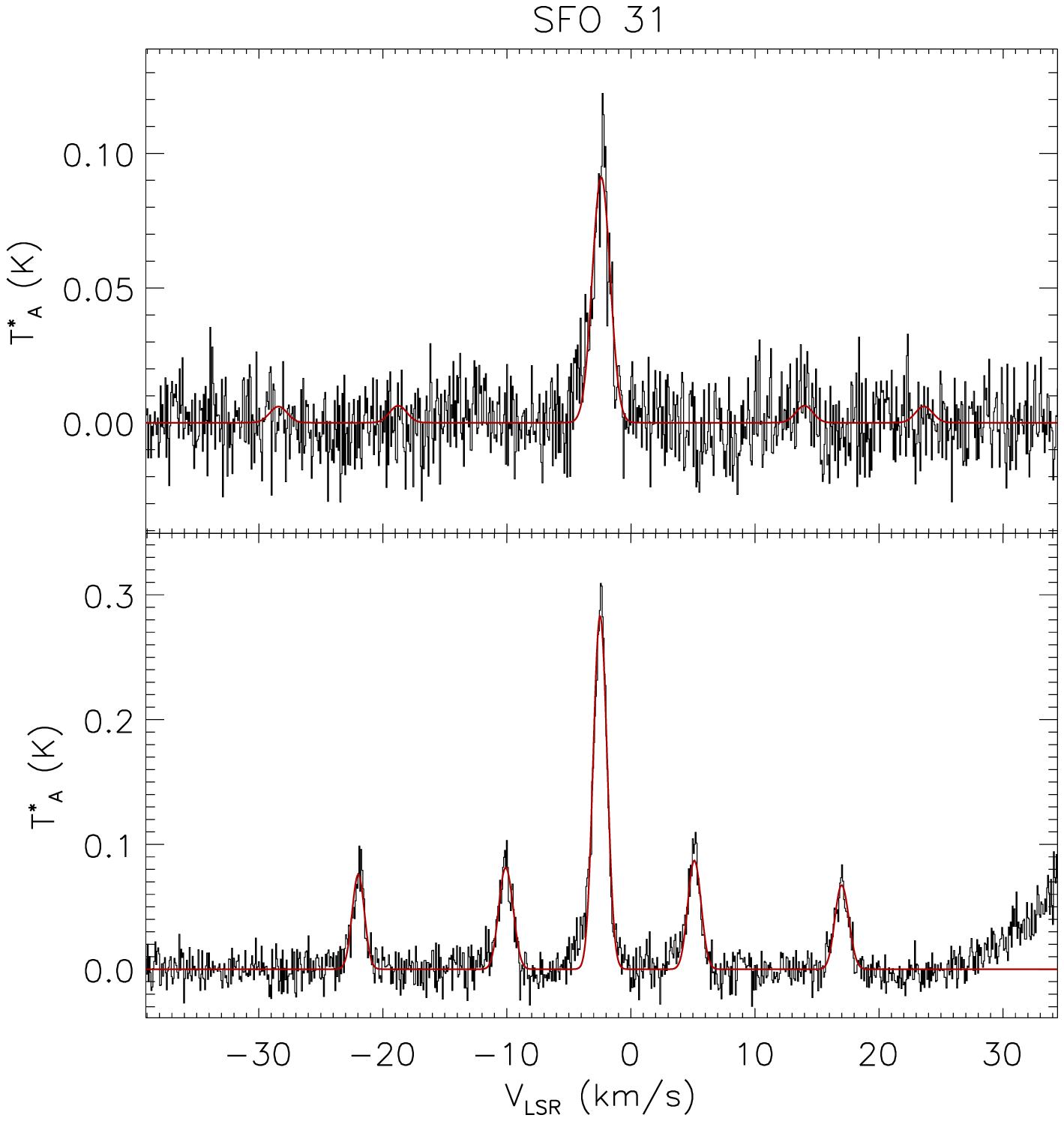}\\
\end{center}
\caption{\label{img:Spectra} Spectral lines are presented with corrected antenna temperature, \tastar, plotted against doppler shifted velocity, \Vlsr. Multiple lines are plotted on the same axis range with the (1,1) transition spectra on the bottom and (2,2) spectra on top. Fitted profiles are overlaid in red, determien.}
\end{figure*}
\end{center}

\addtocounter{figure}{-1}

\begin{center}
\begin{figure*}
\begin{center}
\includegraphics*[width=0.24\textwidth]{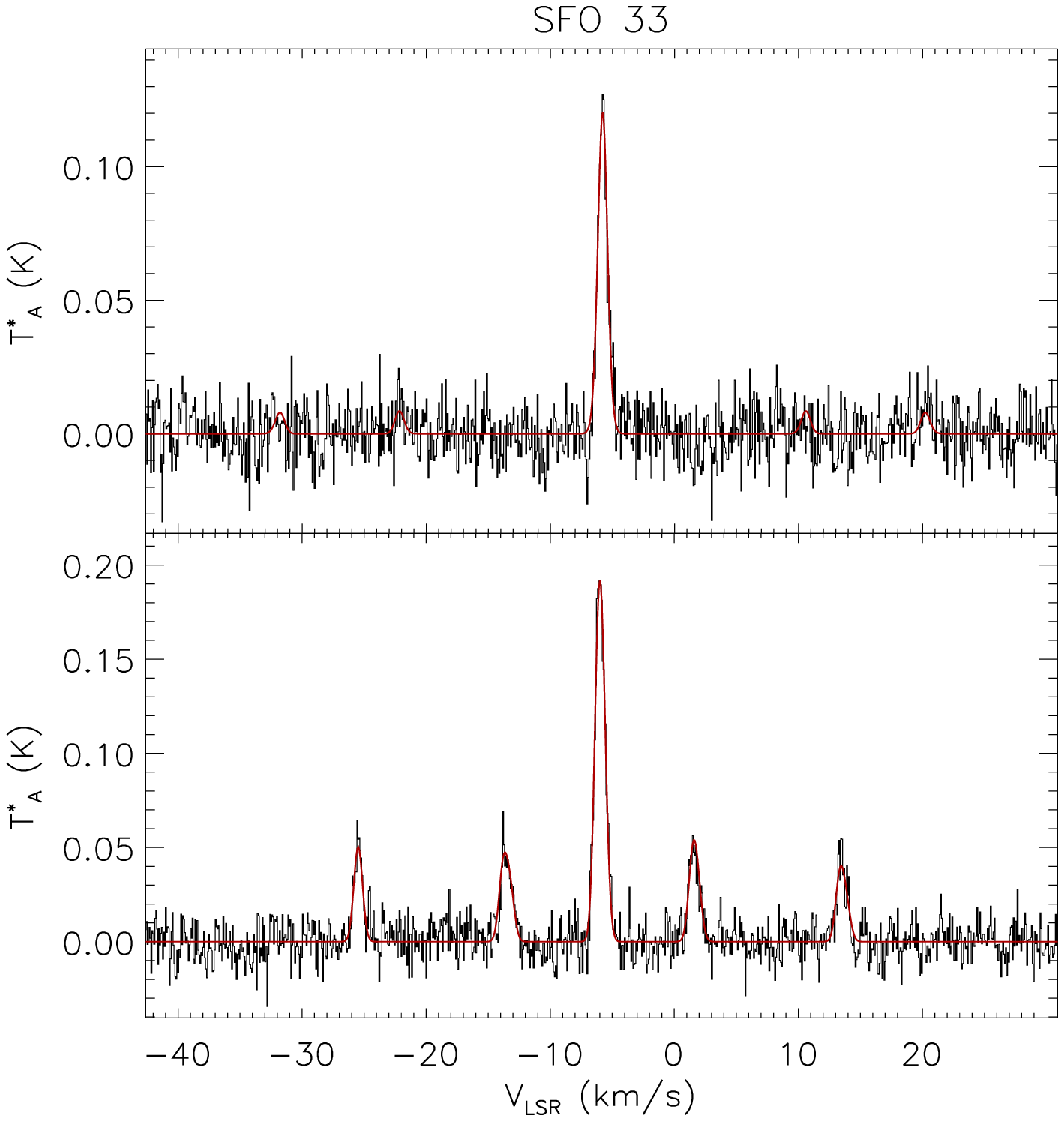}
\includegraphics*[width=0.24\textwidth]{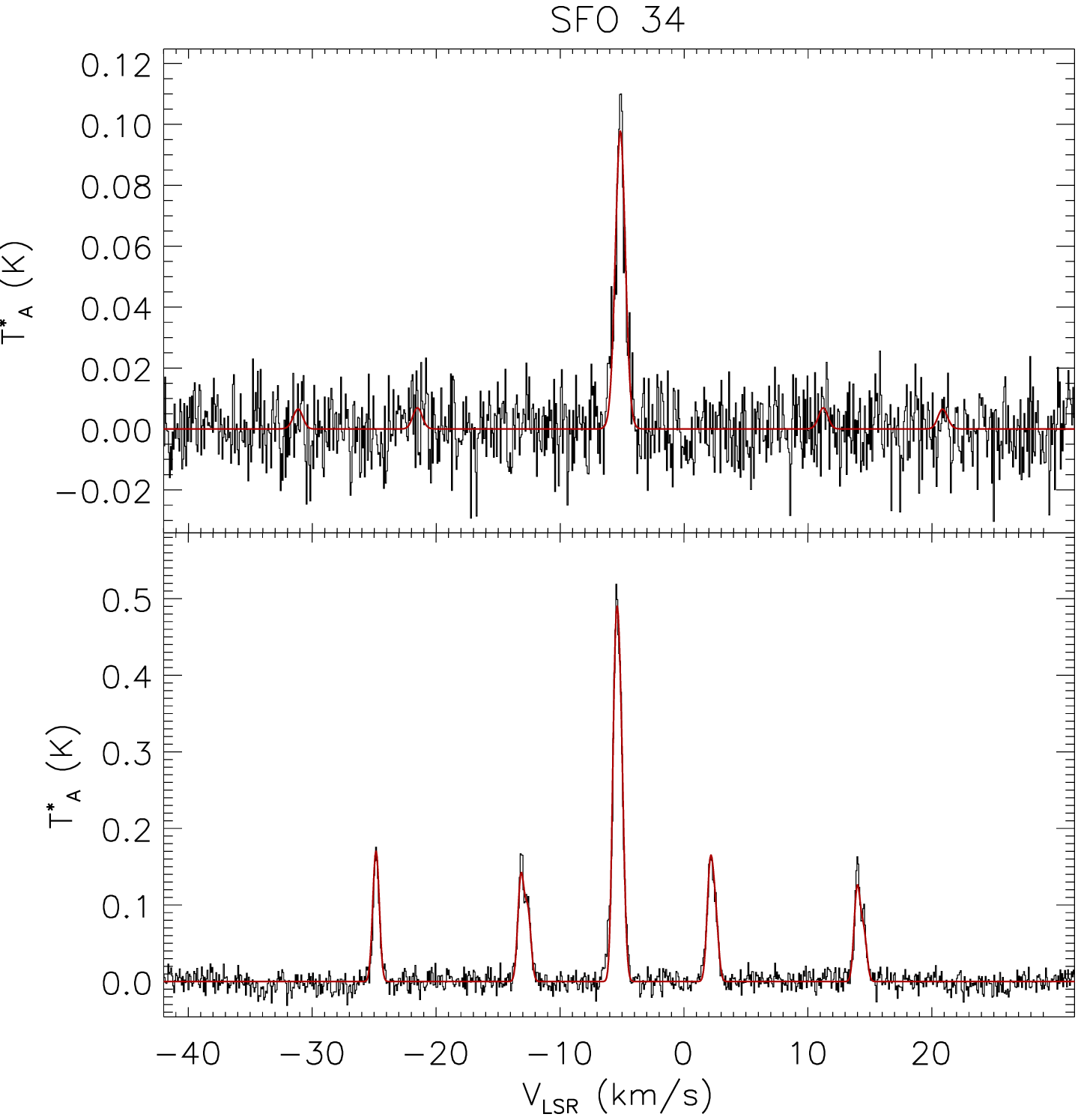}
\includegraphics*[width=0.24\textwidth]{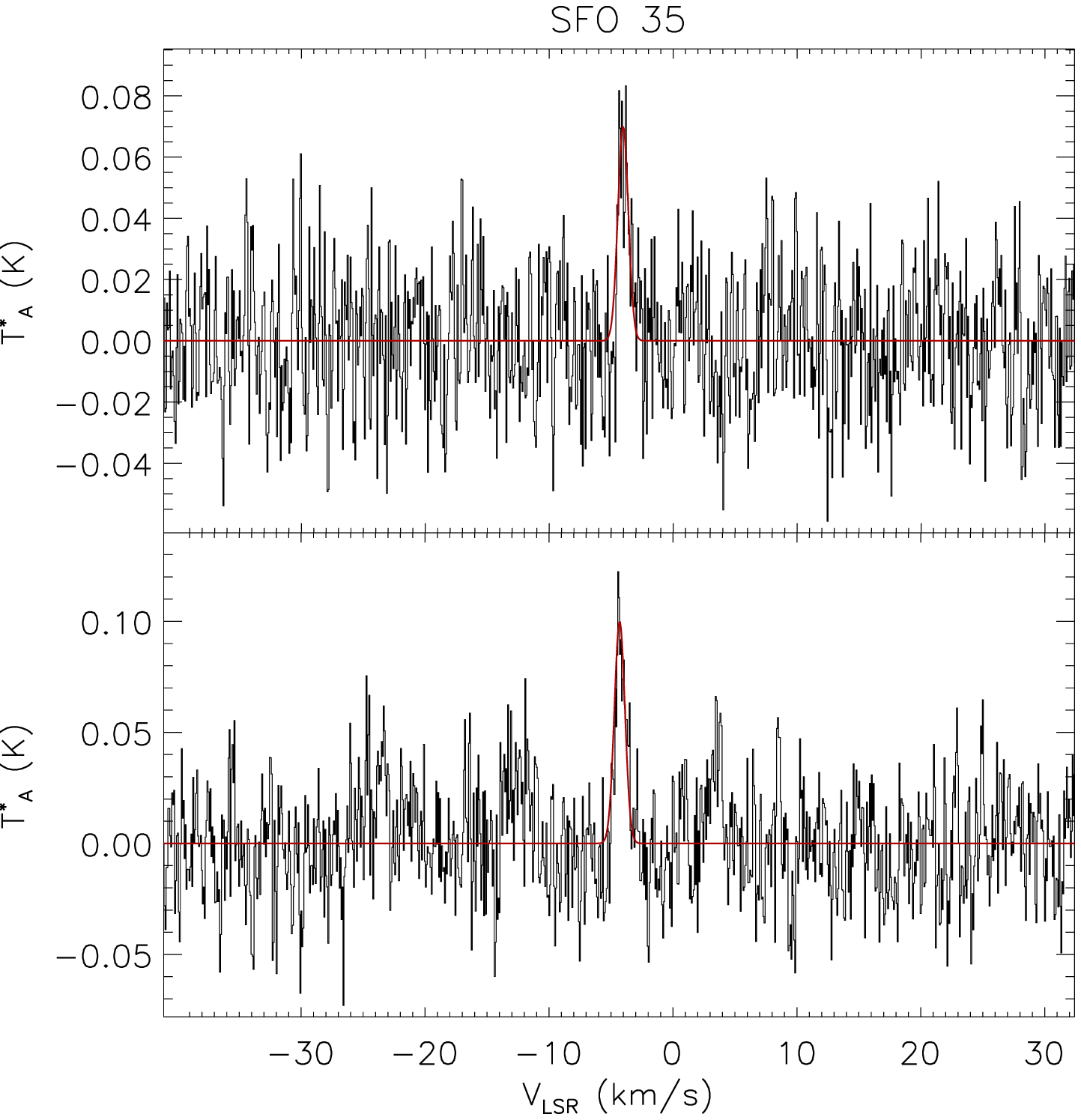}
\includegraphics*[width=0.24\textwidth]{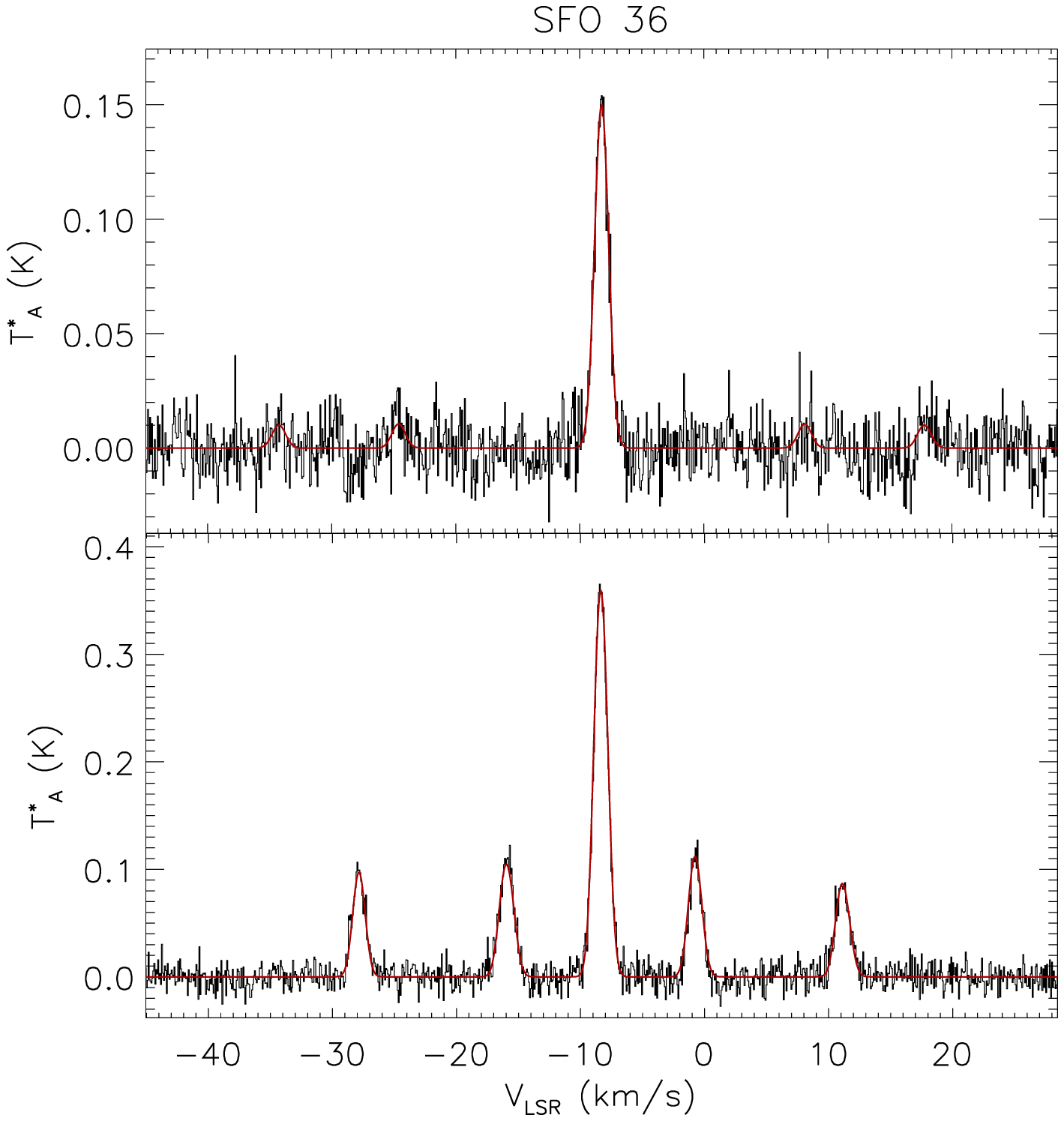}\\
\includegraphics*[width=0.24\textwidth]{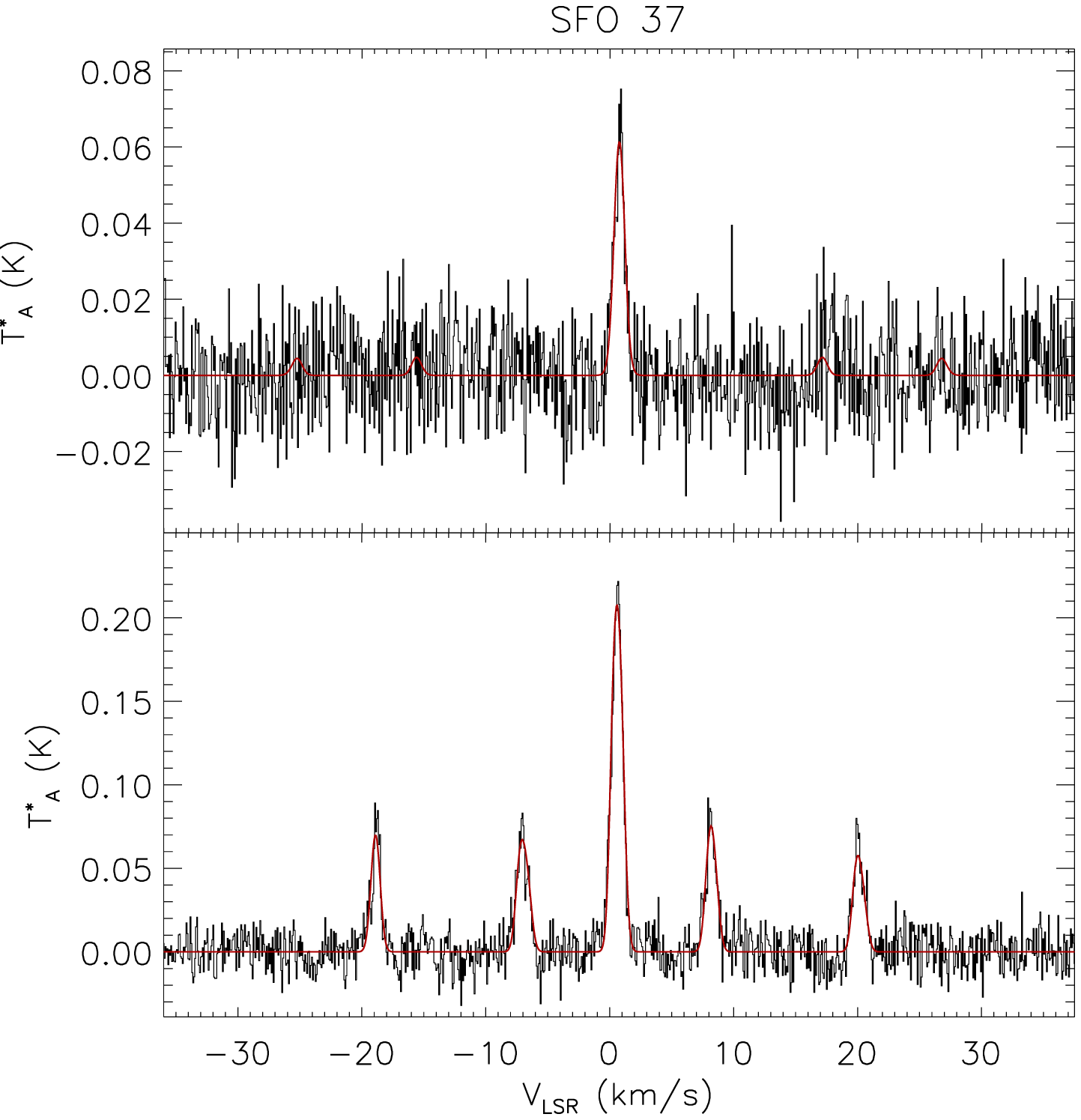}
\includegraphics*[width=0.24\textwidth]{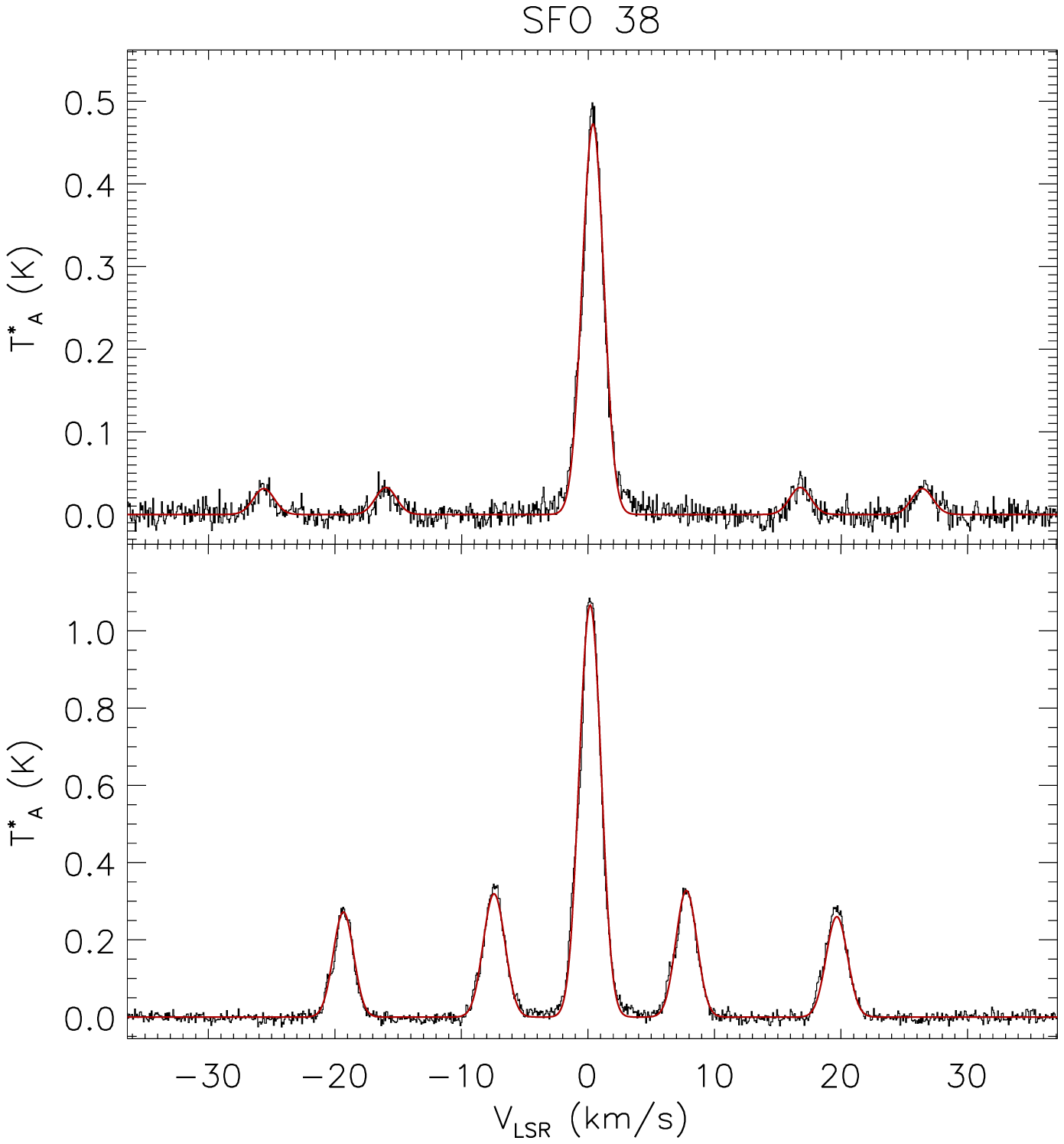}
\includegraphics*[width=0.24\textwidth]{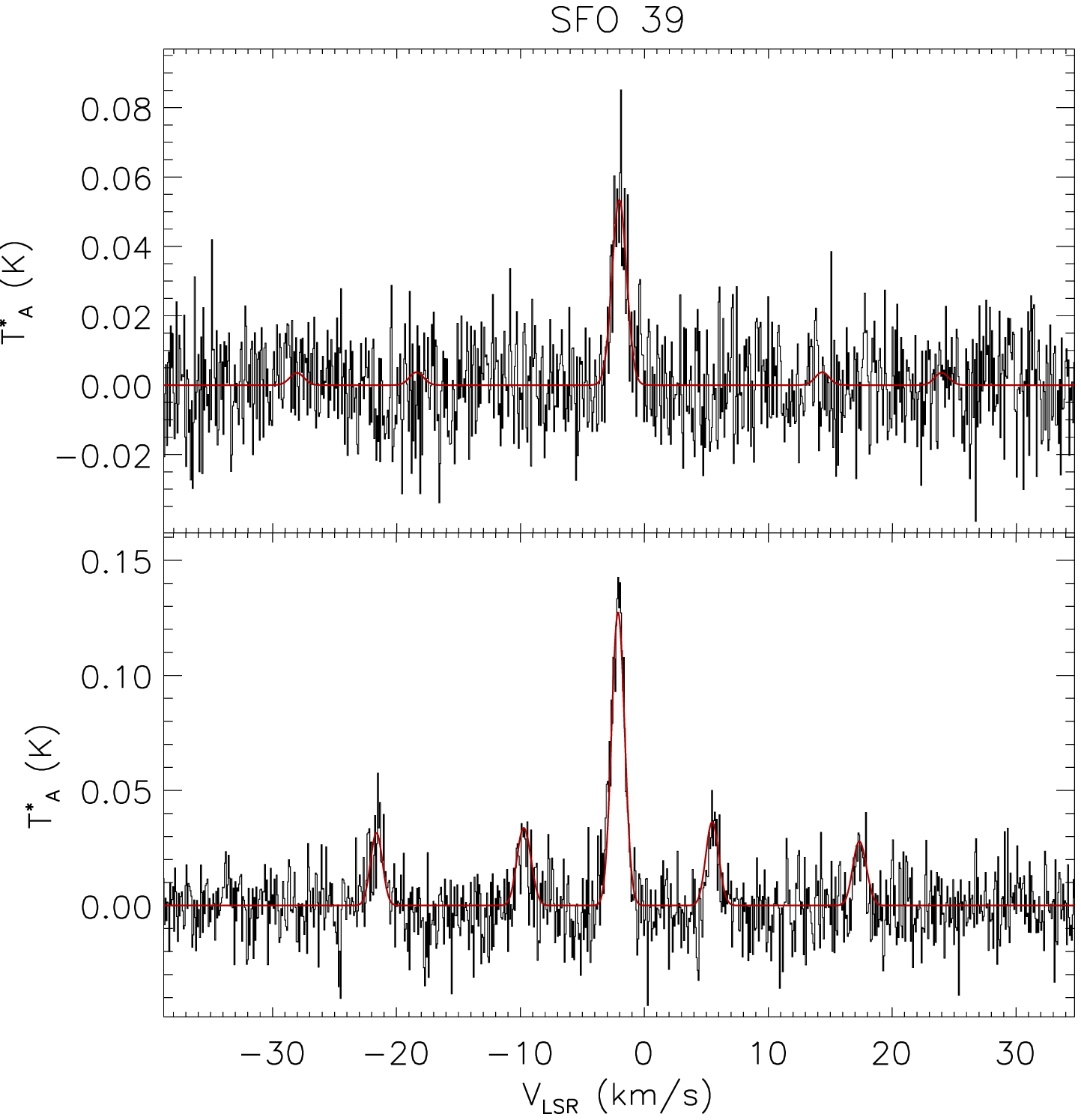}
\includegraphics*[width=0.24\textwidth]{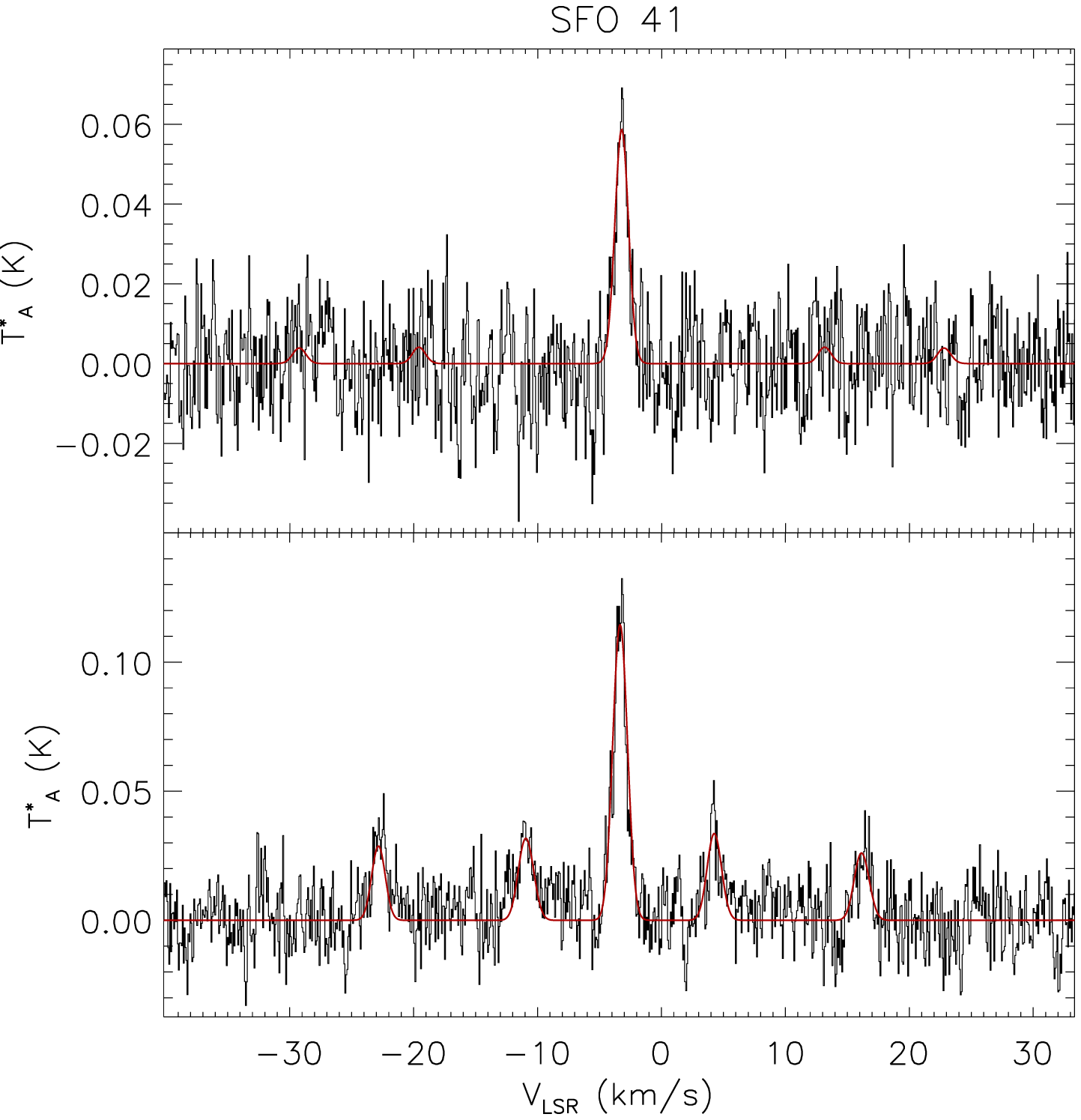}\\
\includegraphics*[width=0.24\textwidth]{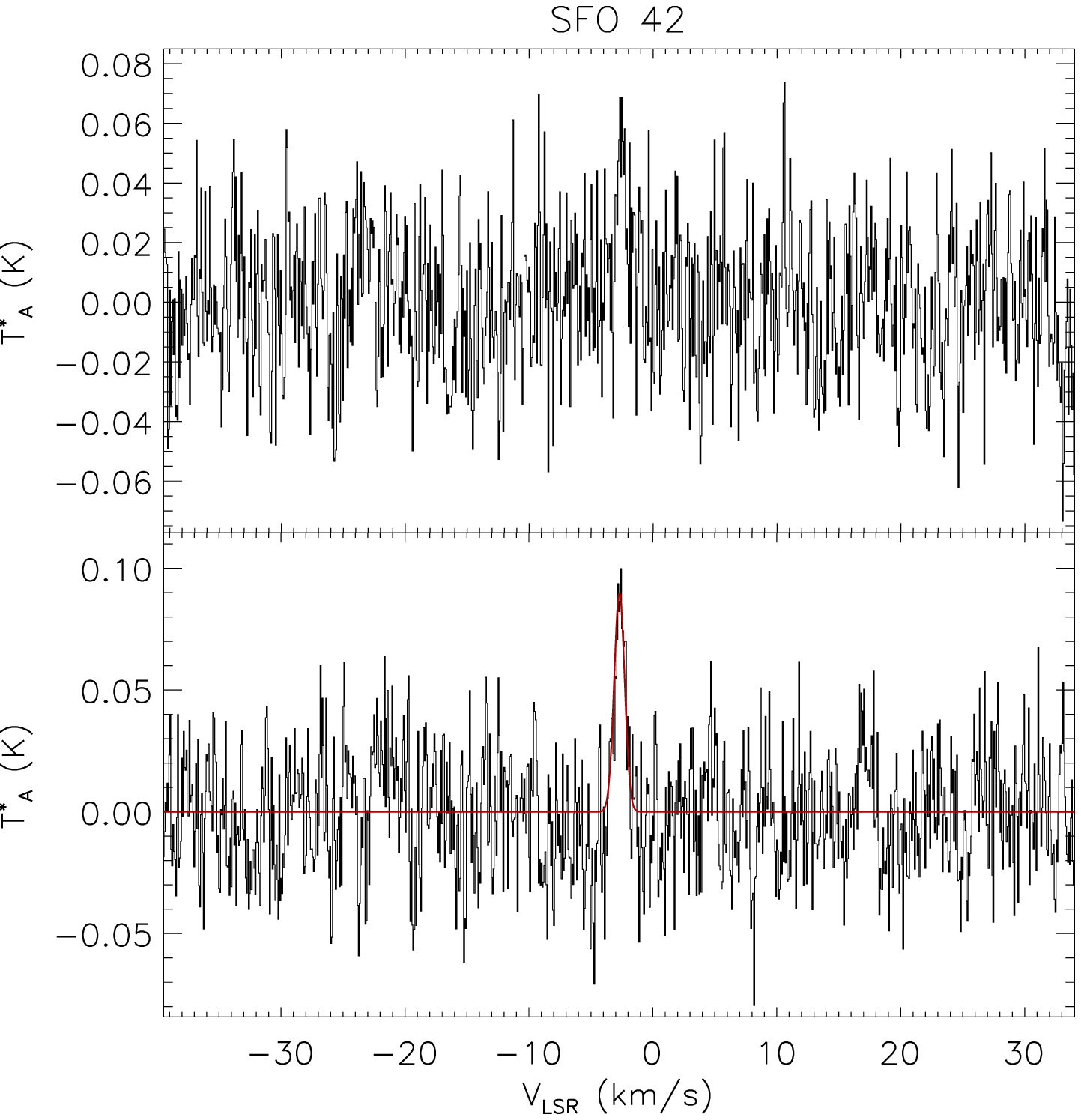}
\includegraphics*[width=0.24\textwidth]{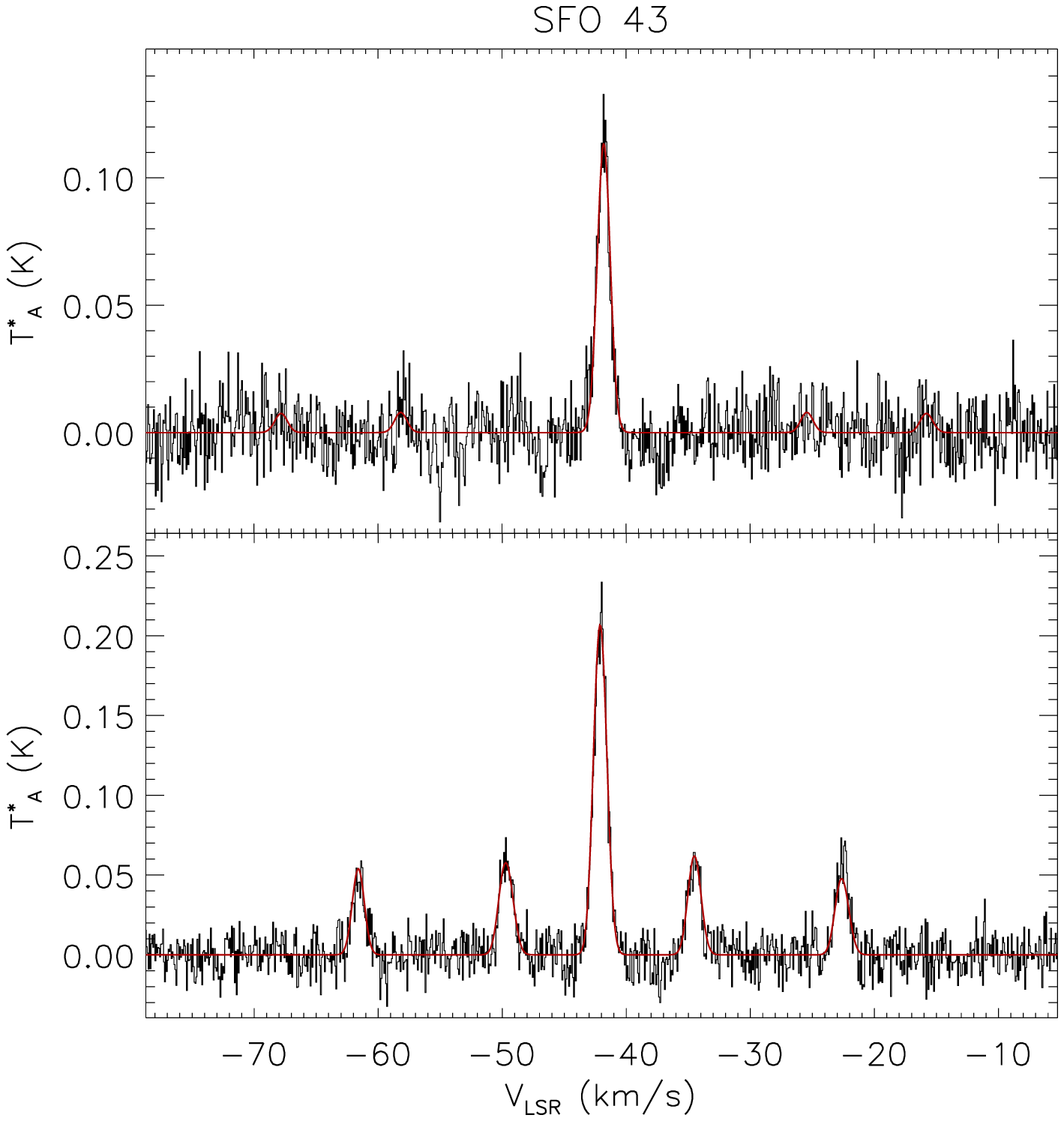}
\includegraphics*[width=0.24\textwidth]{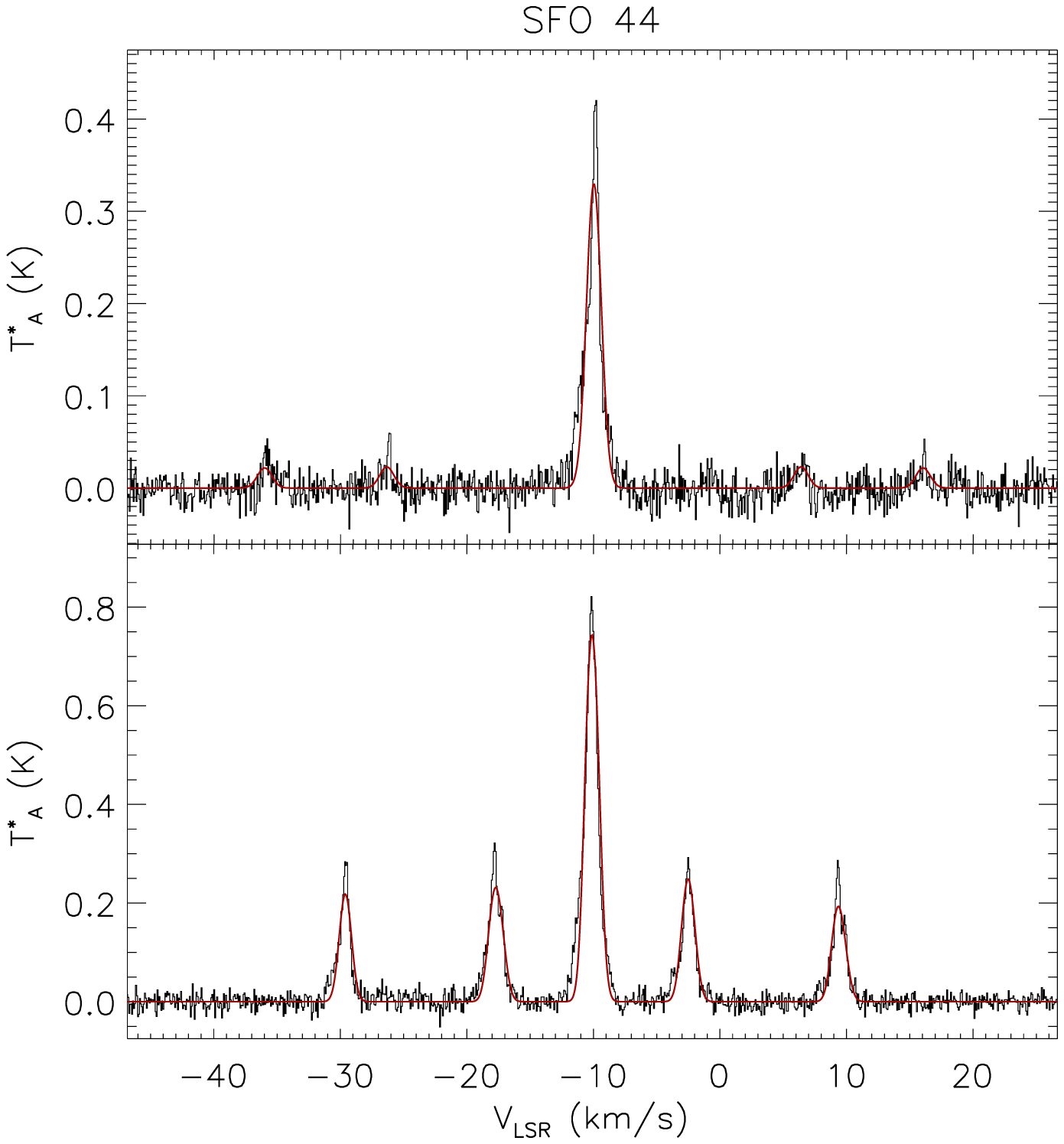}
\end{center}
\caption{(cont.) Spectral lines are presented with corrected antenna temperature, \tastar, plotted against doppler shifted velocity, \Vlsr. Multiple lines are plotted on the same axis range with the (1,1) transition spectra on the bottom and (2,2) spectra on top.}
\end{figure*}
\end{center}

\appendix
\section{(3,3) and (4,4) Detections}
\label{ap}
\begin{center}
\begin{figure*}
\begin{center}
\includegraphics*[width=0.24\textwidth]{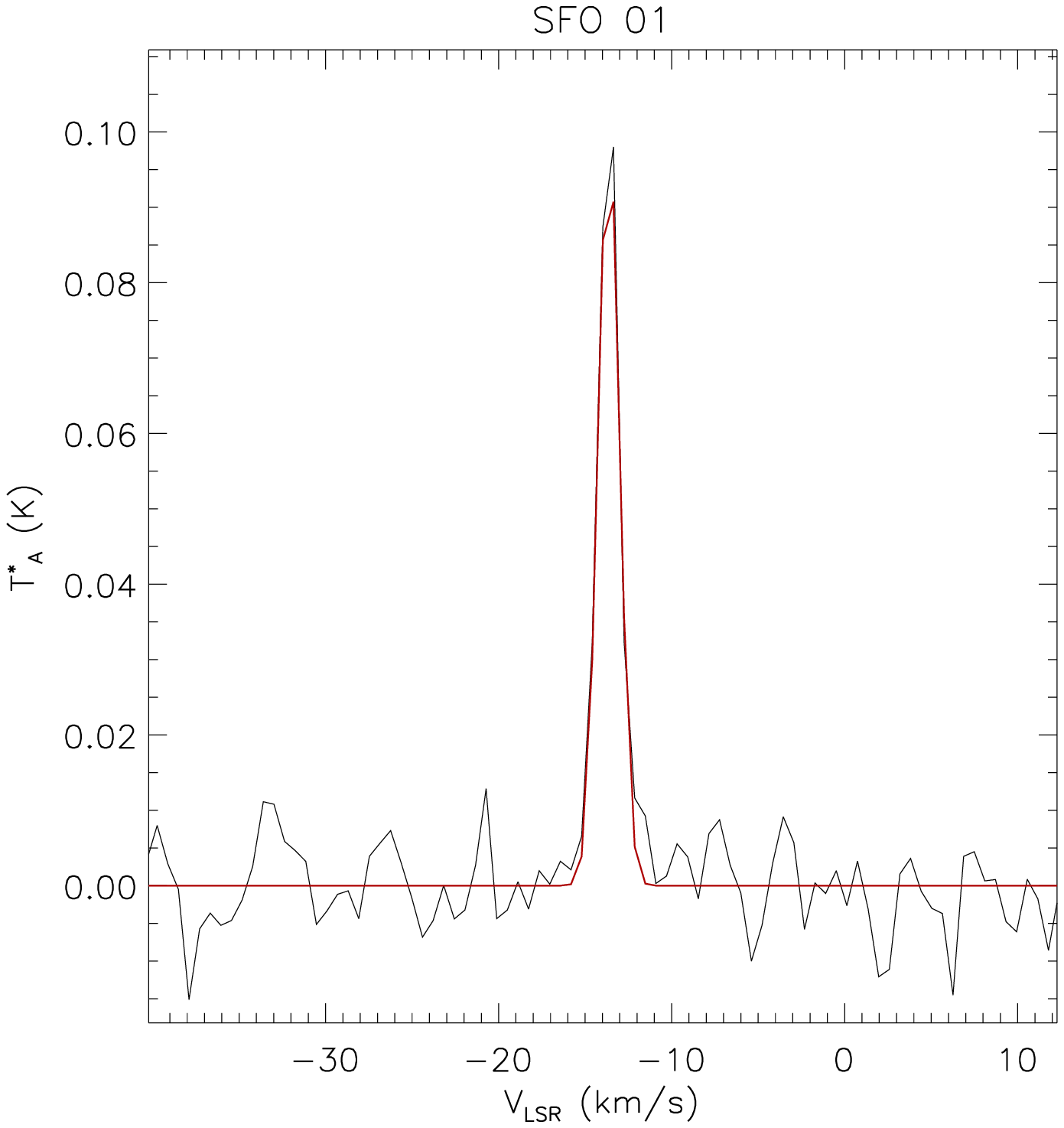}
\includegraphics*[width=0.24\textwidth]{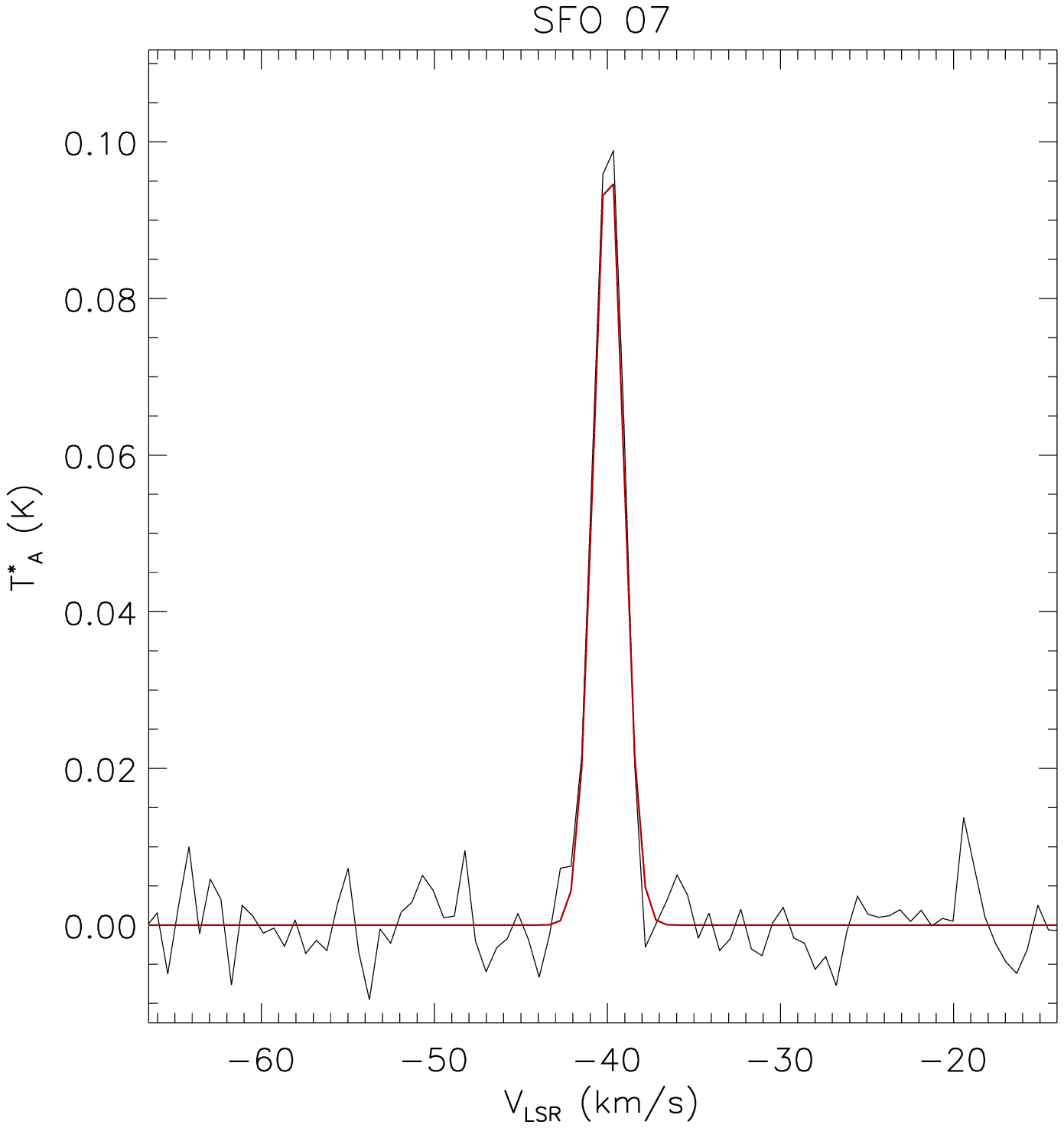}
\includegraphics*[width=0.24\textwidth]{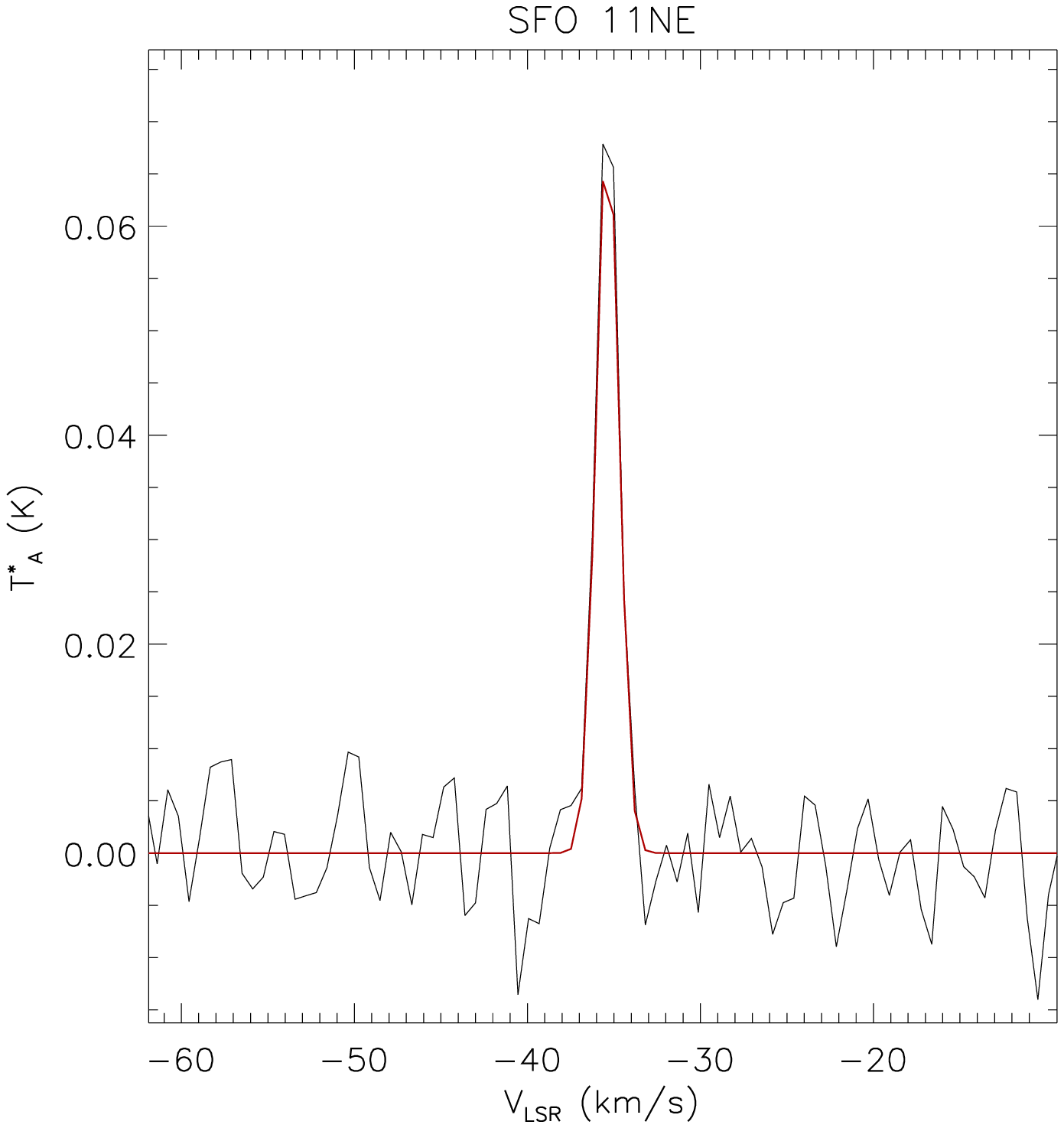}
\includegraphics*[width=0.24\textwidth]{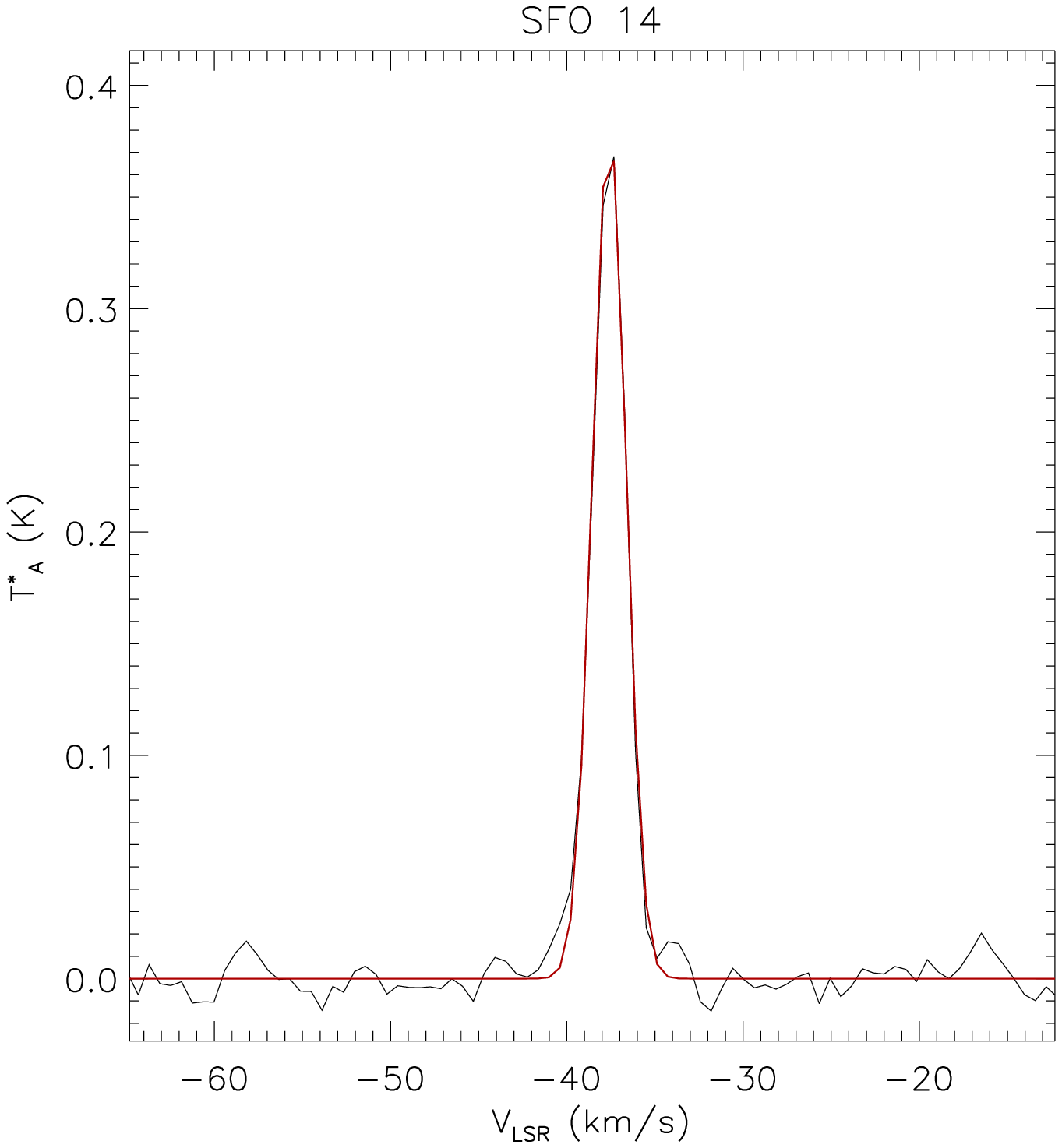}\\
\includegraphics*[width=0.24\textwidth]{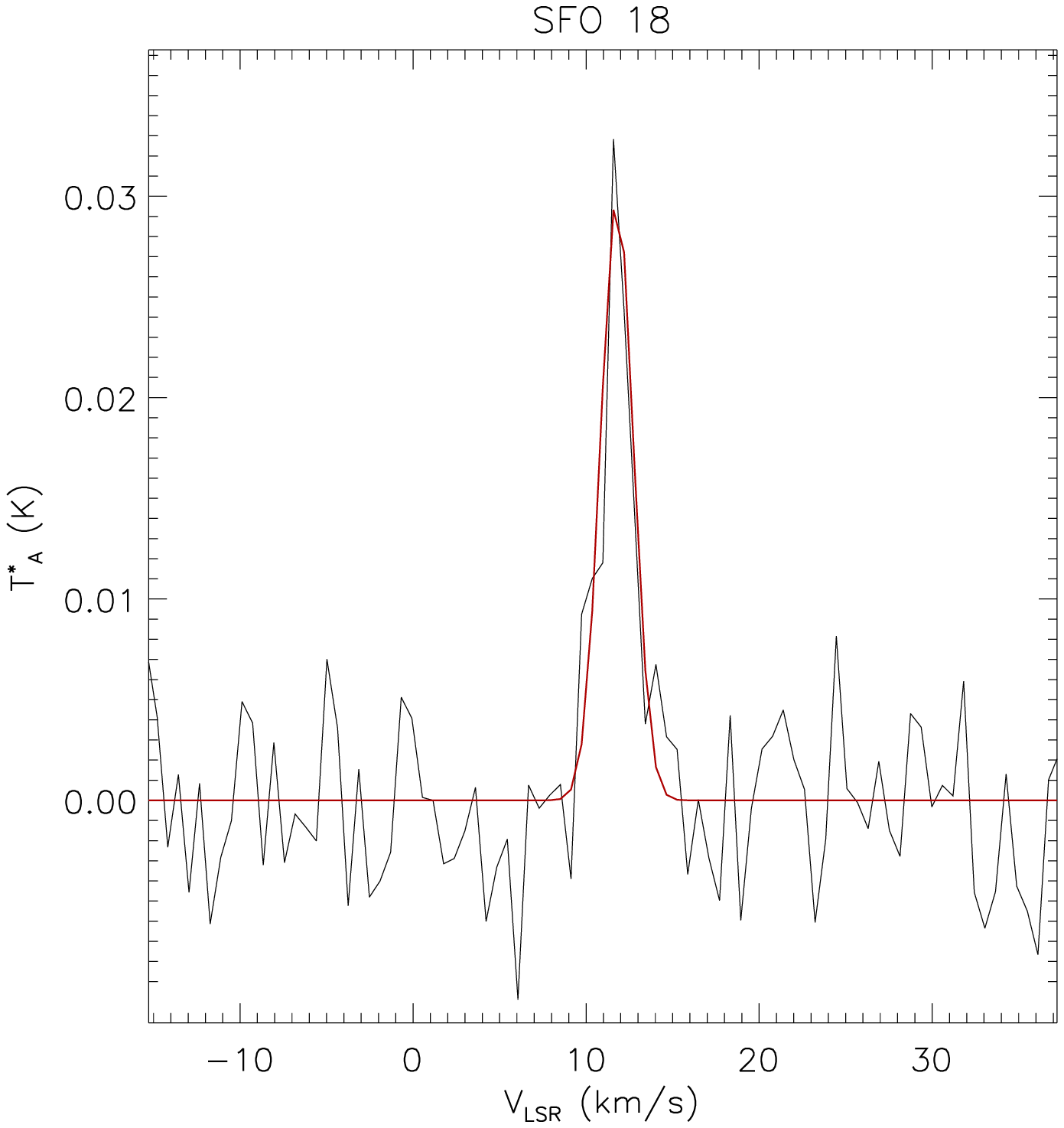}
\includegraphics*[width=0.24\textwidth]{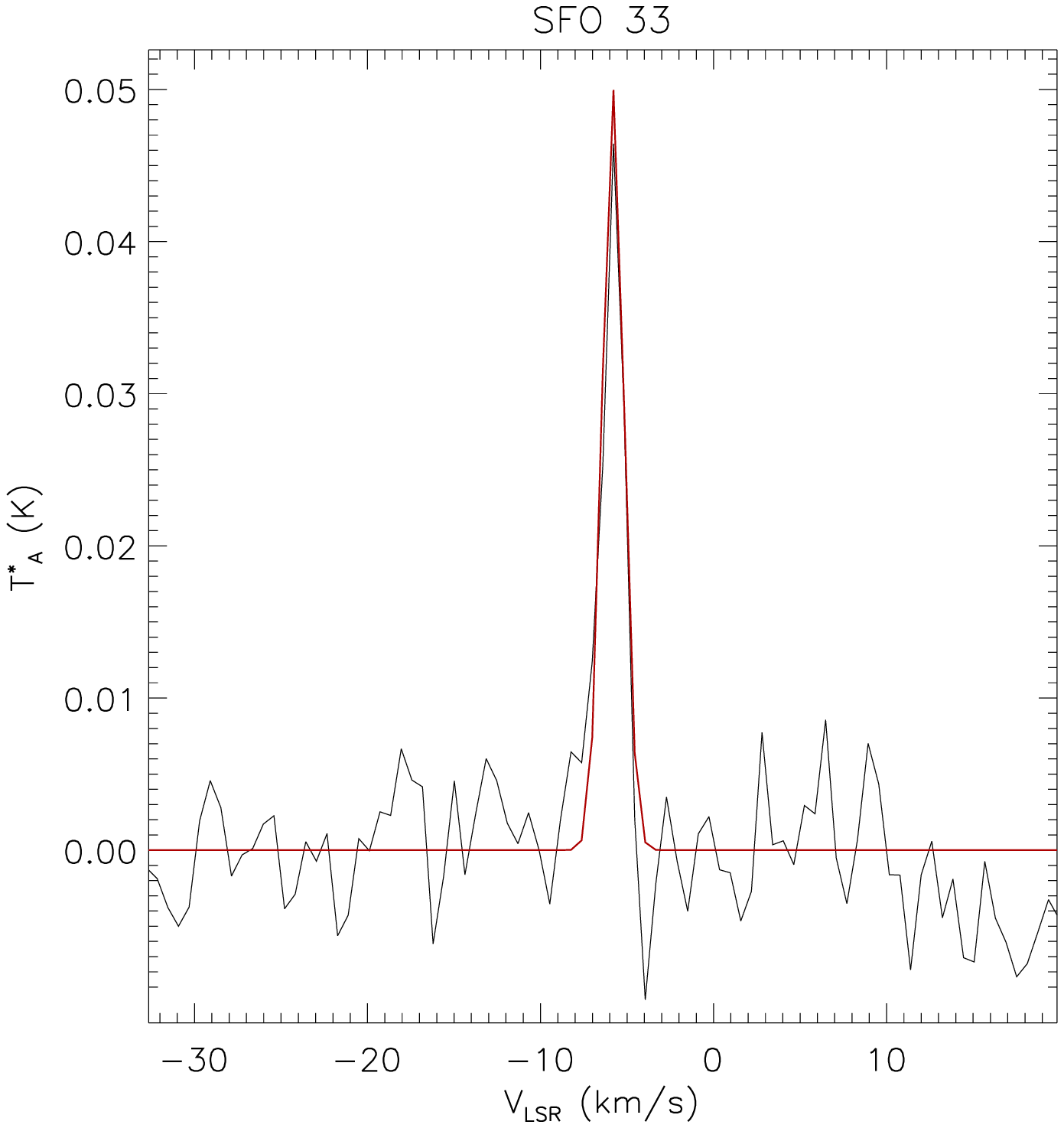}
\includegraphics*[width=0.24\textwidth]{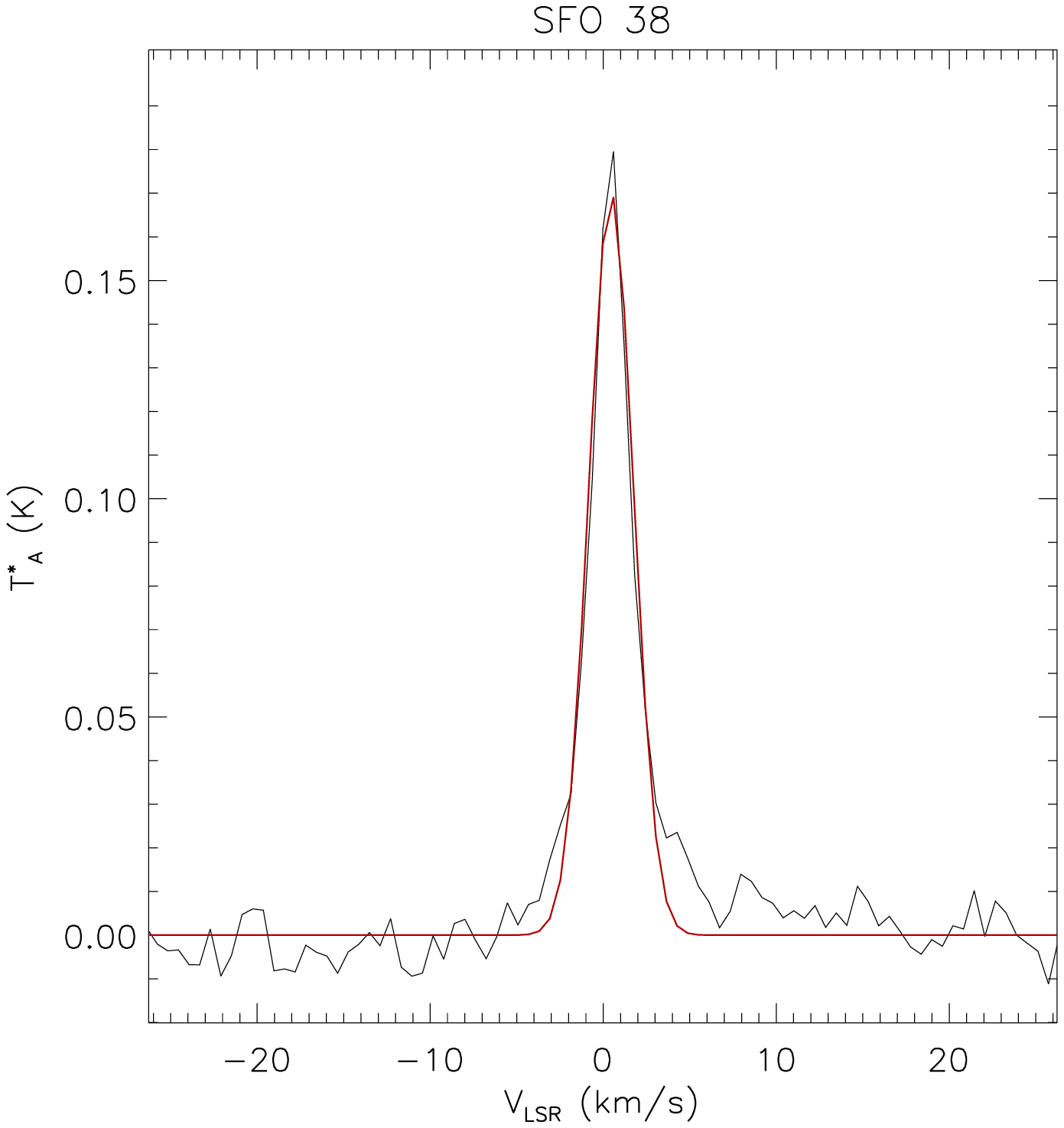}
\includegraphics*[width=0.24\textwidth]{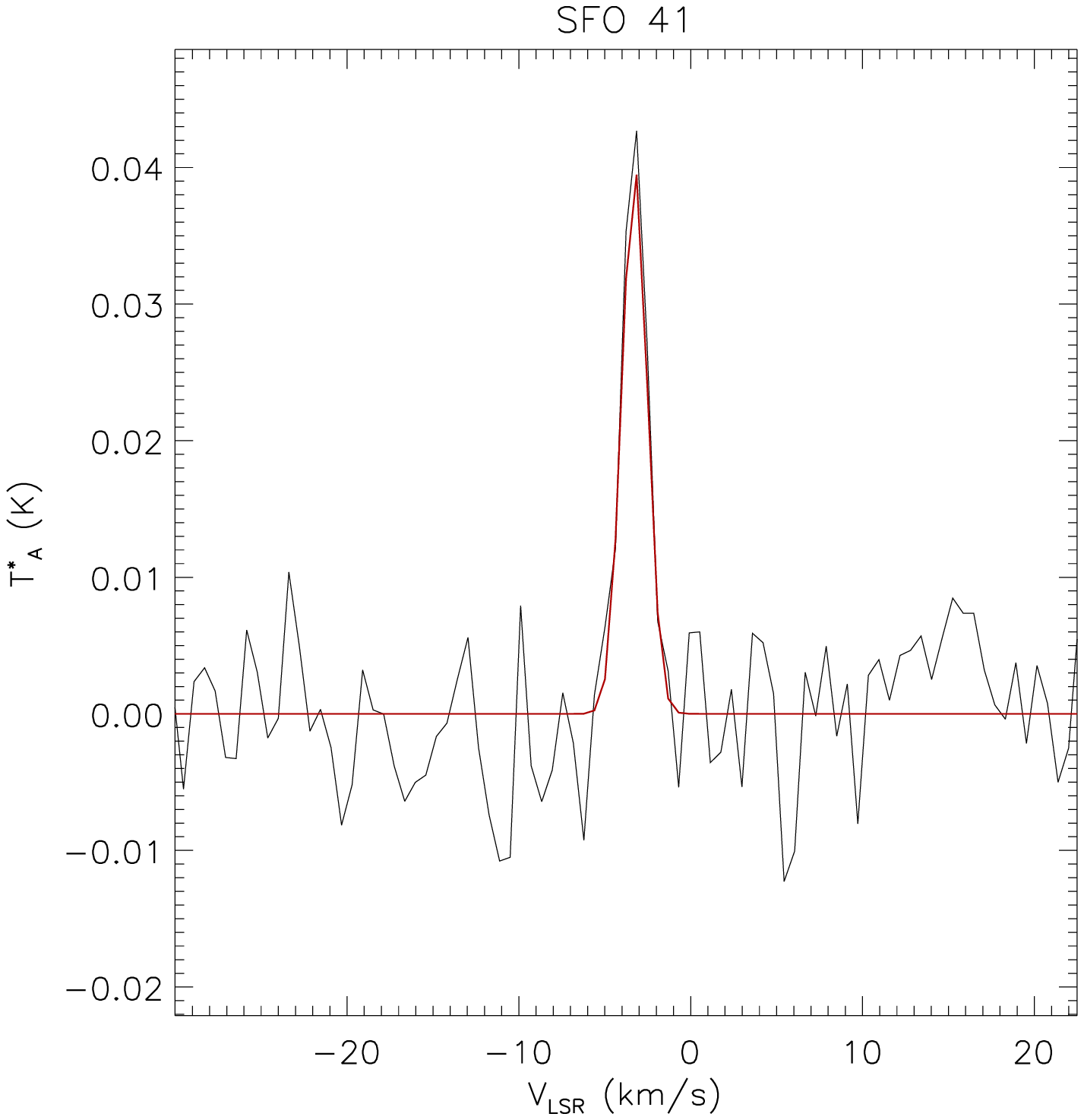}\\
\end{center}
\caption{\label{img:33Spectra} \nh\ (3,3) transition spectral lines. \tastar, plotted against doppler shifted velocity, \Vlsr, with GBTIDL fitted Gaussian profiles overlaid in red.}
\end{figure*}
\end{center}

\begin{center}
\begin{figure*}
\begin{center}
\includegraphics*[width=0.24\textwidth]{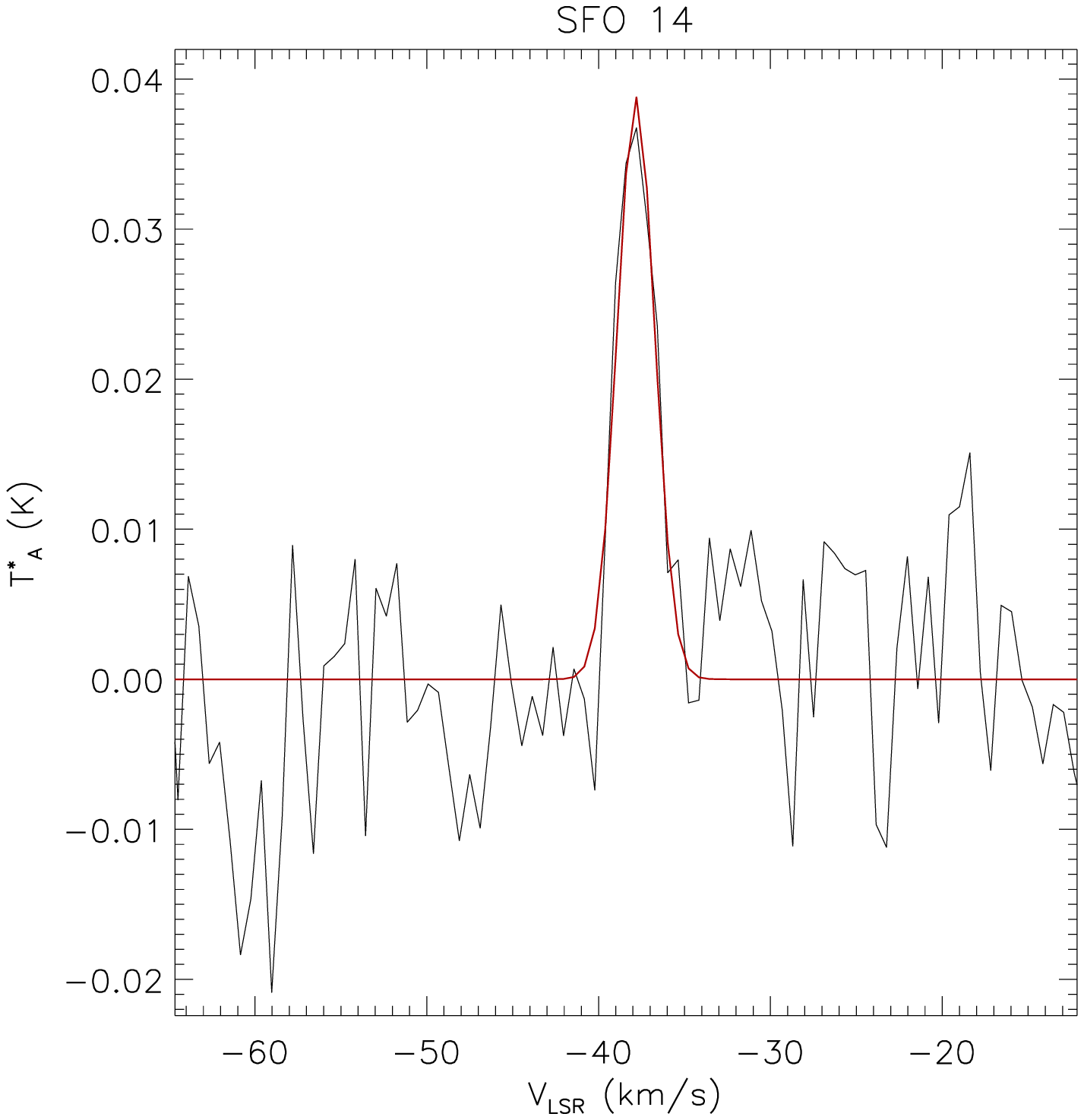}\\
\end{center}
\caption{\label{img:44Spectra} \nh\ (4,4) transition spectral line observed towards SFO 14. \tastar, plotted against doppler shifted velocity, \Vlsr, with GBTIDL fitted Gaussian profiles overlaid in red.}
\end{figure*}
\end{center}

Several sources were detected in the (3,3) and (4,4) rotational transitions of ammonia. Spectra of these sources are presented in Figs.~\ref{img:33Spectra} and \ref{img:44Spectra} and relevant parameters are listed in Table~\ref{tbl:3344}, RMS values of these observations are presented in Table~\ref{tbl:non-detections3344}.\\

Those sources detected in the (3,3) line of ammonia were examined for signs of masing. If masing is occuring in this transition then the brightness temperature of that source would be larger than the brightness temperature in the (1,1) transition and comparable to the (1,1) kinetic temperature (e.g. \citealt{Kuiper1995}). However, none of our sources show such symptoms. The (4,4) line of ammonia may often be observed toward warmer cores (e.g. \citealt{Longmore2007}), only one of our sources provided a detection in this line, SFO 14. This core is indeed one of the warmer sources in our sample at \Tk~$\sim$26K.  

\begin{table}
\caption{\label{tbl:3344} Observationally measured parameters of (3,3) and (4,4) detections.}
\begin{tabular}{lcc} 
\hline\hline
{}	 & $T_{\mathrm{A}}^{*}$   & {$\Delta v$ } \\
{Source} & (K) & {(km~s$^{-1}$)}                  \\
\hline
\multicolumn{3}{c}{\textbf{(3,3) Sources}} \\
\hline														      
SFO 01		& 0.10$\pm$0.006 & 1.45 \\
SFO 07		& 0.10$\pm$0.004 & 2.04 \\
SFO 11NE	& 0.07$\pm$0.005 & 1.55 \\
SFO 14  	& 0.38$\pm$0.007 & 2.24 \\
SFO 18  	& 0.03$\pm$0.004 & 2.20 \\
SFO 33  	& 0.05$\pm$0.003 & 1.45 \\
SFO 38  	& 0.17$\pm$0.005 & 3.03 \\
SFO 41  	& 0.04$\pm$0.005 & 1.73 \\
\hline
\multicolumn{3}{c}{\textbf{(4,4) Sources}} \\
\hline														      
SFO 14  	& 0.04$\pm$0.008 & 2.56  \\
\hline\\
\end{tabular}
\end{table}	

\begin{table}
\begin{center}
\caption{\label{tbl:non-detections3344}RMS values of our (3,3) and (4,4) transition observations.}
\begin{tabular}{lcc}
\hline
\hline
{Source} & (3,3) RMS (mK) & (4,4) RMS (mK) \\
\hline
SFO 01   & 13.3 & 4.7      \\
SFO 04   & 13.1 & $\cdots$ \\
SFO 05   & 12.0 & $\cdots$ \\
SFO 06   &  4.4 & 4.2      \\
SFO 07   & 14.8 & 4.6      \\
SFO 08   & 10.1 & 9.3      \\
SFO 09   & 14.3 & $\cdots$ \\
SFO 10   & 12.6 & $\cdots$ \\
SFO 11E  & 10.1 & 9.9      \\
SFO 11NE & 10.3 & 5.3      \\
SFO 12   &  4.4 & $\cdots$ \\
SFO 13   & 11.4 & $\cdots$ \\
SFO 15   & 13.7 & $\cdots$ \\
SFO 16   &  4.4 & 4.7      \\
SFO 17   & 11.0 & 11.7     \\
SFO 18   &  5.5 & 4.0      \\
SFO 19   & 16.0 & 14.1     \\
SFO 20   &  7.3 & 6.6      \\
SFO 21   &  9.6 & 8.5      \\
SFO 22   &  9.7 & 8.7      \\
SFO 23   &  4.5 & 3.9      \\
SFO 24   &  7.2 & $\cdots$ \\
SFO 25   &  5.7 & $\cdots$ \\
SFO 26   & 10.0 & 9.4      \\
SFO 28   &  7.9 & 8.2      \\
SFO 29   &  7.3 & 6.2      \\
SFO 30   &  7.9 & $\cdots$ \\
SFO 31   &  3.8 & $\cdots$ \\
SFO 33   &  6.7 & 5.0      \\
SFO 34   &  6.4 & 5.2      \\
SFO 35   &  9.4 & 9.8      \\
SFO 36   &  4.5 & $\cdots$ \\
SFO 37   &  5.2 & $\cdots$ \\
SFO 38   & 28.0 & 5.2      \\
SFO 39   &  4.4 & $\cdots$ \\
SFO 41   &  7.7 & 4.6      \\
SFO 42   & 13.4 & 10.5     \\
SFO 43   &  4.8 & $\cdots$ \\
SFO 44   &  5.4 & $\cdots$ \\
\hline\\
\end{tabular}\\
$\cdots$ Source not observed in this transition.
\end{center}
\end{table}	

\bsp

\label{lastpage}

\end{document}